\documentclass[a4paper,11pt]{article}
\usepackage[english]{babel}
\usepackage{jheppub}
\usepackage{graphicx}
\usepackage{cancel}
\usepackage{MnSymbol}
\usepackage{bbold}
\usepackage{amsmath,amssymb,amsfonts}
\usepackage{slashed}
\usepackage{caption}
\usepackage{appendix}
\usepackage{hyperref} 
\usepackage{multirow}
\hypersetup{colorlinks=true, linkcolor=black, urlcolor=blue}
\usepackage{comment}
\usepackage{float}
\usepackage{mathtools}

\usepackage{amstext} 
\usepackage{array}   
\newcolumntype{L}{>{$}l<{$}} 

\usepackage{booktabs}

\usepackage{geometry}
\usepackage{tikz}
\usetikzlibrary{calc}
\def\GridSize{3}

\newcommand*{\ColoredCells}[1]{
  \foreach \row/\column in {#1} {
    \node [fill=red, draw=none, thick, minimum size=1cm] 
      at (\column-.5,\GridSize+0.5-\row) {};
    }%
}%

\newcommand*{\Griddd}[1]{
	\begin{tikzpicture}[scale=1.0]
    \begin{scope}[thick,local bounding box=name]
        \ColoredCells{#1}
        \draw (0, 0) grid (\GridSize, \GridSize);
    \end{scope}
	\end{tikzpicture}
}

\newcommand*{\Gridd}[2]{
	\begin{tikzpicture}[scale=1.0]
    \begin{scope}[thick,local bounding box=name]
        \ColoredCells{#1}
        \draw (0, 0) grid (2, 3);
        \draw (2, 0) grid (3, 1);
        \draw (2, 2) grid (3, 3);
        \node at (+2.5,+1.5) {#2};
    \end{scope}
	\end{tikzpicture}
}

\newcommand*{\SubgroupDiagram}[3]{
	\scalebox{0.25}{
	    \begin{tabular}{lll}
	    \Gridd{#1}{{\LARGE L}} & \Gridd{#2}{{\LARGE R}} & \Griddd{#3}\\
	    \end{tabular}
	}
}

\newcommand*{\SubgroupDiagramDefinition}{
	\scalebox{0.8}{
	\begin{tabular}{lll}
	\begin{tikzpicture}[scale=1.0]
    \begin{scope}[thick,local bounding box=name]
        \draw (0, 0) grid (2, 3);
        \draw (2, 0) grid (3, 1);
        \draw (2, 2) grid (3, 3);
        \node at (+0.5,+2.5) {{\Large $t^{1}_L$}};
        \node at (+1.5,+2.5) {{\Large $t^{2}_L$}};
        \node at (+2.5,+2.5) {{\Large $t^{3}_L$}};
        \node at (+0.5,+1.5) {{\Large $t^{4}_L$}};
        \node at (+1.5,+1.5) {{\Large $t^{5}_L$}};
        \node at (+0.5,+0.5) {{\Large $t^{6}_L$}};
        \node at (+1.5,+0.5) {{\Large $t^{7}_L$}};
        \node at (+2.5,+0.5) {{\Large $t^{8}_L$}};
    \end{scope}
	\end{tikzpicture}
	&
	\begin{tikzpicture}[scale=1.0]
    \begin{scope}[thick,local bounding box=name]
        \draw (0, 0) grid (2, 3);
        \draw (2, 0) grid (3, 1);
        \draw (2, 2) grid (3, 3);
        \node at (+0.5,+2.5) {{\Large $t^{1}_R$}};
        \node at (+1.5,+2.5) {{\Large $t^{2}_R$}};
        \node at (+2.5,+2.5) {{\Large $t^{3}_R$}};
        \node at (+0.5,+1.5) {{\Large $t^{4}_R$}};
        \node at (+1.5,+1.5) {{\Large $t^{5}_R$}};
        \node at (+0.5,+0.5) {{\Large $t^{6}_R$}};
        \node at (+1.5,+0.5) {{\Large $t^{7}_R$}};
        \node at (+2.5,+0.5) {{\Large $t^{8}_R$}};
    \end{scope}
	\end{tikzpicture}
	&
	\begin{tikzpicture}
	\begin{scope}[thick,local bounding box=name]
		\draw (0,0) grid (3,3);
		\node at (+0.5,+2.5) {{\Large $t^{\alpha}{}_{11}$}};
		\node at (+1.5,+2.5) {{\Large $t^{\alpha}{}_{12}$}};
		\node at (+2.5,+2.5) {{\Large $t^{\alpha}{}_{13}$}};
		\node at (+0.5,+1.5) {{\Large $t^{\alpha}{}_{21}$}};
		\node at (+1.5,+1.5) {{\Large $t^{\alpha}{}_{22}$}};
		\node at (+2.5,+1.5) {{\Large $t^{\alpha}{}_{23}$}};
		\node at (+0.5,+0.5) {{\Large $t^{\alpha}{}_{31}$}};
		\node at (+1.5,+0.5) {{\Large $t^{\alpha}{}_{32}$}};
		\node at (+2.5,+0.5) {{\Large $t^{\alpha}{}_{33}$}};
	\end{scope}
	\end{tikzpicture}
	\end{tabular}
	}
}

\usepackage{wasysym}

\usepackage{color}
\definecolor{gray}{rgb}{0.5,0.5,0.5}

\def\Re{\mbox{Re}\,}

\def\SU{\mathrm{SU}}
\def\SO{\mathrm{SO}}
\def\UU{\mathrm{U}(1)}
\def\HSPACEPLOT{\hspace{0.3cm}}

\newcommand\lsim{\mathrel{\rlap{\lower4pt\hbox{\hskip1pt$\sim$}}
    \raise1pt\hbox{$<$}}}
\newcommand\gsim{\mathrel{\rlap{\lower4pt\hbox{\hskip1pt$\sim$}}
    \raise1pt\hbox{$>$}}}

\newcommand{\beq}{\begin{equation}}
\newcommand{\eeq}{\end{equation}}
\newcommand{\bea}{\begin{eqnarray}}
\newcommand{\eea}{\end{eqnarray}}
\newcommand{\bem}{\begin{pmatrix}}
\newcommand{\eem}{\end{pmatrix}}

\newcommand{\ben}{\begin{enumerate}}
\newcommand{\een}{\end{enumerate}}

\def\RED{\textcolor{red}}
\def\BLUE{\textcolor{blue}}

\usepackage{comment}

\definecolor{myGreen}{rgb}{0.39, 0.65, 0.46}
\definecolor{myBlue}{rgb}{0.17, 0.26, 0.65}
\definecolor{myRed}{rgb}{0.85, 0.0, 0.0}

\makeatletter
\gdef\@fpheader{}
\makeatother

\begin{document}

\def\PSI{\Psi}
\def\THETA{\Theta}
\def\LAMBDA{\lambda}
\def\LAMBDAP{\LAMBDA'}
\def\LAMBDAPP{\LAMBDA''}
\def\MP{m'}
\def\MPP{m''}
\def\FERMION{\psi}
\def\YPSI{Y_{\PSI}}
\def\YPSIBF{\mathbf{Y}_{\PSI}}
\def\YTHETA{Y_{\THETA}}
\def\YTHETABF{\mathbf{Y}_{\THETA}}
\def\XI{\xi}
\def\HSPACE{\ }

\title{A realistic theory of $\mathbf{E_{6}}$ unification through novel intermediate symmetries}

\author[a]{K.S.~Babu,}
\author[b]{Borut Bajc}
\author[c]{and Vasja Susič}

\affiliation[a]{Department of Physics, Oklahoma State University, Stillwater, OK, 74078, USA}
\affiliation[b]{Jožef Stefan Institute, Jamova cesta 39, 1000 Ljubljana, Slovenia}
\affiliation[c]{Institute of Particle and Nuclear Physics,
    Faculty of Mathematics and Physics, Charles University,\\ V Hole\v{s}ovi\v{c}k\'{a}ch 2, 180 00 Prague 8, Czech Republic}

\emailAdd{babu@okstate.edu}
\emailAdd{borut.bajc@ijs.si}
\emailAdd{vasja.susic@matfyz.cuni.cz}

\abstract{
We propose a non-supersymmetric $\mathrm{E}_{6}$ GUT with the scalar sector consisting of $\mathbf{650}\oplus \mathbf{351'} \oplus \mathbf{27}$. Making use of the first representation for the initial symmetry breaking to an intermediate stage, and the latter two representations for second-stage breaking to the Standard Model and a realistic Yukawa sector, this theory represents the minimal $\mathrm{E}_{6}$ GUT that proceeds through one of the intermediate stages that are novel compared to 
$\mathrm{SU(5)}$ or $\mathrm{SO}(10)$ GUT: trinification $\mathrm{SU}(3)_C\times \mathrm{SU}(3)_L\times \mathrm{SU}(3)_R$, $\mathrm{SU}(6)\times \mathrm{SU}(2)$ and flipped $\mathrm{SO}(10)\times\mathrm{U}(1)$. 
We analyze these possibilities under the choice of vacuum that preserves a $\mathbb{Z}_{2}$ ``spinorial parity'', which disentangles the chiral and vector-like fermions of $\mathrm{E}_{6}$ and provides a dark matter candidate in the form of a (scalar) inert doublet. Three cases are shown to consistently unify under the extended survival hypothesis
(with minimal fine-tuning): trinification symmetry $\mathrm{SU}(3)_C\times \mathrm{SU}(3)_L\times \mathrm{SU}(3)_R$ with either $LR$ or $CR$ parity, and $\mathrm{SU}(6)_{CR}\times\mathrm{SU}(2)_L$. Although the successful cases give a large range for proton lifetime estimates, all of them include regions consistent with current experimental bounds and within reach of forthcoming experiments. The scenario investigated in this paper essentially represents the unique (potentially) viable choice in the class of $\mathrm{E}_{6}$ GUTs proceeding through a novel-symmetry intermediate stage, since non-minimal alternatives seem to be intrinsically non-perturbative.
}

\maketitle

\section{Introduction and motivation}

Soon after the first proposal of quark-lepton unification by Pati and Salam~\cite{Pati:1974yy} and a Grand Unified Theory (GUT) based on $\SU(5)$ by Georgi and Glashow~\cite{Georgi:1974sy}, the exceptional group $\mathrm{E}_{6}$ was recognized as a possible symmetry of unification~\cite{Gursey:1975ki}. 

Indeed, demanding that the unified group be \textit{simple}, contain the Standard Model (SM) gauge group $\SU(3)\times \SU(2)\times\UU$, and to have complex representations (to accommodate the chiral nature of the SM), the space of candidates in four space-time dimensions is essentially reduced to one of three types~\cite{Slansky:1981yr}: $\SU(n)$ for $n\geq 5$, $\SO(4m+2)$ for $m\geq 2$, and $\mathrm{E}_{6}$. The minimal candidates of each type can be arranged into the following hierarchy of inclusions:
\begin{align}
    \SU(5) \subset \SO(10) \subset \mathrm{E}_{6}.
\end{align}

The SM gauge group\footnote{Note on abbreviated notation for gauge groups: we use $[n]\equiv \SU(n)$ for $n\neq 1$, $[1]\equiv\UU$ and $(10)\equiv \SO(10)$, where subscripts such as $C$ (color), $L$ (left) or $R$ (right) may indicate additional information about the embedding. The subscript replaces the square brackets when present.} $\SU(3)\times \SU(2)\times\UU\equiv 3_C\, 2_L\, 1_Y$ is a maximal subgroup of $\SU(5)$, hence there is no possibility of symmetry breaking in multiple stages and for an intermediate symmetry between SM and GUT to arise in such a case. Symmetry breaking to intermediate stages, such as the well-studied possibilities of Pati-Salam $4_C\,2_L\,2_R$~\cite{Pati:1974yy} or left-right symmetry $3_C\,2_L\,2_R\,1_{B-L}$~\cite{Mohapatra:1974gc,Senjanovic:1975rk}, may first arise in the larger $\SO(10)$ group~\cite{Rizzo:1981su,Rizzo:1981dm,Rizzo:1981jr,Caswell:1982fx,Chang:1983fu,Gipson:1984aj,Chang:1984qr,Deshpande:1992au,Deshpande:1992em,Bertolini:2009qj,Bertolini:2009es,Bertolini:2010ng,Babu:2015bna,Graf:2016znk,Jarkovska:2021jvw,Jarkovska:2023zwv}. 

The still bigger $\mathrm{E}_{6}$ offers further possibilities for intermediate stages, the most intriguing being its maximal subgroups $\SU(3)^3\equiv [3][3][3]$ (trinification), see ref.~\cite{DeRujula:trinification,Babu:1985gi,Babu:2017xlu,Babu:2021hef}, $\SU(6)\times\SU(2)\equiv [6][2]$, see ref.~\cite{Dimopoulos:1985xs,Faraggi:2014bla}, and $\SO(10)\times\UU\equiv(10)[1]$, cf.~e.g.~\cite{Ranfone:1995di,Maekawa:2003wm,Bertolini:2010yz} for the flipped case in SUSY context, collectively referred to from now on as the \textit{novel symmetries}. This paper deals with building and studying a realistic model utilizing these novel symmetries as an intermediate stage; this model is essentially unique in its class, as argued below.\footnote{A rather general unification analysis of this class of models was studied before~\cite{Huang:2017uli}, although with an emphasis on the effect of Planck-induced operators and not in regard to perturbativity or explicit construction of a model.}

The only sensible way to break $\mathrm{E}_{6}$ to any of the novel symmetries is by use of the representation $\mathbf{650}$. Namely, it is the smallest irreducible representation of $\mathrm{E}_{6}$ with singlets under the novel symmetries, while the next-smallest is $\mathbf{2430}$, cf.~e.g.~\cite{Slansky:1981yr}, which we consider as prohibitively large and intrinsically non-perturbative. The indispensability of the $\mathbf{650}$ for the novel symmetries was already pointed out in our previous work~\cite{Babu:2023zsm}, where we analyzed the minima of the scalar potential involving one such representation.

To go beyond the first stage of symmetry breaking and arrive to the Standard Model, Michel's conjecture on symmetry breaking by a single irreducible representation (irrep) to only those subgroups that are maximal little groups of the irrep (see~\cite{Michel:1971th,Slansky:1981yr} and references therein) requires further scalar representations, as does the implementation of a realistic Yukawa sector. Embedding the SM fermions of each family into the $\mathbf{27}$ of $\mathrm{E}_{6}$, the most suitable candidates can be found in the symmetric product of two such fermionic representations, suggesting the addition of scalar representations $\mathbf{27}\oplus \mathbf{351'}$. This choice is very economical, since the representations simultaneously enable the next-stage breaking down to the SM (containing a combined $7$ complex SM-singlet scalars), as well as a realistic Yukawa sector with two symmetric matrices.\footnote{This $\mathrm{E}_{6}$ setup includes and is analogous to the common $\SO(10)$ setup, where the relevant $\SO(10)\subset\mathrm{E}_{6}$ inclusions are $\mathbf{16}\subset \mathbf{27}$ for fermions, and $\mathbf{10}\subset \mathbf{27}$ and $\mathbf{126}\subset \mathbf{351'}$ for scalars~\cite{Bajc:2005zf}. Curiously, the $\SO(10)$ case with a a complex $\mathbf{10}$, which is necessary for realistic fermion masses, leads to $3$ symmetric Yukawa matrices, reduced to $2$ if Peccei-Quinn (PQ) symmetry is invoked, while $\mathrm{E}_{6}$ implies the more predictive $2$-matrix case regardless of PQ symmetry.  
} 

The above discussion suggests the choice of scalar sector $\mathbf{27}\oplus\mathbf{351'}\oplus \mathbf{650}$ to be the minimal choice for a realistic $\mathrm{E}_{6}$ GUT with a novel symmetry at an intermediate stage. The large irreducible representations of $\mathrm{E}_{6}$ greatly restrict alternative routes for model building, since one expects an upper limit of field degrees of freedom (DOF) for the model
to be amenable to perturbation theory methods. For concreteness, one can consider the one-loop beta function $a$ for the renormalization group (RG) running of the unified gauge coupling as a function of the number of copies $N_R$ of various scalar representations $R$ one might consider:
\begin{align}
    a&= -38 + N_{27} + 2\, N_{78} + 25\, N_{351} + 28\, N_{351'} + 25\, N_{650} + 160\, N_{1728} + 135\, N_{2430}. \label{eq:Acoefficient-general}
\end{align}
In the above equation we already assumed the presence of three families of fermions in the fundamental $\mathbf{27}$, took the self-conjugate representations of scalars to have real DOF, and considered the smallest $\mathrm{E}_{6}$ representations in ascending order up to $\mathbf{2430}$ (the first alternative to $\mathbf{650}$ for novel symmetries). Our scalar sector leads to a still quite reasonable $a=16$, while a Landau pole is for example reached within one-order of magnitude above the GUT scale for $a\approx 109$, disfavoring any additions or modifications to our choice other than possibly adding $\mathbf{27}$s or $\mathbf{78}$s; we shall see later in this paper that the perturbative limitations are even more strict as stated above, hence the claim of essential uniqueness of the model.

The above considerations motivate our study of the model with the scalar sector $\mathbf{27}\oplus\mathbf{351'}\oplus \mathbf{650}$ in this paper. We organize our systematic investigation as follows:
\begin{center}
\begin{tabular}{lp{14cm}}
    Section~\ref{sec:model}: &
        we give the $\mathrm{E}_{6}$ model's definition, scalar potential, and the Yukawa sector. 
        \\[2pt]
    Section~\ref{sec:embeddings}: &
        we determine the possible embeddings of the SM group into novel symmetries, obtaining a single possible embedding of trinification $3_C\,3_L\,3_R$, three possible embeddings $6_{CL}\,2_R$, $6_{CL}\,2_{R'}$ and $6_{CR}\,2_{L}$ for $[6][2]$, and two embeddings $(10)\,1_{\psi}$ and $(10)'\,1_{\psi'}$ for $(10)[1]$.  
        \\[2pt]
    Section~\ref{sec:minimal-models}: &
        for embedding cases of interest, i.e.~those which do not unify the gauge couplings already at the intermediate stage, we determine the ``minimal'' field content of intermediate theories consistent with the extended survival hypothesis and spinorial parity, with the latter enabling a dark matter (DM) candidate.
        \\[2pt]
    Section~\ref{sec:analysis-unification}: &
        we analyze each minimal intermediate setup in terms of unification of gauge couplings in $\mathrm{E}_{6}$, and find there are three viable scenarios under minimal fine-tuning: trinification with $LR$ or $CR$ parity, and $6_{CR}\,2_{L}$.
        \\[2pt]
    Section~\ref{sec:analysis-proton}: &
        we analyze proton decay for various cases of intermediate symmetry and show that the viable cases from Section~\ref{sec:analysis-unification} admit scenarios consistent with experiment.
        \\
\end{tabular}
\end{center}
\noindent
Finally, we conclude then in Sec.~\ref{sec:conclusions}, followed by technical Appendices~\ref{app:state-definitions}, \ref{app:masses}, \ref{app:embeddings} and \ref{app:proton-decay-gamma-factors}. 

\section{The model with $\mathbf{27}\oplus\mathbf{351'}\oplus\mathbf{650}$ scalars \label{sec:model}}

We consider a non-supersymmetric $\mathrm{E}_{6}$ GUT with the following field content:
\begin{align}
    \text{fermions:} &\qquad 3\times \mathbf{27}_F, \\
    \text{scalars:} &\qquad \mathbf{27}\oplus \mathbf{351'} \oplus \mathbf{650}.
\end{align}
The fermionic representations, labeled by the subscript F, each contain all the SM particles of a single generation in the $\mathbf{27}$, along with two extra singlet neutrinos and a vector like pair of exotics, as discussed in detail later. 

Regarding the purpose of scalar representations, the $\mathbf{650}$ is envisioned to trigger the first stage of symmetry breaking to a novel intermediate symmetry, while the pair $\mathbf{27}\oplus \mathbf{351'}$ allows for a realistic Yukawa sector and some of its light remnants trigger the second-stage breaking to the SM. Our breaking set-up can thus be summarized by the diagram
\begin{align}
    \mathrm{E}_6 \quad\xrightarrow{\langle \mathbf{650}\rangle}\quad G  \quad\xrightarrow{\langle \mathbf{351'},\mathbf{27}\rangle}\quad \SU(3)_C\times\SU(2)_L\times\UU_{Y},
    \label{eq:breaking-pattern}
\end{align}
where the ``novel'' candidate groups of interest for the intermediate symmetry $G$ are 
\begin{align}
        G&\in \{\ \SU(3)\times\SU(3)\times\SU(3),
        \quad \SU(6)\times\SU(2), 
        \quad \SO(10)\times\UU\ \}.
        \label{eq:G-choices}
\end{align}
One more breaking step, namely the usual electroweak symmetry breaking in the SM, is assumed but not explicitly written in Eq.~\eqref{eq:breaking-pattern}.

Note that the representations $\mathbf{27}$ and $\mathbf{351'}$ are complex, while the representation $\mathbf{650}$ is real. Furthermore, they can be written tensorially as 
\begin{align}
    \mathbf{27}&\sim \PSI^{i},
    & \mathbf{351'} &\sim \THETA^{ij},
    & \mathbf{650} &\sim X^{i}{}_{j}, \label{eq:irrep-labels}
\end{align}
where all $\mathrm{E}_{6}$-related indices $i,j$ run from $1$ to $27$, with upper (lower) indices considered to be (anti)-fundamental. The latter two representations can thus be written as complex matrices $\mathbf{\THETA}$ and $\mathbf{X}$, where $\mathbf{\THETA}$ is symmetric and $\mathbf{X}$ is Hermitian, and they must satisfy additional irreducibility conditions~\cite{Babu:2023zsm}:
\begin{align}
    d_{ijk}\,\THETA^{jk}&=0, &
    \mathrm{Tr}(\mathbf{X})&=0, &
    \mathrm{Tr}(\mathbf{X}\,\mathbf{T}^{A})&=0,
\end{align}
where $d_{ijk}$ is the invariant tensor of $\mathrm{E}_{6}$, and $\mathbf{T}^{A}=(T^{A})^{i}{}_{j}$ are generator matrices of $\mathrm{E}_{6}$ in the fundamental representation $\mathbf{27}$, where the adjoint index $A$ runs from $1$ to $78$. These $\mathrm{E}_{6}$ generators are best arranged according to their transformation properties under trinification: the adjoint decomposes under $\mathrm{E}_{6}\to \SU(3)_C\times\SU(3)_L\times\SU(3)_R$ as
\begin{align}
    \mathbf{78}&= 
        \underbrace{(\mathbf{8},\mathbf{1},\mathbf{1})}_{t^{A}_{C}}\oplus  
        \underbrace{(\mathbf{1},\mathbf{8},\mathbf{1})}_{t^{A}_{L}}\oplus 
        \underbrace{(\mathbf{1},\mathbf{1},\mathbf{8})}_{t^{A}_{R}}\oplus
        \underbrace{(\mathbf{3},\mathbf{\bar{3}},\mathbf{\bar{3}})}_{t^{\alpha}{}_{aa'}} \oplus 
        \underbrace{(\mathbf{\bar{3}},\mathbf{3},\mathbf{3})}_{\bar{t}_{\alpha}{}^{aa'}}, 
        \label{eq:e6-adjoint}
\end{align}
where the labeling conventions for the corresponding generators are indicated below the braces, with $A$ now denoting a $SU(3)$-adjoint index that runs from $1$ to $8$, while $(\alpha,a,a')$ correspond to the fundamental indices that each go from $1$ to $3$ and are associated to the respective $\SU(3)$-factors $(C,L,R)$. For further details on the generators and the use of tensor methods in $\mathrm{E}_{6}$ more generally, we refer the reader to~\cite{Kephart:1981gf,Babu:2023zsm,Deppisch:2016xlp}.

\subsection{Scalar potential}

The scalar potential of the theory is considered at the renormalizable level and is a function of the fields from Eq.~\eqref{eq:irrep-labels}. We use the program LiE~\cite{Leeuwen1992LiE} to compute the number of independent invariants for every possible combination of representations or their conjugates. The number of invariants of each type and the labels used for the associated couplings are summarized in Table~\ref{tab:numbers-of-invariants}.

Due to the assumed two step breaking pattern of Eq.~\eqref{eq:breaking-pattern}, the scalar potential can be split for the sake of convenience into separate pieces, based on whether the terms contain only the representations from the first or second stage, or a mixture of both. It takes the explicit form

\begin{align}
V(X,\PSI,\THETA)&=V_{650}(X)+V_{27,351'}(\PSI,\THETA)+V_{\text{mix}}(X,\PSI,\THETA), \label{eq:potential-all}
\end{align}
where 
\begingroup
\allowdisplaybreaks
\begin{align}
V_{650}(X)=& 
    \phantom{+}\; M_{X}^2\cdot \mathrm{Tr}(\mathbf{X}^2) 
    \nonumber\\
    & \; + m_1\cdot \mathrm{Tr}(\mathbf{X}^3) 
    \quad + m_2\cdot X^{i}{}_{l}\,X^{j}{}_{m}\,X^{k}{}_{n}\; d^{lmn}d_{ijk} 
    \nonumber\\
    & \; +\LAMBDA_1\cdot (\mathrm{Tr}(\mathbf{X}^2))^2 
    \quad +\LAMBDA_2\cdot \mathrm{Tr}(\mathbf{X}^4) 
    \quad +\LAMBDA_3\cdot (\mathbf{X}^2)^{k}{}_{i}\;(\mathbf{X}^2)^{l}{}_{j}\;D^{ij}{}_{kl}  
    \nonumber\\
    & \qquad +\LAMBDA_4\cdot X^{i}{}_{i'}\,X^{j}{}_{j'}\,X^{k}{}_{k'}\,X^{l}{}_{l'}\;D^{i'j'}{}_{kl}\;D^{k'l'}{}_{ij} 
    \quad +\LAMBDA_5\cdot X^{i}{}_{l}\,X^{j}{}_{m}\,(\mathbf{X}^2)^{k}{}_{n}\; d^{lmn}d_{ijk}, \label{eq:potential-650}
\end{align}
\endgroup

\begingroup
\allowdisplaybreaks
\begin{align}
V_{27,351'}(\PSI,\THETA)=
    &\phantom{+}\; M_{\PSI}^2\cdot \PSI^{i}\,\PSI_{i}^\ast
     \nonumber\\
    &\; + M_{\THETA}^2 \cdot \THETA^{ij}\,\THETA_{ij}^{\ast}
    \nonumber\\
    &\; +\MP_1 \cdot \PSI^{i}\,\PSI^{j}\,\PSI^{k}\;d_{ijk} 
    \HSPACE +h.c. \nonumber\\
    &\; +\MP_2 \cdot \THETA^{ii'}\,\THETA^{jj'}\,\THETA^{kk'}\;d_{ijk}d_{i'j'k'}
    \HSPACE +h.c. \nonumber\\
    &\; +\MP_3 \cdot \PSI^{i}\,\PSI^{j}\,\THETA^{\ast}_{ij}
    \HSPACE +h.c. \nonumber\\
    &\; +\LAMBDAP_{1} \cdot (\PSI^{i}\,\PSI^\ast_{i})^2
    \quad +\LAMBDAP_{2} \cdot \PSI^{i}\,\PSI^{j}\,\PSI^\ast_{k}\,\PSI^\ast_{l}\;D^{kl}{}_{ij}
    \nonumber\\
    &\; +\LAMBDAP_{3} \cdot (\THETA^{ij}\,\THETA^{\ast}_{ij})^2
    \quad +\LAMBDAP_{4} \cdot \THETA^{ij}\,\THETA^{\ast}_{jk}\,\THETA^{kl}\,\THETA^\ast_{li}
    \quad + \LAMBDAP_{5}\cdot \THETA^{im}\,\THETA^{jn}\,\THETA^\ast_{km}\,\THETA^\ast_{ln}\,D^{kl}{}_{ij}
    \nonumber\\
    &\qquad + \LAMBDAP_{6}\cdot \THETA^{ii'}\,\THETA^{jj'}\,\THETA^{\ast}_{kk'}\,\THETA^\ast_{ll'}\;D^{kl}{}_{ij}\,D^{k'l'}{}_{i'j'}
    \nonumber\\
    &\; +\LAMBDAP_{7} \cdot (\PSI^{i}\,\PSI^\ast_{i})(\THETA^{kl}\,\THETA^\ast_{kl})
    \quad +\LAMBDAP_{8} \cdot \PSI^{i}\,\THETA^\ast_{ij}\,\THETA^{jk}\,\PSI^\ast_k
    \quad +\LAMBDAP_{9} \cdot \PSI^{i}\,\THETA^{jm}\,\PSI^\ast_{k}\,\THETA^\ast_{lm}\;D^{kl}{}_{ij}
    \nonumber\\
    &\; +\LAMBDAP_{10} \cdot \PSI^{i}\,\PSI^{j}\,\PSI^{\ast}_{l}\,\THETA^{kl}\,d_{ijk}
    \HSPACE +h.c. \nonumber\\
    &\; + \LAMBDAP_{11}\cdot \THETA^{il}\,\THETA^{jm}\,\PSI^{k}\,\PSI^{l}\;d_{ijk}\,d_{lmn}
    \HSPACE +h.c. \nonumber\\
    &\; + \LAMBDAP_{12}\cdot \THETA^{im}\,\THETA^{jn}\,\THETA^\ast_{mn}\,\PSI^{k}\,d_{ijk}
    \HSPACE +h.c., \label{eq:potential-27-351}
\end{align}
\endgroup
and
\begingroup
\allowdisplaybreaks
\begin{align}
V_{\text{mix}}(X,\PSI,\THETA)=
    &\phantom{+}\; \MPP_1\cdot \PSI^\ast_i\,X^{i}{}_{j}\,\PSI^j
    \nonumber\\
    & \; + \MPP_2\cdot \THETA^\ast_{ij}\,X^{j}{}_{k}\,\THETA^{ik}
    \nonumber\\
    &\; + \MPP_3\cdot \THETA^{il}\,X^{j}{}_{l}\,\PSI^{k}\,d_{ijk}
    \HSPACE +h.c. \nonumber\\
    &\; + \LAMBDAPP_{1} \cdot \PSI^{i}\,\PSI^{j}\,\PSI^{l}\,X^{k}{}_{l}\,d^{ijk}
    \HSPACE +h.c. \nonumber\\
    &\; + \LAMBDAPP_{2} \cdot \PSI^{i}\,\PSI^{k}\,X^{j}{}_{k}\,\THETA^\ast_{ij}
    \quad + \LAMBDAPP_{3} \cdot \PSI^{i}\,\PSI^{j}\,X^{m}{}_{l}\,\THETA^\ast_{km}\,D^{kl}{}_{ij}
    \HSPACE +h.c. \nonumber\\
    &\; + \LAMBDAPP_{4}\cdot \THETA^{il}\,\THETA^{km}\,\PSI^\ast_{m}\,X^{j}{}_{l}\;d_{ijk}
    \quad + \LAMBDAPP_{5}\cdot \THETA^{il}\,\THETA^{jm}\,\PSI^{\ast}_{n}\,X^{k}{}_{o}\;d_{ijk}\,D^{no}{}_{lm}
    \HSPACE +h.c. \nonumber \\
    &\; + \LAMBDAPP_{6} \cdot \THETA^{il}\,\THETA^{jm}\,\THETA^{kp}\,X^{n}{}_{p}\,d_{ijk}\,d_{lmn}
    \HSPACE +h.c.  \nonumber\\
    &\; + \LAMBDAPP_{7} \cdot \THETA^{il}\,(\mathbf{X}^2)^{j}{}_{l}\,\PSI^{k}\;d_{ijk}
    \quad + \LAMBDAPP_{8} \cdot \THETA^{lm}\,X^{i}{}_{l}\,X^{j}{}_{m}\,\PSI^{k}\;d_{ijk}
    \nonumber\\
    &\qquad + \LAMBDAPP_{9} \cdot \THETA^{il}\,X^{j}{}_{l}\,\PSI^{m}\,X^{k}{}_{m}\;d_{ijk}
    \quad + \LAMBDAPP_{10}\cdot \THETA^{il}\,\Psi^{m}\,X^{j}{}_{n}\,X^{k}{}_{o}\;d_{ijk}\,D^{no}{}_{lm}
    \HSPACE +h.c. \nonumber\\
    &\; + \LAMBDAPP_{11}\cdot (\PSI^i\,\PSI^\ast_i)\;\mathrm{Tr}(\mathbf{X}^2)
    \quad + \LAMBDAPP_{12}\cdot \PSI^\ast_{i} (\mathbf{X}^2)^{i}{}_{j}\,\PSI^{j}
    \quad + \LAMBDAPP_{13}\cdot \PSI^{i}\,\PSI^{\ast}_{k}\,(\mathbf{X}^2)^{j}{}_{l}\;D^{kl}{}_{ij}
    \nonumber\\
    &\qquad + \LAMBDAPP_{14}\cdot \PSI^{i}\,\PSI^{\ast}_{j}\,X^{k}{}_{l}\,X^{m}{}_{n}\;D^{jl}{}_{mo}\,D^{on}{}_{ik}
    \nonumber\\
    &\; + \LAMBDAPP_{15}\cdot (\THETA^{ij}\,\THETA^\ast_{ij}) \;\mathrm{Tr}(\mathbf{X}^2)
    \quad + \LAMBDAPP_{16}\cdot \THETA^\ast_{ij}\,X^{i}{}_{k}\,X^{j}{}_{l}\,\THETA^{kl}
    \quad + \LAMBDAPP_{17}\cdot \THETA^\ast_{ik}\,(\mathbf{X}^2)^{k}{}_{j}\,\THETA^{ij}
    \nonumber\\
    &\qquad + \LAMBDAPP_{18}\cdot \THETA^\ast_{km}\,X^{m}{}_{l}\,\THETA^{in}\,X^{j}{}_{n}\;D^{kl}{}_{ij}
    \quad + \LAMBDAPP_{19}\cdot \THETA^{im}\,\THETA^\ast_{km}\,(\mathbf{X}^2)^{j}{}_{l}\;D^{kl}{}_{ij}
    \nonumber\\
    &\qquad + \LAMBDAPP_{20}\cdot \THETA^{ii'}\,\THETA^\ast_{kk'}\,X^{j}{}_{l'}\,X^{j'}{}_{l}\,D^{kl}{}_{ij}\,D^{k'l'}{}_{i'j'}
    \quad + \LAMBDAPP_{21}\cdot \; \THETA^{io}\,\THETA^\ast_{lo}\,X^{j}{}_{m}\,X^{k}{}_{n} \;d_{ijk}\,d^{lmn}. \label{eq:potential-mix}
\end{align}
\endgroup

The number of terms of each type in Eqs.~\eqref{eq:potential-650}--\eqref{eq:potential-mix} agrees with Table~\ref{tab:numbers-of-invariants}, and furthermore the terms have been checked to be linearly independent by explicit computation. When writing the terms, each new line corresponds to invariants with different powers of irreps, with the continuation of the line shown as an indentation if we run out of space; also, an explicit $+\,h.c.$ at the end of lines involving complex couplings serves as a reminder that a conjugate terms should be added for each term in that line.
\begin{table}[htb]
    \caption{Number (denoted by $\#$) of $\mathrm{E}_{6}$ invariants of the type $\mathbf{27}^{n_1}\cdot \mathbf{27}^{\ast n_2}\cdot \mathbf{351'}^{n_3}\cdot \mathbf{351'}^{\ast n_4}\cdot \mathbf{650}^{n_5}$. For each case the labels of the associated real/complex ($\mathbb{R}/\mathbb{C}$) couplings are specified. The tables on the left- and right-hand side list the cubic and quartic invariants, respectively. \label{tab:numbers-of-invariants}}
\begin{minipage}[t]{.5\linewidth}
    \centering
    \caption*{(a) Cubic invariants and associated couplings.}
    \begin{tabular}[t]{c@{\ }c@{\ }c@{\ }c@{\ }c@{\quad}ccl}
    \toprule
    $n_1$&$n_2$&$n_3$&$n_4$&$n_5$&inv.~$\#$&$\mathbb{R}/\mathbb{C}$&labels\\
    \midrule
    0&0&0&0&3 & 2 &$\mathbb{R}$&$m_1$, $m_2$\\[5pt]
    3&0&0&0&0 & 1 &$\mathbb{C}$&$\MP_1$\\
    0&0&3&0&0 & 1 &$\mathbb{C}$&$\MP_2$\\
    2&0&0&1&0 & 1 &$\mathbb{C}$&$\MP_3$\\[5pt]
    1&1&0&0&1 & 1 &$\mathbb{R}$&$\MPP_1$\\
    0&0&1&1&1 & 1 &$\mathbb{R}$&$\MPP_2$\\
    1&0&1&0&1 & 1 &$\mathbb{C}$&$\MPP_3$\\
    \bottomrule
    \end{tabular}
\end{minipage}\hspace{0.1cm}
\begin{minipage}[t]{.5\linewidth}
    \centering
    \caption*{(b) Quartic invariants and associated couplings.}
    \begin{tabular}[t]{c@{\ }c@{\ }c@{\ }c@{\ }c@{\quad}ccl}
    \toprule
    $n_1$&$n_2$&$n_3$&$n_4$&$n_5$&inv.~$\#$&$\mathbb{R}/\mathbb{C}$&labels\\
    \midrule
    0&0&0&0&4 & 5 &$\mathbb{R}$ &$\LAMBDA_1$ \ldots $\LAMBDA_5$ \\[5pt]
    2&2&0&0&0 & 2 &$\mathbb{R}$ &$\LAMBDAP_1$, $\LAMBDAP_2$ \\
    0&0&2&2&0 & 4 &$\mathbb{R}$ &$\LAMBDAP_3$ \ldots $\LAMBDAP_6$\\
    1&1&1&1&0 & 3 &$\mathbb{R}$ &$\LAMBDAP_7$ \ldots $\LAMBDAP_9$\\
    2&1&1&0&0 & 1 &$\mathbb{C}$ &$\LAMBDAP_{10}$\\
    2&0&2&0&0 & 1 &$\mathbb{C}$ &$\LAMBDAP_{11}$\\
    1&0&2&1&0 & 1 &$\mathbb{C}$ &$\LAMBDAP_{12}$\\[5pt]
    3&0&0&0&1 & 1 &$\mathbb{C}$ &$\LAMBDAPP_{1}$ \\
    2&0&0&1&1 & 2 &$\mathbb{C}$ &$\LAMBDAPP_{2}$, $\LAMBDAPP_{3}$\\
    0&1&2&0&1 & 2 &$\mathbb{C}$ &$\LAMBDAPP_{4}$, $\LAMBDAPP_{5}$\\
    0&0&3&0&1 & 1 &$\mathbb{C}$ &$\LAMBDAPP_{6}$\\
    1&0&1&0&2 & 4 &$\mathbb{C}$ &$\LAMBDAPP_{7}$ \ldots $\LAMBDAPP_{10}$\\
    1&1&0&0&2 & 4 &$\mathbb{R}$ &$\LAMBDAPP_{11}$ \ldots $\LAMBDAPP_{14}$\\
    0&0&1&1&2 & 7 &$\mathbb{R}$ &$\LAMBDAPP_{15}$ \ldots $\LAMBDAPP_{21}$\\
    \bottomrule
    \end{tabular}
\end{minipage}
\end{table}

The invariants were written by seamlessly transitioning between matrix and index notations, e.g.~between $\mathbf{X}$ and $X^{i}{}_{j}$. The construction of invariants involves the primitive invariant tensor $d_{ijk}$, cf.~Appendix~A of Ref.~\cite{Babu:2023zsm}, and the composite invariant tensor
\begin{align}
D^{ij}{}_{kl}&:=d^{ijm}d_{mkl}.
\end{align}

As regards the naming of the couplings, the parts $(V_{650},V_{27,351'},V_{\text{mix}})$ label the cubic couplings respectively by $(m_{i},\MP_{i},\MPP_{i})$ and the quartic couplings by $(\LAMBDA_i,\LAMBDAP_i,\LAMBDAPP_i)$. Altogether there are $3$ $\mathbb{R}$-quadratic, $4$ $\mathbb{R}$-cubic, $4$ $\mathbb{C}$-cubic,  $25$ $\mathbb{R}$-quartic and $13$ $\mathbb{C}$-quartic invariants. Note that the part $V_{650}$, including the invariant definitions and the coupling labels, corresponds to the potential given in Ref.~\cite{Babu:2023zsm}, except for making the replacement $M^{2}\mapsto M^{2}_{X}$.

\subsection{Two stage breaking and scalar masses \label{sec:two-stage-breaking-and-masses}}

The scalar potential of Eqs.~\eqref{eq:potential-all}--\eqref{eq:potential-mix} is completely general. For a suitable choice of parameters, however, the breaking pattern of Eq.~\eqref{eq:breaking-pattern} can be achieved, i.e.~if the mass scales involved in $V_{650}$ are much higher than those in $V_{27,351'}$.

Under these circumstances, the first-stage breaking then involves the minimization only of $V_{650}$, which was studied in \cite{Babu:2023zsm}. For each choice of intermediate symmetry $G$ from Eq.~\eqref{eq:G-choices}, there is a corresponding vacuum expectation value (VEV) direction $\langle\mathbf{650}\rangle_G=\tilde{s}_G$. A special situation happens for $G=3_C\,3_L\,3_R$: there are three discrete vacua corresponding to preserved $LR$, $CL$ and
$CR$ parities, whose VEVs are respectively labeled by $s$, $s'$ and $s''$. These differ within $\mathbf{650}$ in direction but not in size, see \cite{Babu:2023zsm}. We give their explicit definitions in Appendix~\ref{app:state-definitions-G-singlets}.

In all cases of $G$, the VEV is a function of the non-primed parameters from $V_{650}$ and determines the GUT-scale masses of the $G$-irreps of the scalar sector. To proceed further, we first consider the decompositions $\mathrm{E}_{6}\to G$ of scalar sector irreps for all choice of $G$ in Eq.~\eqref{eq:G-choices}. The resulting branching rules\footnote{In our convention the $\mathbf{351'}^\ast$ from \cite{Slansky:1981yr} is denoted as $\mathbf{351'}$, so that $\mathbf{351'}\subset \mathbf{27}\otimes\mathbf{27}$, analogous to $\mathbf{6}\subset\mathbf{3}\otimes\mathbf{3}$ in $\SU(3)$.}, cf.~\cite{Slansky:1981yr}, for $\mathrm{E}_6\to \SU(3)\times\SU(3)\times\SU(3)$ are
\begingroup
\allowdisplaybreaks
\begin{align}
    \mathbf{27}&= 
        (\mathbf{3},\mathbf{3},\mathbf{1}) \;\oplus\; 
        (\mathbf{\bar{3}},\mathbf{1},\mathbf{\bar{3}}) \;\oplus\; 
        (\mathbf{1},\mathbf{\bar{3}},\mathbf{3}), \label{eq:decomposition-to333-27}\\[3pt]
    \mathbf{351'} &=
        (\mathbf{\bar{3}},\mathbf{\bar{3}},\mathbf{1}) \;\oplus\; 
        (\mathbf{3},\mathbf{1},\mathbf{3}) \;\oplus\; 
        (\mathbf{1},\mathbf{3},\mathbf{\bar{3}}) \;\oplus\; 
        (\mathbf{6},\mathbf{6},\mathbf{1}) \;\oplus\; 
        (\mathbf{\bar{6}},\mathbf{1},\mathbf{\bar{6}}) \;\oplus\; 
        (\mathbf{1},\mathbf{\bar{6}},\mathbf{6}) \nonumber \\
        &\quad \oplus\; (\mathbf{\bar{3}},\mathbf{\bar{3}},\mathbf{8}) \;\oplus\; 
        (\mathbf{3},\mathbf{8},\mathbf{3}) \;\oplus\; 
        (\mathbf{8},\mathbf{3},\mathbf{\bar{3}}), \label{eq:decomposition-to333-351p}\\[3pt]
    \mathbf{650} &= 
        2\times (\mathbf{1},\mathbf{1},\mathbf{1}) \;\oplus\;
        2\times (\mathbf{3},\mathbf{\bar{3}},\mathbf{\bar{3}}) \;\oplus\; 
        2\times (\mathbf{\bar{3}},\mathbf{3},\mathbf{3}) \;\oplus\;
        (\mathbf{8},\mathbf{1},\mathbf{1}) \;\oplus\; 
        (\mathbf{1},\mathbf{8},\mathbf{1}) \;\oplus\; 
        (\mathbf{1},\mathbf{1},\mathbf{8}) \nonumber\\
        &\quad \oplus\;(\mathbf{3},\mathbf{6},\mathbf{\bar{3}}) \;\oplus\; 
        (\mathbf{3},\mathbf{\bar{3}},\mathbf{6}) \;\oplus\;  
        (\mathbf{6},\mathbf{3},\mathbf{3}) \;\oplus\;
        (\mathbf{1},\mathbf{8},\mathbf{8}) \;\oplus\; 
        (\mathbf{8},\mathbf{1},\mathbf{8}) \;\oplus\; 
        (\mathbf{8},\mathbf{8},\mathbf{1})\nonumber\\
        &\quad \oplus\;(\mathbf{\bar{3}},\mathbf{\bar{6}},\mathbf{3}) \;\oplus\;  
        (\mathbf{\bar{3}},\mathbf{3},\mathbf{\bar{6}}) \;\oplus\;  
        (\mathbf{\bar{6}},\mathbf{\bar{3}},\mathbf{\bar{3}}), \label{eq:decomposition-to333-650}
\end{align}
\endgroup
for $\mathrm{E}_6\to \SU(6)\times\SU(2)$ the decompositions are
\begingroup
\allowdisplaybreaks
\begin{align}
    \mathbf{27}&=
        (\mathbf{15},\mathbf{1}) \;\oplus\;
        (\mathbf{\bar{6}},\mathbf{2}), \label{eq:decomposition-to62-27}\\[3pt]
    \mathbf{351'}&=
        (\mathbf{\overline{15}},\mathbf{1}) \;\oplus\;
        (\mathbf{\overline{105'}},\mathbf{1}) \;\oplus\;
        (\mathbf{84},\mathbf{2}) \;\oplus\;
        (\mathbf{\overline{21}},\mathbf{3}), \label{eq:decomposition-to62-351p}\\[3pt]
    \mathbf{650}&=
        (\mathbf{1},\mathbf{1}) \;\oplus\;
        (\mathbf{35},\mathbf{1}) \;\oplus\;
        (\mathbf{189},\mathbf{1}) \;\oplus\;
        (\mathbf{20},\mathbf{2}) \;\oplus\;
        (\mathbf{70},\mathbf{2}) \;\oplus\;
        (\mathbf{\overline{70}},\mathbf{2}) \;\oplus\;
        (\mathbf{35},\mathbf{3}),\label{eq:decomposition-to62-650}
\end{align}
\endgroup
and for $\mathrm{E}_6\to \SO(10)\times\UU$ the decompositions are 
\begingroup
\allowdisplaybreaks
\begin{align}
    \mathbf{27}&= 
        (\mathbf{1},+4) \;\oplus\; 
        (\mathbf{10},-2)\;\oplus\;  
        (\mathbf{16},+1), \label{eq:decomposition-to101-27}\\[3pt]
    \mathbf{351'}&= 
        (\mathbf{1},+8) \;\oplus\;  
        (\mathbf{10},+2) \;\oplus\;
        (\mathbf{16},+5) \;\oplus\;
        (\mathbf{54},-4) \;\oplus\;
        (\mathbf{126},+2) \;\oplus\;
        (\mathbf{\overline{144}},-1), \label{eq:decomposition-to101-351p}\\[3pt]
    \mathbf{650}&=
        (\mathbf{1},0) \;\oplus\; 
        (\mathbf{10},+6) \;\oplus\; 
        (\mathbf{10},-6) \;\oplus\; 
        (\mathbf{16},-3) \;\oplus\; 
        (\mathbf{\overline{16}},+3) \;\oplus\; 
        (\mathbf{45},0) \;\oplus\; 
        (\mathbf{54},0)  \nonumber\\
        &\quad \oplus\; (\mathbf{144},-3) \;\oplus\; 
        (\mathbf{\overline{144}},+3) \;\oplus\; 
        (\mathbf{210},0). \label{eq:decomposition-to101-650}
\end{align}
\endgroup

Notice that for all cases of $G$, there is no common $G$-irrep between the list in $\mathbf{650}$ and the list in $\mathbf{27}\oplus\mathbf{351'}$. This means the first-stage breaking does not mix the irreps in $\mathbf{650}$ and those in $\mathbf{27}\oplus\mathbf{351'}$. 

Consequently, the heavy (first-stage) masses of $G$-irreps in the $\mathbf{650}$ are determined solely by the $V_{650}$ part, with results already reported in~\cite{Babu:2023zsm}. The heavy masses of $G$-irreps in $\mathbf{27}\oplus\mathbf{351'}$, on the other hand, are independently determined from $V_{\text{mix}}$ (and from pure mass terms $M^{2}_\PSI$ and $M^{2}_\THETA$ in $V_{27,351'}$). Their corresponding expressions are relegated to Appendix~\ref{app:masses-scalars}, but the key observation is that they depend on the double-primed parameters from $V_{\text{mix}}$ and all masses are independent (except for respecting parity in the cases with trinification). 

Some of the $G$-irrep pieces then need to be tuned to the intermediate scale: some pieces are involved in the second stage of symmetry breaking, while others are needed for the Yukawa sector (they contain those SM-doublets that posses a SM-Higgs admixture). The minimal situation fulfilling both requirements is known as the \textit{extended survival hypothesis}~\cite{Mohapatra:1982aq,Dimopoulos:1984ha}.
Crucially, the independent nature of the mass expressions allows for a free choice of irreps to be pushed to the intermediate scale.

\subsection{Yukawa sector}

We denote the fermionic representations by $\FERMION_I$, written as Weyl fermions, where the family index $I$ goes from $1$ to $3$. The notation for scalars is that of Eq.~\eqref{eq:irrep-labels}.

The renormalizable Yukawa sector of the model is written explicitly as 
\begin{align}
    \mathcal{L}_{\text{Yuk}} &=
        \YPSI^{IJ}\; \FERMION^{i}{}_{I}\, \FERMION^{j}{}_{J}\, \PSI^{k}\, d_{ijk}
        +\YTHETA^{IJ}\; \FERMION^{i}{}_{I}\, \FERMION^{j}{}_{J}\, \THETA^{\ast}{}_{ij}, \label{eq:Yukawa-sector}
\end{align}
where $\{I,J\}$ are family indices and $\{i,j,k\}$ are $\mathrm{E}_{6}$ indices, with all repeated indices summed over. The Lorentz structure (spinorial indices in fermions) has been suppressed in the above notation. 
The Yukawa sector thus consists of two symmetric $3\times 3$ matrices $\YPSIBF$ and $\YTHETABF$, corresponding to the coupling of the scalar representations $\mathbf{27}$ and $\mathbf{351'}$ to the fermions, respectively.

The fermionic $\mathbf{27}\sim\FERMION$ of each generation contains two singlet neutrinos $\nu^{c},n$ and vector-like exotics $d'\oplus d'^{c}$ and $L'\oplus L'^{c}$ alongside the SM fields:
\begin{align}
    \mathbf{27}=Q\oplus u^{c}\oplus e^{c} \oplus d^{c} \oplus L \oplus \nu^c \oplus n \oplus d' \oplus d'^{c} \oplus L' \oplus L'^{c}, \label{eq:fermion-content-27}
\end{align}
where the usual SM fermions in each generation transform under $3_C\,2_L\,1_Y$ as
    \begin{align}
        Q&\sim (\mathbf{3},\mathbf{2},+\tfrac{1}{6}),&
        u^c&\sim (\mathbf{\bar{3}},\mathbf{1},-\tfrac{2}{3}), \nonumber\\
        e^c&\sim (\mathbf{1},\mathbf{1},+1), &
        d^c&\sim (\mathbf{\bar{3}},\mathbf{1},+\tfrac{1}{3}), \nonumber\\
        L&\sim (\mathbf{1},\mathbf{2},-\tfrac{1}{2}), \label{eq:SM-fermions}
    \end{align}
and the exotic fermions transform as 
    \begin{align}
        L'=\left(\begin{smallmatrix} \nu'\\ e'\\ \end{smallmatrix}\right)&\sim (\mathbf{1},\mathbf{2},-\tfrac{1}{2}), &
            d'&\sim (\mathbf{3},\mathbf{1},-\tfrac{1}{3}), \nonumber\\
        L'^{c}=\left(\begin{smallmatrix} -e'^{c}\\ \nu'^{c} \\ \end{smallmatrix}\right)&\sim (\mathbf{1},\mathbf{2},+\tfrac{1}{2}), &
            d'^{c} & \sim (\mathbf{\bar{3}},\mathbf{1},+\tfrac{1}{3}), \label{eq:exotic-fermions} \\
        \nu^{c},n&\sim (\mathbf{1},\mathbf{1},0). \nonumber
    \end{align}
The entire fermionic irrep $\mathbf{27}_F$ is then most conveniently presented when arranged into trinification $3_C\,3_L\,3_R$ irreps under Eq.~\eqref{eq:decomposition-to333-27}, cf.~Appendix~C in \cite{Babu:2023zsm} and also~\cite{Babu:2021hef}:

\begin{align}
	(\mathbf{3},\mathbf{3},\mathbf{1})\sim \mathbf{L}&=
		\begin{pmatrix} 
			u_{1} & d_{1} & d'_{1}\\ 
			u_{2} & d_{2} & d'_{2}\\ 
			u_{3} & d_{3} & d'_{3}\\ 
		\end{pmatrix}, &
	(\mathbf{1},\mathbf{\bar{3}},\mathbf{3})\sim \mathbf{M}&=
		\begin{pmatrix}
			\nu'^{c} & e' & e \\
			e'^{c} & -\nu' & -\nu\\
			e^{c} & \nu^{c} & n\\
		\end{pmatrix}, &
	(\mathbf{\bar{3}},\mathbf{1},\mathbf{\bar{3}})\sim \mathbf{N}&= 
		\begin{pmatrix}
			u^{c}_{1} & u^{c}_{2} & u^{c}_{3} \\
			-d^{c}_{1} & -d^{c}_{2} & -d^{c}_{3} \\
			d'^{c}_{1} & d'^{c}_{2} & d'^{c}_{3} \\
		\end{pmatrix}. \label{eq:particles-in-27}
\end{align} 

Clearly, the Yukawa sector of Eq.~\eqref{eq:Yukawa-sector} provides masses to all the fields from Eq.~\eqref{eq:particles-in-27}. Since further explicit analysis requires a number of prior definitions, we postpone it to Section~\ref{sec:minimal-models-Yukawa}. 

As a final note on the Yukawa sector, we recapitalute our remarks from the Introduction that the $2$-Yukawa situation in $\mathrm{E}_6$ is more predictive than the $\mathbf{10}_{\mathbb{C}}\oplus\mathbf{126}$ renormalizable Yukawa sector of $\SO(10)$, which has $3$ symmetric Yukawa matrices unless PQ-symmetry is invoked \cite{Peccei:1977hh,Bajc:2005zf}. Furthermore, $\mathbf{27}\oplus\mathbf{351'}$ play a double role: beside accommodating the SM Higgs, they also trigger the 2nd stage of symmetry breaking. 

For some studies of the $\mathrm{E}_{6}$ Yukawa sector in other contexts, usually in supersymmetry, we refer the reader to~\cite{Bando:1999km,Bando:2000gs,Bando:2001bj}.

\section{Embedding the Standard Model and intermediate symmetries \label{sec:embeddings}}

\subsection{The SM and assorted small subgroups in $\mathrm{E}_6$}

The SM gauge group $3_C\,2_L\,1_Y$ is a subgroup of $\mathrm{E}_{6}$, with a single inequivalent embedding corresponding to the usual branching rule for fermions in the $\mathbf{27}$, cf.~e.g.~\cite{Slansky:1981yr}. Conjugating all elements of $3_C\,2_L\,1_Y$ with $A\in\mathrm{E}_{6}$ results however in an equivalent embedding, i.e.~under $Ad_A: \mathrm{E}_{6}\to\mathrm{E}_{6}$, $B\mapsto ABA^{-1}$, the image of an embedded SM subgroup is an equivalent embedding.

We remove this freedom in the embedding of the SM group by fixing the SM generators using the following convention:
\begin{align}
    \SU(3)_C:&\qquad \{\; t^{A}_{C};\ A=1,\ldots,8\;\}, \label{eq:embedding-su3c}\\
    \SU(2)_L:&\qquad \{\; t^{1}_{L},\ t^{2}_{L},\ t^{3}_L\;\}, \label{eq:embedding-su2l}\\
    \UU_{Y}: &\qquad \{\; t_{Y} \;\}, \label{eq:embedding-u1y}
\end{align}
where the hypercharge generator $t_{Y}$ is defined by
\begin{align}
    t_{Y}:= \tfrac{1}{\sqrt{3}}t^{8}_{L}+\tfrac{1}{\sqrt{3}}t^{8}_{R}+t^{3}_{R}.\label{eq:generator-Y}
\end{align}
We used the notation for $\mathrm{E}_{6}$ generators from Eq.~\eqref{eq:e6-adjoint}, and the hypercharge is normalized so that  $t_Y(e^{c})=+1$ for the charged lepton singlet $e^{c}$ in SM. The electromagnetic charge $t_{Q}$ is then given by
\begin{align}
    t_{Q}&=t_{L}^{3}+t_{Y}.\label{eq:generator-Q}
\end{align}
It proves convenient to define also several other $\UU$ charges found in $\mathrm{E_{6}}$ in common use:
\begin{align}
    t_{B-L}&:=\tfrac{2}{\sqrt{3}}(t_{L}^{8}+t_{R}^{8}),\label{eq:generator-BL}\\
    t_{\chi}&:=4\,t_{R}^{3}-3\,t_{B-L}, \label{eq:generator-chi}\\
    t_{\psi}&:= 2\sqrt{3}(t_{L}^{8}-t_{R}^{8}), \label{eq:generator-psi}\\
    t_{\psi'}&:= 2\sqrt{3}t^8_L+3t^{3}_R+\sqrt{3}t^{8}_R, \label{eq:generator-psip} \\
    t_{R'}&:= -\tfrac{1}{2} t^{3}_{R}+\tfrac{\sqrt{3}}{2}t^{8}_{R}, \label{eq:generator-Rp}
\end{align}
where $B-L$ is the usual baryon-minus-lepton number gauged in e.g.~Pati-Salam, the $\chi$-charge is the one defined via the standard $\SO(10)\supset\SU(5)\times\UU_{\chi}$ decomposition
    \begin{align}
        \mathbf{16}&=(\mathbf{1},-5)\oplus (\mathbf{\bar{5}},+3)\oplus(\mathbf{10},-1),
    \end{align}
while $\psi$ and $\psi'$ are the $\UU$ charges in the decompositions of Eq.~\eqref{eq:decomposition-to101-27}--\eqref{eq:decomposition-to101-650} to the standard $\SO(10)\times\UU_{\psi}$ and flipped $\SO(10)'\times\UU_{\psi'}$. 

Furthermore, we define two different $\SU(2)$ subgroups of the trinification factor $\SU(3)_R$ by specifying the constituent generators:
\begin{align}
    \SU(2)_{R}:&\qquad \{\; t^{1}_{R},\ t^{2}_{R},\ t^{3}_{R} \;\}, \label{eq:embedding-su2r}\\
    \SU(2)_{R'}:&\qquad \{\; t^{6}_{R},\ t^{7}_{R},\ t_{R'} \;\}. \label{eq:embedding-su2rp}
\end{align}
The diagonal generator in $\SU(2)_{R'}$ is the $R'$-charge from Eq.~\eqref{eq:generator-Rp}. 

Intuitively, the subgroup $\SU(2)_R$ rotates between the 1st and 2nd component of an $\SU(3)_R$ triplet, while $\SU(2)_{R'}$ rotates between the 2nd and 3rd components. In the embedding of fermions into the fundamental representation $\mathbf{27}$ of Eq.~\eqref{eq:particles-in-27}, $\SU(2)_R$ performs the $1$-$2$ rotation of columns in $\mathbf{M}$ and rows in $\mathbf{N}$, while $\SU(2)_{R'}$ performs analogous $1$-$3$ rotations.

Phenomenologically, the $\SU(2)_R$ is thus the same group found in the group of left-right symmetry $\SU(3)_C\times\SU(2)_L\times\SU(2)_R\times\mathrm{U}_{B-L}$.  Notice that combining Eqs.~\eqref{eq:generator-Y} and \eqref{eq:generator-Q} gives the expression 
    \begin{align}
        t_{Q}&=(t^{3}_{L}+t^{3}_R)+\tfrac{2}{\sqrt{3}}(t_{L}^{8}+t_{R}^{8})
    \end{align}
for the EM charge, which is left-right symmetric, as it should be. On the other hand, $\SU(2)_{R'}$ rotates between the SM and exotic vector-like fermions that have the same SM quantum numbers, i.e.~it rotates in the space $(d,d')$ and in the space $(L,L')$, see Eqs.~\eqref{eq:SM-fermions} and \eqref{eq:exotic-fermions}. In other words, it rotates between the $\mathbf{\bar{5}}$ parts of $\SU(5)$ located in the representations $\mathbf{16}$ and $\mathbf{10}$ of the standard $\SO(10)$ GUT. 

Furthermore, it is clear that the entire $\SU(3)_R$ commutes with both the color (C) and weak (L) interaction of the SM group, but not with hypercharge. The $\SU(2)_{R'}$ subgroup is exactly that part of $\SU(3)_R$ which commutes also with $1_Y$ and hence the entire SM group. The embedding 
\begin{align}
    \SU(3)_C\times\SU(2)_L\times\UU_Y\times\SU(2)_{R'} &\subset \mathrm{E}_6 \label{eq:SM-group-with-Rp}
\end{align}
embeds the SM group together with a non-Abelian factor into the unified group. This is an interesting novelty of $\mathrm{E}_{6}$, since such a non-Abelian factor cannot be accommodated in the case of the smaller GUT groups $\SU(5)$ or $\SO(10)$. Since $\SU(2)'_{R}$ commutes with $3_C\,2_L\,1_Y$, its freedom can be used without changing the embedding of the SM group into $\mathrm{E}_{6}$ from Eqs.~\eqref{eq:embedding-su3c}--\eqref{eq:embedding-u1y}.

\subsection{Embeddings of intermediate symmetry groups \label{sec:embeddings-intermediate}}

Having defined our embedding of SM into $\mathrm{E}_{6}$, we turn to the question of embedding the intermediate symmetry groups $G$ from Eq.~\eqref{eq:G-choices}.

Each $G$ is a maximal subgroup in $\mathrm{E}_{6}$ and has has a unique embedding up to $\mathrm{E}_{6}$-equivalency~\cite{Feger:2019tvk}. Since the resulting low energy theory should be the SM and we keep the embedding of $3_C\,2_L\,1_Y$ fixed by our convention in Eqs.~\eqref{eq:embedding-su3c}--\eqref{eq:embedding-u1y}, we are interested in classifying embeddings of $G$ only up to $3_C\,2_L\,1_Y\,2_{R'}$-equivalency, since this group from Eq.~\eqref{eq:SM-group-with-Rp} is exactly the one that preserves the SM embedding. As we shall see, this will lead to multiple inequivalent choices of how the SM group is embedded into $G$.

\begin{figure}
    \begin{center}
    \SubgroupDiagramDefinition
    \caption{The pattern of squares corresponding to generators that define a subalgebra diagram. \label{figure:subgroup-diagram-definition}}
    \end{center}
\end{figure}

To simplify the explicit identification of $G$, we define a \textit{subalgebra diagram}, a visual tool to specify a regular maximal-rank subalgebra of $\mathrm{E}_{6}$ that contains the SM. We arrange the relevant set of $\mathrm{E}_{6}$ generators from Eq.~\eqref{eq:e6-adjoint} into the pattern of Figure~\ref{figure:subgroup-diagram-definition} and color those squares that correspond to the generators of $G$. The reasoning behind the pattern in the subalgebra diagram is the following: 
\begin{itemize}
    \item The color generators $t^{A}_{C}$ are part of the SM group $3_C\,2_Y\,1_Y$, so they are automatically part of $G$ and need not be indicated.
    \item The first and second panel of the diagram specify which generators from $\SU(3)_L$ and $\SU(3)_R$ are part of $G$. 
    \item In the last panel of the diagram, we specify which of the generators $t^{\alpha}{}_{aa'}$ are part of $G$. The same pattern applies for $\bar{t}_{\alpha}{}^{aa'}$, since $t$ and $\bar{t}$ are complex generators analogous to $t^{\pm}=t^{1}\pm i t^{2}$ in $\SU(2)$. 
    Also, notice that the generators $t^{\alpha}{}_{aa'}$ for a fixed $a$ and $a'$ come in color triplets (index $\alpha$), where the entire triplet is either present or absent from $G$. This allows the relevant information to be conveyed compactly by specifying only which $(a,a')$ to include into $G$.
\end{itemize}

Although one can build a good intuition for colored patterns in subalgebra diagrams, we relegate that discussion to Appendix~\ref{app:subalgebra-diagram}, and now use the diagrams merely as a way to explicitly define $G$. The possible embeddings of $G$ are summarized in Table~\ref{tab:embeddings} and motivated as follows:
\begin{enumerate}
    \item For $G=[3][3][3]$, there is only one embedding up to a choice of which $\SU(3)$-factor of trinification contains $3_C$ and which contains $2_L$. A fixed definition for SM generators determines these factors unambiguously.
    
    \item For $G=[6][2]$ there are three inequivalent embeddings given a fixed SM: the \textit{standard}, \textit{flipped} and \textit{LR-flipped}. The possibilities reflect that the $2_L$ of SM can be chosen to be part of the $\SU(2)$ (LR-flipped) or $\SU(6)$ factor, and in the latter case one can further choose whether $1_Y$ of SM is entirely in $\SU(6)$ (flipped) or whether it partially extends to the $\SU(2)$ (standard).
    
    \item For $G=(10)[1]$ there are two embeddings for a fixed SM, namely the \textit{standard} and \textit{flipped} case. They differ in whether the $1_Y$ of SM is entirely in $\SO(10)$ (standard) or whether it also has an admixture of the $\UU$ factor (flipped). 
\end{enumerate}
A more rigorous way to derive the completeness of this classification is to write an ansatz for the hypercharge in terms of diagonal generators of $G$, and determine the solutions that yield the known fields from the $\mathbf{27}$ of $\mathrm{E}_{6}$. Additional intuition for each embedding of $G$ can be gained by considering how the fermions in $\mathbf{27}$ of $\mathbf{E}_{6}$ distribute among $G$-representations; we relegate this discussion to Appendix~\ref{app:embeddings-fermions}. For the purpose of computing decompositions of irreps under $G\supset 3_{C}\,2_{L}\,1_{Y}$, projection matrices are given in Appendix~\ref{app:embeddings-projection-matrices}.

\begin{table}[htb]
\begin{center}
\caption{Classification of all embeddings of $G$ from Eq.~\eqref{eq:G-choices} into $E_{6}$ given a fixed embedding of the Standard Model group.
    \label{tab:embeddings}
}
\vskip 0.0cm
\begin{tabular}{llllrr}
\toprule
embedding of $G$&label&name&subalgebra diagram&$D\subseteq D_{3}$&$D\subset G$?\\
\midrule
$\SU(3)_C\times\SU(3)_L\times\SU(3)_R$&$3_C\,3_L\,3_R$&trinification&\SubgroupDiagram{1/1,1/2,1/3,2/1,2/2,3/1,3/2,3/3}{1/1,1/2,1/3,2/1,2/2,3/1,3/2,3/3}{}
&$D_{3}$&no\\\addlinespace[18pt]
$\SU(6)_{CL}\times\SU(2)_{R}$&$6_{CL}\,2_{R}$&standard&
\SubgroupDiagram{1/1,1/2,1/3,2/1,2/2,3/1,3/2,3/3}{1/1,1/2,1/3,3/3}{1/3,2/3,3/3}&
$\mathbb{Z}_{2}^{CL}$&no\\\addlinespace[9pt]
$\SU(6)_{CL}\times\SU(2)_{R'}$&$6_{CL}\,2_{R'}$&flipped&\SubgroupDiagram{1/1,1/2,1/3,2/1,2/2,3/1,3/2,3/3}{1/3,3/1,3/2,3/3}{1/1,2/1,3/1}&$\mathbb{Z}_{2}^{CL}$&yes\\\addlinespace[9pt]
$\SU(6)_{CR}\times\SU(2)_{L}$&$6_{CR}\,2_{L}$&LR-flipped&
\SubgroupDiagram{1/1,1/2,1/3,3/3}{1/1,1/2,1/3,2/1,2/2,3/1,3/2,3/3}{3/1,3/2,3/3}&$\mathbb{Z}_{2}^{CR}$&no \\\addlinespace[18pt]
$\SO(10)\times \UU_{\psi}$&$(10)\,1_{\psi}$&standard&
\SubgroupDiagram{1/1,1/2,1/3,3/3}{1/1,1/2,1/3,3/3}{1/1,1/2,2/1,2/2,3/3}&$\mathbb{Z}_{2}^{LR}$&
yes\\\addlinespace[9pt]
$\SO(10)'\times \UU_{\psi'}$&$(10)'\,1_{\psi'}$&flipped&
\SubgroupDiagram{1/1,1/2,1/3,3/3}{1/3,3/1,3/2,3/3}{1/2,1/3,2/2,2/3,3/1}&/&/\\
\bottomrule
\end{tabular}
\end{center}
\end{table}

The labeling of different cases leverages subscripts on various group factors to identify their generator content. 
Specifically, the subscripts $L,R,R'$ for $\SU(2)$ identify it as one of the cases in Eqs.~\eqref{eq:embedding-su2l}, \eqref{eq:embedding-su2r} or \eqref{eq:embedding-su2rp}; 
the double index on $\SU(6)$ specifies the identity of the two factors in its $[3][3]$ subgroup in terms of the factors $3_C$, $3_L$ and $3_R$;
$\psi$ or $\psi'$ on the $\UU$-factor identifies it as the charge defined in Eq.~\eqref{eq:generator-psi} or \eqref{eq:generator-psip}. 

Beside the label, name and diagram for each possibility of $G$, Table~\ref{tab:embeddings} also specifies information about a discrete group $D$ that is associated with each $G$. To elaborate, the unified gauge group $\mathrm{E}_{6}$ contains a discrete subgroup $D_{3}$, which acts as a permutation group on the three factors $C,L,R$ of trinification,
cf.~Appendix~B in \cite{Babu:2023zsm} for an explicit definition and detailed discussion. It is generated by three parities, each defined by the pair of trinification factors it exchanges: $LR$, $CL$ and $CR$. 
The discrete group $D$ is then defined as the subgroup of $D_{3}$ that preserves the set of generators of $G$. It can be directly determined from the subalgebra diagram, as explained in Appendix~\ref{app:subalgebra-diagram}. 

The significance of $D$ is that its action on $G$ closes, and as such $D$ may be preserved 
under first-stage spontaneous symmetry breaking $\mathrm{E}_{6}\to G$.\footnote{The ultimate fate of $D$ after the second stage is always to be broken, since the action of no part of $D$ closes on the SM group and thus any VEV breaking to the SM will automatically break $D$ completely.} If $D$ is part of $G$, then it is preserved automatically and is not of particular interest. The last column of Table~\ref{tab:embeddings} indicates three cases of interest where $D$ is not part of $G$: $3_C\,3_L\,3_R$, $6_{CL}\,2_R$ and $6_{CR}\,2_L$. The (non)preservation of $D$ then depends on the $\mathrm{E}_{6}$-irrep used for first-stage breaking and the details of the vacuum. For our case of $\mathbf{650}$, the results are as follows:
\begin{itemize}
    \item For trinification $3_C\,3_L\,3_R$, it was shown in \cite{Babu:2023zsm} that there are three discrete vacua of $V_{650}$, each preserving one of the parities $LR$, $CL$ and $CR$.
    \item For the standard $6_{CL}\,2_{R}$ and LR-flipped $6_{CR}\,2_{L}$ cases, explicit computation shows that they preserve the $CL$ and $CR$ parity, respectively, cf.~Table~\ref{tab:G-singlets-for-embeddings} in Appendix~\ref{app:state-definitions-G-singlets}.
\end{itemize}

The above discussion on parities can be further elucidated with an analogy. Consider the breaking of $\SO(10)$ to Pati-Salam $4_C\,2_L\,2_R$. The group $\SO(10)$ contains a discrete $D$-parity~\cite{Kibble:1982dd,Chang:1983fu,Chang:1984uy},\footnote{The $LR$-parity in $\mathrm{E}_{6}$ is in fact identified with $D$-parity of the standard embedding $\SO(10)\subset \mathrm{E}_{6}$. Note that it is unrelated to charge conjugation parity.} whose action on Pati-Salam closes but is not part of Pati-Salam itself. Thus a Pati-Salam vacuum may or may not preserve $D$-parity. As is well known, the Pati-Salam singlet in $\mathbf{210}$ of $\SO(10)$ breaks $D$-parity, while the one in $\mathbf{54}$ preserves it.

\section{Intermediate scenarios of interest \label{sec:minimal-models}}

In Section~\ref{sec:embeddings-intermediate} we essentially determined the gauge symmetry $G$ of possible effective theories after the first-stage breaking in Eq.~\eqref{eq:breaking-pattern} in a way that also accounts for the SM embedding. The possibilities are listed in  Table~\ref{tab:embeddings} and they are the starting point for determining intermediate scenarios of interest that are realistic.

We make the following considerations for effective theory scenarios after the first-stage breaking: 
\begin{enumerate}
\item
Notice that two of the cases, namely the standard $(10)\,1_{\psi}$ and the flipped $6_{CL}\,2_{R'}$, already contain the entire SM group in the first factor. The SM interactions thus unify already at the intermediate scale. Such a scenario is not consistent with RG running of SM gauge couplings without some intervention, such as introducing one more stage of breaking or invoking large threshold corrections at the intermediate scale to unify already there. These scenarios could be considered already under the umbrella of $\SO(10)$ and $\SU(6)$ GUTs, where the intermediate scale must be sufficiently high not to be excluded by proton decay. In such cases the ultimate unification into $\mathrm{E}_{6}$ only has implications on which $\SO(10)$ or $\SU(6)$ irreps can be present in an intermediate-scale GUT, but has no phenomenological significance for gauge coupling running or proton decay. We thus consider such cases of lesser interest for the purposes of this work and will not pursue them further.
\item
Some of the cases lead to a vacuum in which a discrete parity $D$ that is outside the group $G$ is preserved, as discussed in Section~\ref{sec:embeddings-intermediate}. In particular, $G=3_C\,3_L\,3_R$ comes with one of three preserved parities $LR$, $CL$ and $CR$. Since this parity impacts the spectrum of the intermediate theory and thus gauge coupling running, each of these cases shall be considered separately. Less importantly, $CL$ and $CR$ parity are preserved in the cases $G=6_{CL}\,2_{R}$ and $6_{CR}\,2_{L}$, respectively. Unlike the trinification case, where parities relate different factors of the semisimple group, parity acts non-trivially only on the $\SU(6)$ part in the $[6][2]$ cases, leading to no additional relations between gauge couplings or constraints on the spectrum. We shall thus often omit any reference to parities in the non-trinification cases. 
\end{enumerate}

The above two considerations lead to a list of $6$ interesting possibilities for the symmetry of the vacuum (including discrete parities) when using the $\mathbf{650}$ of $\mathrm{E}_{6}$ for first-stage breaking, as summarized in the first two columns of Table~\ref{tab:minimal-models}.

\begin{table}[htb]
\centering
\caption{The list of scenarios of interest for the symmetry of the $G$-vacuum after first-stage breaking in Eq.~\eqref{eq:breaking-pattern} (including a discrete parity, if preserved), as well as the fermionic $\mathbf{F}_{b}$ and scalar $\mathbf{S}_{a}$ content in each scenario consistent with $\text{ESH}+\mathbb{Z}_{2}^{\psi}$ (extended survival hypothesis and spinorial parity, see main text) as determined in Sections~\ref{sec:minimal-models-breaking}-\ref{sec:minimal-models-Yukawa}. The subscripts on irrep labels are $F$ for fermions and $\mathbb{C}$ for complex scalars, with the latter sometimes listed as conjugated relative to Eqs.~\eqref{eq:decomposition-to333-27}--\eqref{eq:decomposition-to101-650} according to convenience. In addition, we also specify the notation for the gauge couplings $\alpha$ for each factor, the parity-induced relation between them (when applicable), and the one- and two-loop beta coefficients $a_{i}$ and $b_{ij}$ of RG running used in Section~\ref{sec:analysis-unification}. We use the GUT normalization $\alpha_{1'}^{-1}=\tfrac{1}{24}\alpha_{\psi'}^{-1}$ in the $(10)'\,1_{\psi'}$ case.
\label{tab:minimal-models}
} 
\begin{tabular}{llp{5cm}p{2cm}cc}
\toprule
case &$G$-vacuum&field content: $\mathbf{F}_{b}$, $\mathbf{S}_{a}$&$\alpha$-labels&$a_{i}$&$b_{ij}$\\
\midrule
\addlinespace[6pt]
$1$&$3_{C}\,3_L\,3_R \rtimes \mathbb{Z}^{LR}_{2}$ &
    \hbox{$3\times(\mathbf{3},\mathbf{3},\mathbf{1})_F$, $3\times(\mathbf{\bar{3}},\mathbf{1},\mathbf{\bar{3}})_F$,}
    \hbox{$3\times(\mathbf{1},\mathbf{\bar{3}},\mathbf{3})_F$,}
    \hbox{$2\times(\mathbf{1},\mathbf{\bar{3}},\mathbf{3})_{\mathbb{C}}$, 
    $(\mathbf{1},\mathbf{\bar{6}},\mathbf{6})_{\mathbb{C}}$}
    &
    $(\alpha_{C}$,$\alpha_L$,$\alpha_{R})$\par $\alpha_L=\alpha_R$ &
   $\begin{pmatrix} -5\\1\\1\\\end{pmatrix}$&
    $\begin{pmatrix} 
        12 & 12 & 12 \\ 12 & 264 & 228 \\ 12 & 228 & 264 \\
    \end{pmatrix}$\\
$2$&$3_{C}\,3_L\,3_R \rtimes \mathbb{Z}^{CL}_{2}$ &
    \hbox{$3\times(\mathbf{3},\mathbf{3},\mathbf{1})_F$, $3\times(\mathbf{\bar{3}},\mathbf{1},\mathbf{\bar{3}})_F$,}
    \hbox{$3\times(\mathbf{1},\mathbf{\bar{3}},\mathbf{3})_F$,}
    \hbox{$2\times (\mathbf{1},\mathbf{\bar{3}},\mathbf{3})_{\mathbb{C}}$, 
    $(\mathbf{1},\mathbf{\bar{6}},\mathbf{6})_{\mathbb{C}}$,}
    \hbox{$2\times(\mathbf{\bar{3}},\mathbf{1},\mathbf{\bar{3}})_{\mathbb{C}}$, 
    $(\mathbf{\bar{6}},\mathbf{1},\mathbf{\bar{6}})_{\mathbb{C}}$}
    &
    $(\alpha_{C}$,$\alpha_L$,$\alpha_{R})$\par $\alpha_C=\alpha_L$ &
    $\begin{pmatrix} 1\\ 1\\ 7\\ \end{pmatrix}$&
    $\begin{pmatrix} 
        264 & 12 & 228 \\ 12 & 264 & 228 \\ 228 & 228 & 516 \\
    \end{pmatrix}$\\
$3$&$3_{C}\,3_L\,3_R \rtimes \mathbb{Z}^{CR}_{2}$ &
    \hbox{$3\times(\mathbf{3},\mathbf{3},\mathbf{1})_F$, $3\times(\mathbf{\bar{3}},\mathbf{1},\mathbf{\bar{3}})_F$,}
    \hbox{$3\times(\mathbf{1},\mathbf{\bar{3}},\mathbf{3})_F$,}
    \hbox{$2\times (\mathbf{1},\mathbf{\bar{3}},\mathbf{3})_{\mathbb{C}}$, 
    $(\mathbf{1},\mathbf{\bar{6}},\mathbf{6})_{\mathbb{C}}$,}
    \hbox{$2\times(\mathbf{3},\mathbf{3},\mathbf{1})_{\mathbb{C}}$, 
    $(\mathbf{6},\mathbf{6},\mathbf{1})_{\mathbb{C}}$}
    &
    $(\alpha_{C}$,$\alpha_L$,$\alpha_{R})$\par $\alpha_C=\alpha_R$ &
    $\begin{pmatrix} 1\\ 7\\ 1\\ \end{pmatrix}$&
    $\begin{pmatrix} 
        264 & 228 & 12 \\ 228 & 516 & 228 \\ 12 & 228 & 264 \\ 
    \end{pmatrix}$\\
$4$&$6_{CL}\, 2_{R}\rtimes \mathbb{Z}_{2}^{CL}$ &
    \hbox{$3\times(\mathbf{15},\mathbf{1})_F$, $3\times(\mathbf{\bar{6}},\mathbf{2})_F$,}
    \hbox{$(\mathbf{15},\mathbf{1})_\mathbb{C}$, $(\mathbf{\overline{21}},\mathbf{3})_\mathbb{C}$,}
    \hbox{$(\mathbf{\bar{6}},\mathbf{2})_\mathbb{C}$, $(\mathbf{84},\mathbf{2})_\mathbb{C}$}
    &
    $(\alpha_{CL}$, $\alpha_{R})$ &
    $\begin{pmatrix}\tfrac{5}{3}\\ \tfrac{83}{3}\\\end{pmatrix}$&
    $\begin{pmatrix}
        \tfrac{9779}{6} & \tfrac{435}{2} \\
        \tfrac{5075}{2} & \tfrac{3691}{6} \\ 
        \end{pmatrix}$\\
$5$&$6_{CR}\, 2_{L} \rtimes \mathbb{Z}_{2}^{CR}$ &
    \hbox{$3\times(\mathbf{15},\mathbf{1})_F$, $3\times(\mathbf{\bar{6}},\mathbf{2})_F$,}
    \hbox{$(\mathbf{15},\mathbf{1})_\mathbb{C}$, $(\mathbf{\overline{105'}},\mathbf{1})_\mathbb{C}$,} 
    \hbox{$(\mathbf{\bar{6}},\mathbf{2})_\mathbb{C}$, $(\mathbf{84},\mathbf{2})_\mathbb{C}$}
    &
    $(\alpha_{CR}$, $\alpha_{L}$) &
    $\begin{pmatrix}\tfrac{25}{3}\\\tfrac{41}{3}\\\end{pmatrix}$&
    $\begin{pmatrix} 
            \frac{16531}{6} & \frac{243}{2} \\
            \frac{2835}{2} & \frac{1339}{6} \\
        \end{pmatrix}$\\
$6$&$(10)'\,1_{\psi'}$ &
    \hbox{$3\times(\mathbf{16},+1)_F$, $3\times(\mathbf{10},-2)_F$,}
    \hbox{$3\times(\mathbf{1},+4)_F$,}
    \hbox{$(\mathbf{16},+1)_\mathbb{C}$, $(\mathbf{126},+2)_\mathbb{C}$, $(\mathbf{10},-2)_\mathbb{C}$}
    &
    $(\alpha_{10'}$, $\alpha_{1'})$ &
    $\begin{pmatrix} -\tfrac{32}{3}\\\tfrac{124}{9}\\\end{pmatrix}$&
    $\begin{pmatrix} 
        \frac{9749}{6} & \frac{155}{6} \\
        \frac{2325}{2} & \frac{355}{18} \\
    \end{pmatrix}$\\
\bottomrule
\end{tabular}
\end{table}

For each of these possibilities, though, there are potentially still many choices of field content in the intermediate theory. As discussed in Section~\ref{sec:two-stage-breaking-and-masses},
some $G$-irreps need to be present around the intermediate scale so as to trigger the second stage of symmetry breaking in Eq.~\eqref{eq:breaking-pattern}, and to also provide a suitable low-energy SM Higgs doublet for the Yukawa sector. We thus want to determine the minimal set of $G$-irreps from the decompositions in Eqs.~\eqref{eq:decomposition-to333-27}--\eqref{eq:decomposition-to101-650} to achieve both these goals, i.e.~we are interested in the extended survival hypothesis (ESH) scenario~\cite{Dimopoulos:1984ha}. Note that the field content must be considered not only in terms of the group $G$, but in terms of its embedding, since the SM content of $G$-irreps differs based on embedding.

Furthermore we limit ourselves in this work to scenarios in which the fermion sector is $\SO(10)$-like, namely the light mass eigenmodes of chiral fermions are entirely in the $\mathbf{16}$ of the standard $\SO(10)$, and the heavy mass eigenmodes of vector-like exotics are in the $\mathbf{10}$. 
Mass mixing between the two irreps can easily be eliminated if all \textit{spinorial VEVs} vanish, i.e.~those SM-singlet VEVs that are part of a spinorial ($\mathbf{16}$ or $\mathbf{144}$) irrep of the standard $\SO(10)$ within $\mathrm{E}_{6}$. Since the $\SO(10)$ part of the symmetry already requires all such spinorial VEVs to be present in pairs in all terms of the scalar potential, taking them to vanish is a self-consistent ansatz (for solving the stationarity conditions). 

The choice of vanishing spinorial VEVs can be conveniently described through symmetry. Notice that under the decompositions $\mathrm{E}_{6}\supset \SO(10)\times\UU$ of Eqs.~\eqref{eq:decomposition-to101-27}--\eqref{eq:decomposition-to101-650}, the spinorial representations of $\SO(10)$ have odd charges, while the non-spinorial ones have even charges.
For the standard $\SO(10)$ embedding, the associated charge factor is $\UU_{\psi}$. The spinorial ansatz is thus equivalent to stating that only fields with even $\psi$-charge can acquire a VEV, i.e.~the \textit{spinorial parity} $\mathbb{Z}_{2}^{\psi}$ is preserved.\footnote{A note of clarification: spinorial parity is always defined by the $\psi$-charge, cf.~Eq.~\eqref{eq:generator-psi}, regardless of the choice of intermediate symmetry $G$. Since every embedding of $G$ in Table~\ref{tab:embeddings} is of maximum rank, $\psi$ must necessarily be part of $G$, even if the basis of diagonal generators most useful to discuss $G$ is not aligned with it. In particular, even in the flipped $(10)'\,1_{\psi'}$ case, spinorial parity is defined by $\psi$. } The spinorial parity of a field can be explicitly defined as $(-1)^{\psi+2S}$, where $S$ is the spin of the particle, and it phenomenologically plays the same role as $T$-parity~\cite{Cheng:2003ju,Babu:2021hef} or $R$-parity~\cite{Aulakh:2000sn,Aulakh:2003kg} in other contexts. 

A straightforward consequence of spinorial parity is that the lightest $\psi$-odd scalar cannot decay into two SM fermions (since they are $\psi$-odd as well), and we thus obtain a dark matter (DM) candidate. In our case we envision this to be a $\psi$-odd scalar doublet $(\mathbf{1},\mathbf{2},\pm 1/2)$ of $3_C\,2_L\,1_Y$, leading to the case of inert doublet DM~\cite{Barbieri:2006dq,Ma:2006km,LopezHonorez:2006gr}.

We note that the concept of spinorial parity and its applicability to DM~\cite{Kadastik:2009cu} was already pointed out in the context of non-supersymmetric $\SO(10)$ GUTs~\cite{Mambrini:2015vna}, where the concept can first be defined. Unlike in $\SO(10)$ GUT, where spinorial parity is useful only if non-spinorial fermionic irreps or spinorial scalar irreps are added to the field content, the $\mathrm{E}_{6}$ case already has both spinorial and non-spinorial components automatically embedded in its irreps. The concept is thus intrinsic to any $\mathrm{E}_{6}$ GUT and thus better structurally motivated. 

Our goal for the remainder of this section is to determine the scalar field content consistent with the \textit{extended survival hypothesis} and \textit{spinorial parity}, abbreviated as the $\text{ESH}+\mathbb{Z}_{2}^{\psi}$ scenario. We analyze the second-stage breaking requirements in Section~\ref{sec:minimal-models-breaking} and the Yukawa requirements in Section~\ref{sec:minimal-models-Yukawa}. The results for the required scalar content $\mathbf{S}_{a}$ of the minimal scenarios are summarized already in the third column of Table~\ref{tab:minimal-models}; we refer to these intermediate effective theories as $G$-theories.\footnote{A further note of clarification on the terminology used throughout the paper: depending on context, we refer to the choice of ``group $G$'' when referring to the possibilities in Eq.~\eqref{eq:G-choices}, to ``$G$-embeddings'' when referring to an embedded $G$ with fixed representation theory from Table~\ref{tab:embeddings}, to a ``$G$-vacuum'' when referring to a choice in the first column of Table~\ref{tab:minimal-models} when discrete parity comes into play, and to a ``$G$-theory'' when referring to the entire row of data in Table~\ref{tab:minimal-models} having the scalar content of the $G$-vacuum fixed under the $\text{ESH}+\mathbb{Z}_{2}^{\psi}$ scenario.}

\subsection{Requirements for second-stage breaking  \label{sec:minimal-models-breaking}}

In accordance with Eq.~\eqref{eq:breaking-pattern}, the second stage of symmetry breaking should break the group $G$ into the SM group. This is achieved by some $G$-irreps that survive down to the intermediate scale, which obtain VEVs in the direction of SM-singlets. Our task is to identify the simplest scenario under which this can occur for each case of $G$-vacuum in Table~\ref{tab:minimal-models}. 

We begin by listing all SM-singlets available to us in the scalar sector $\mathbf{27}\oplus\mathbf{351'}\oplus\mathbf{650}$. We label the $2$ complex SM-singlets in $\mathbf{27}$ by $V_{i}$, the $5$ complex SM-singlets in $\mathbf{351'}$ by $W_{i}$, and the $11$ real SM-singlets in $\mathbf{650}$ by\footnote{In the notation used for SM-singlet VEVs, we use capital letters for complex VEVs and small letters for real VEVs, thus $X_{3}$ and $Y_{3}$ contain $2$ real VEVs each.}
\begin{align}
s,\ a,\ z,\ x_{1},\ x_{2},\ X_{3},\ y_{1},\ y_{2},\ Y_{3}.\label{eq:SM-vevs-650}
\end{align}
Note that the VEV notation of Eq.~\eqref{eq:SM-vevs-650} is identical to the one used in~\cite{Babu:2023zsm}. The exact definitions of all VEVs in $\mathbf{27}\oplus\mathbf{351'}\oplus\mathbf{650}$ are given in Appendix~\ref{app:state-definitions-SMsinglets}. 

According to the breaking scheme in Eq.~\eqref{eq:breaking-pattern}, the second stage should be triggered by $G$-irreps in $\mathbf{27}\oplus\mathbf{351'}$, so we first focus on the VEVs $V_{i}$ and $W_{i}$, and discuss the possible involvement of the $G$-irreps in $\mathbf{650}$ afterwards.

We first determine the location of VEVs $V_{i}$ and $W_{i}$ in terms of $G$-irreps of all possible $G$-embeddings from Table~\ref{tab:embeddings}; the results are shown in Table~\ref{tab:vevs-27-and-351p}.
Curiously, it is non-trivial that all VEVs can be defined in such a way that they are contained in a single $G$-irrep for all $G$-embeddings.\footnote{Indeed, the usual situation is reflected by SM-singlets in $\mathbf{650}$, cf.~Appendix~\ref{app:state-definitions-SMsinglets}, where the VEVs adapted to $G$-irreps depend on $G$.}

Since we are investigating the $\text{ESH}+\mathbb{Z}_{2}^{\psi}$ scenarios, the parity-odd VEVs under $\mathbb{Z}_{2}^{\psi}$ vanish. These VEVs can be identified by their inclusion into a spinorial ($\mathbf{16}$ or $\mathbf{144}$) irrep of the standard $(10)\,1_{\psi}$ embedding. This implies 
\begin{align}
    V_{1}=W_{1}=W_{4}&=0, \label{eq:ansatz-spinorial-VEVs}
\end{align}
indicating that only $V_{2}$, $W_{2}$, $W_{3}$ and $W_{5}$ are possibly non-vanishing.

The requirements for second-stage breaking are illuminated further by considering what happens to the rank of the groups in the breaking pattern of Eq.~\eqref{eq:breaking-pattern}. The group $G$ for all cases in Table~\ref{tab:minimal-models} is a maximal regular subgroup of $\mathrm{E}_{6}$ and as such has rank $6$, while the SM group has rank $4$. This requires the second stage VEVs to break rank by $2$, namely the charges $\UU_{\chi}$ and $\UU_{\psi}$ must both be broken. The $(\chi,\psi)$ charges of SM-singlets are specified in Table~\ref{tab:vevs-27-and-351p}. We immediately notice that $V_{2}$, $W_{2}$ and $W_{5}$ all preserve the $\chi$-charge, which must then necessarily be broken by the only remaining non-vanishing VEV $W_{3}$.\footnote{If $W_{3}$ is chosen at the $\mathrm{TeV}$ scale, one can obtain a light $Z'$ corresponding to the $\chi$-charge.}

This insight can be cross-checked by explicit computation. In Appendix~\ref{app:masses-gauge} we provide the mass expressions for $\mathrm{E}_{6}$ gauge bosons in terms of the non-spinorial SM-singlet VEVs in $\mathbf{27}\oplus\mathbf{351'}$ and the $G$-singlet VEV in $\mathbf{650}$. For any of the three pairs $\{W_{3},V_{2}\}$, $\{W_{3},W_{2}\}$, or $\{W_{3},W_{5}\}$, the gauge boson spectrum confirms that only SM gauge boson masses vanish. If one instead takes $W_{3}=0$, rank cannot be broken by $2$, as seen from the mass matrix of SM-singlets $(\mathbf{1},\mathbf{1},0)$.  The second stage of the breaking thus indeed requires $W_{3}$ and at least one VEV from the list $\{V_{2},W_{2},W_{5}\}$ to be non-vanishing. 

Although it is possible to analyze the breaking in further detail for each case, the above conclusion will already prove sufficient once the considerations from the Yukawa sector come into play.

\begin{table}[htb]
    \centering
    \caption{Location of SM VEVs from $\mathbf{27}\oplus\mathbf{351'}$ in $G$-irreps for all embeddings of $G$ in Table~\ref{tab:embeddings}. Also, we specify for each VEV its $(\chi,\psi)$ charges from Eqs.~\eqref{eq:generator-chi} and \eqref{eq:generator-psi}.\label{tab:vevs-27-and-351p}}
    \begin{tabular}{lrrrrrrrrr}
    \toprule
    vev& $\mathrm{E}_{6}$ &$\chi$&$\psi$&$3_C\,3_L\,3_R$ & $6_{CL}\,2_{R}$ & $6_{CL}\,2_{R'}$ & $6_{CR}\,2_L$ & $(10)\,1_{\psi}$ & $(10)'\,1_{\psi'}$\\
    \midrule
    $V_{1}$&$\mathbf{27}$& $-5$&$+1$&
        $(\mathbf{1},\mathbf{\bar{3}},\mathbf{3})$&
        $(\mathbf{\bar{6}},\mathbf{2})$& 
        $(\mathbf{\bar{6}},\mathbf{2})$& 
        $(\mathbf{15},\mathbf{1})$& 
        $(\mathbf{16},+1)$& 
        $(\mathbf{16},+1)$\\
    $V_{2}$&$\mathbf{27}$& $0$&$+4$&
        $(\mathbf{1},\mathbf{\bar{3}},\mathbf{3})$& 
        $(\mathbf{15},\mathbf{1})$& 
        $(\mathbf{\bar{6}},\mathbf{2})$& 
        $(\mathbf{15},\mathbf{1})$& 
        $(\mathbf{1},+4)$& 
        $(\mathbf{16},+1)$\\
    $W_{1}$&$\mathbf{351'}$& $+5$&$-1$&
        $(\mathbf{1},\mathbf{3},\mathbf{\bar{3}})$&
        $(\mathbf{84},\mathbf{2})$& 
        $(\mathbf{84},\mathbf{2})$& 
        $(\mathbf{\overline{15}},\mathbf{1})$& 
        $(\overline{\mathbf{144}},-1)$& 
        $(\overline{\mathbf{144}},-1)$\\
    $W_{2}$&$\mathbf{351'}$& $0$&$-4$&
        $(\mathbf{1},\mathbf{3},\mathbf{\bar{3}})$&
        $(\mathbf{\overline{15}},\mathbf{1})$& 
        $(\mathbf{84},\mathbf{2})$& 
        $(\mathbf{\overline{15}},\mathbf{1})$& 
        $(\mathbf{54},-4)$& 
        $(\overline{\mathbf{144}},-1)$\\
    $W_{3}$&$\mathbf{351'}$& $-10$&$+2$&
        $(\mathbf{1},\mathbf{\bar{6}},\mathbf{6})$&
        $(\mathbf{\overline{21}},\mathbf{3})$& 
        $(\mathbf{\overline{21}},\mathbf{3})$& 
        $(\overline{\mathbf{105'}},\mathbf{1})$& 
        $(\mathbf{126},+2)$& 
        $(\mathbf{126},+2)$\\
    $W_{4}$&$\mathbf{351'}$& $-5$&$+5$&
        $(\mathbf{1},\mathbf{\bar{6}},\mathbf{6})$&
        $(\mathbf{84},\mathbf{2})$& 
        $(\mathbf{\overline{21}},\mathbf{3})$& 
        $(\overline{\mathbf{105'}},\mathbf{1})$& 
        $(\mathbf{16},+5)$& 
        $(\mathbf{126},+2)$\\
    $W_{5}$&$\mathbf{351'}$& $0$&$+8$&
        $(\mathbf{1},\mathbf{\bar{6}},\mathbf{6})$&
        $(\overline{\mathbf{105'}},\mathbf{1})$& 
        $(\mathbf{\overline{21}},\mathbf{3})$& 
        $(\overline{\mathbf{105'}},\mathbf{1})$& 
        $(\mathbf{1},+8)$& 
        $(\mathbf{126},+2)$\\
    \bottomrule
    \end{tabular}
\end{table}

We conclude the discussion on second-stage breaking by commenting on the possibility that some $G$-irrep in $\mathbf{650}$ is involved, i.e.~that the VEVs in Eq.~\eqref{eq:SM-vevs-650} are involved also in the second-stage breaking. An immediate observation is that some of the VEVs in $\mathbf{27}\oplus\mathbf{351'}$ must be involved regardless, otherwise we break to the SM with only a single irrep $\mathbf{650}$ in violation of Michel's conjecture. Furthermore, the irrep $\mathbf{650}$ is real and it contains only two complex VEVs $X_{3}$ and $Y_{3}$ charged under $(\chi,\psi)$. As seen from Table~\ref{tab:SM-singlets-in-650} in Appendix~\ref{app:state-definitions-SMsinglets}, however, these VEVs are $\psi$-odd and thus $X_{3}=Y_{3}=0$ under our assumption of $\mathbb{Z}_{2}^{\psi}$ preservation. Since the rank-breaking requirements for the VEVs in $\mathbf{27}\oplus\mathbf{351'}$ automatically lead to a complete breakdown to the SM group anyway, any addition of real VEVs from $\mathbf{650}$ is redundant for the breaking and thus not part of the minimal scenario under ESH. We thus see that the non-involvement of $\mathbf{650}$ in the second stage, cf.~Eq.~\eqref{eq:breaking-pattern}, is justified under the $\text{ESH}+\mathbb{Z}_{2}^{\psi}$ scenario.

\subsection{Requirements for realistic Yukawa sector \label{sec:minimal-models-Yukawa}}

The Yukawa interactions of the model, written in $\mathrm{E}_{6}$ formalism, are given in Eq.~\eqref{eq:Yukawa-sector}. We now study these terms in detail and determine the required scalar content for each $G$-vacuum case of Table~\ref{tab:minimal-models} that allows realistic fermion masses under the assumptions of the $\text{ESH}+\mathbb{Z}_{2}^{\psi}$ scenario.  

We first determine the general form Yukawa terms take after EW symmetry breaking. The fermionic $\psi\sim\mathbf{27}$ of each family consists of SM irreps given in Eqs.~\eqref{eq:SM-fermions} and \eqref{eq:exotic-fermions}, which in the EW-broken phase with symmetry $3_C\,1_{Q}$ further group in the following way: 
\def\FERU{\mathbf{u}}
\def\FERUC{\mathbf{\bar{u}}}
\def\FERD{\mathbf{d}}
\def\FERDC{\mathbf{\bar{d}}}
\def\FERE{\mathbf{e}}
\def\FEREC{\mathbf{\bar{e}}}
\def\FERN{\mathbf{\boldsymbol{\nu}}}
\def\FERNC{\bar{\boldsymbol{\nu}}}
\def\MFU{\mathbf{M}_U}
\def\MFD{\mathbf{M}_D}
\def\MFE{\mathbf{M}_E}
\def\MFN{\mathbf{M}_N}
\def\MFR{\mathbf{M}_R}
\def\MGUT{M_U}
\def\MI{M_I}
\def\MZ{M_Z}
\def\MPL{M_{Pl}}
\newcommand\matrixp[2]{
    \begin{pmatrix}
        #1 \\ #2 \\
    \end{pmatrix}
}
\begin{align}
    \FERU&:= (u_I), &
        \FERUC &:= (u^{c}_{I}), \label{eq:fermion-basis-U}\\
    \FERD&:= (d_I, d'_I),&
        \FERDC&:= (d^{c}_I, d'^{c}_I), \label{eq:fermion-basis-D}\\
    \FERE&:= (e_{I},e'_{I}),&
        \FEREC &:= (e^{c}_{I},e'^{c}_{I}), \label{eq:fermion-basis-E}\\
    \FERN &:= (\nu_I,\nu'_{I}),&
        \FERNC &:= (\nu^{c}_{I},n_{I},\nu'^{c}_{I}), \label{eq:fermion-basis-N}
\end{align}
where $I=1,2,3$ is a family index. The generated mass terms in the EW-broken phase can then be written in matrix notation as
\begin{align}
    \mathcal{L}_{Y}&\supset 
        \FERU^{T}\,\MFU\,\FERUC \;+\; 
        \FERDC^{T}\,\MFD\,\FERD \;+\;
        \FERE^{T}\,\MFE\,\FEREC \;+\;
        \FERN^{T}\,\MFN\,\FERNC \;+\;
        \tfrac{1}{2}\,\FERNC^{T}\,\MFR\,\FERNC, \label{eq:Yukawa-terms-explicit}
\end{align}
where the subscripts $\{U,D,E,N\}$ in mass matrices denote respectively the up, down, charged lepton and neutrino (Dirac) mass matrices, while $\MFR$ includes some Majorana type masses. We omitted the type II seesaw contributions that should also be present for neutrinos, since they will not be relevant for the subsequent discussion.

Two different types of SM irreps in the scalar $\PSI\sim\mathbf{27}$ and $\THETA\sim\mathbf{351'}$ acquire VEVs:
\begin{enumerate}
    \item \underline{SM singlets $(\mathbf{1},\mathbf{1},0)$}: they acquire intermediate-scale VEVs in the second-stage breaking. These are listed in Table~\ref{tab:vevs-27-and-351p}. 
    \item \underline{Weak doublets $(\mathbf{1},\mathbf{2},\pm\tfrac{1}{2})$:} one linear combination is the SM Higgs doublet, and so the doublets acquire VEVs of EW scale and proportional to the admixture of the Higgs inside them. We label the VEVs of the EM neutral component for doublets with hypercharge $+1/2$ by $\delta_{i}$. There are $11$ such doublets in the representations $\mathbf{27}$ and $\mathbf{351'}$ or their conjugates. Their location in terms of $G$-representations has been determined explicitly and is shown in Table~\ref{tab:SM-doublets}, while their precise definition can be found in Appendix~\ref{app:state-definitions-SMdoublets}. 
\end{enumerate}
After both types of VEVs are engaged, the following explicit form for the mass matrices in Eq.~\eqref{eq:Yukawa-terms-explicit} is obtained (with all entries representing $3\times 3$ blocks):
\begingroup
\allowdisplaybreaks
\begin{align}
\MFU&= 
    -\YPSIBF \BLUE{\delta_{7}}+\tfrac{1}{\sqrt{15}}\YTHETABF \BLUE{\delta_{8}}, \label{eq:fermion-mass-U}\\[4pt]
\MFD&= 
    \begin{pmatrix}
        \YPSIBF \,\BLUE{\delta_{1}^{\ast}}-\frac{1}{\sqrt{15}}\YTHETABF  \BLUE{\delta_{2}^{\ast}} & \YPSIBF \RED{V_{1}}+\frac{1}{\sqrt{15}}\YTHETABF \,\RED{W_{1}}^{\ast} \\
        -\YPSIBF  \BLUE{\delta_{4}^{\ast}}+\frac{1}{\sqrt{15}}\YTHETABF \BLUE{\delta_{5}^{\ast}} & 
        -\YPSIBF \RED{V_{2}}+\frac{1}{\sqrt{15}}\YTHETABF \,\RED{W_{2}}^{\ast} \\
    \end{pmatrix}, \label{eq:fermion-mass-D}\\[4pt]
\MFE&= 
    \begin{pmatrix}
        -\YPSIBF  \BLUE{\delta_{1}^{\ast}}-\frac{1}{2} \sqrt{\frac{3}{5}} \YTHETABF  
            \BLUE{\delta_{2}^{\ast}}+\frac{1}{2}\YTHETABF  \BLUE{\delta_{3}^{\ast}} & 
        \YPSIBF \RED{V_{1}}-\frac{1}{2} \sqrt{\frac{3}{5}} \YTHETABF \RED{W_{1}}^{\ast} \\
        \YPSIBF  \BLUE{\delta_{4}^{\ast}}+\frac{1}{2} \sqrt{\frac{3}{5}} \YTHETABF  
            \BLUE{\delta_{5}^{\ast}}+\frac{1}{2}\YTHETABF  \BLUE{\delta_{6}^{\ast}} & 
        - \YPSIBF \RED{V_{2}}-\frac{1}{2} \sqrt{\frac{3}{5}} \YTHETABF \RED{W_{2}}^{\ast} \\
    \end{pmatrix}, \label{eq:fermion-mass-E}\\[4pt]
\MFN&= 
    \begin{pmatrix}
        \YPSIBF  \BLUE{\delta_{7}}+\frac{1}{2} \sqrt{\frac{3}{5}} \YTHETABF  \BLUE{\delta_{8}}- 
            \frac{1}{2}\YTHETABF  \BLUE{\delta_{9}} & 
        -\frac{1}{\sqrt{2}} \YTHETABF  \BLUE{\delta_{11}} & 
        \YPSIBF \RED{V_{1}} -\frac{1}{2} \sqrt{\frac{3}{5}} \YTHETABF \RED{W_{1}}^{\ast}  \\
        -\frac{1}{\sqrt{2}} \YTHETABF  \BLUE{\delta_{10}} & 
        -\YPSIBF  \BLUE{\delta_{7}}-\frac{1}{2} \sqrt{\frac{3}{5}} \YTHETABF  \BLUE{\delta_{8}}-
            \frac{1}{2}\YTHETABF  \BLUE{\delta_{9}} & 
        -\YPSIBF \RED{V_{2}} -\frac{1}{2} \sqrt{\frac{3}{5}} \YTHETABF \RED{W_{2}}^{\ast}  \\
    \end{pmatrix}, \label{eq:fermion-mass-N}\\[4pt]
\MFR &= 
    \begin{pmatrix}
        \YTHETABF \RED{W_{3}}^{\ast}  & 
        \frac{1}{\sqrt{2}} \YTHETABF \RED{W_{4}}^{\ast} & 
        -\YPSIBF  \BLUE{\delta_{4}^{\ast}}-\frac{1}{2} \sqrt{\frac{3}{5}} \YTHETABF 
            \BLUE{\delta_{5}^{\ast}}+\frac{1}{2} \YTHETABF  \BLUE{\delta_{6}^{\ast}} \\
        \frac{1}{\sqrt{2}} \YTHETABF \RED{W_{4}}^{\ast} & 
        \YTHETABF \RED{W_{5}}^{\ast} & 
        \YPSIBF  \BLUE{\delta_{1}^{\ast}}+\frac{1}{2} \sqrt{\frac{3}{5}} \YTHETABF  
            \BLUE{\delta_{2}^{\ast}}+\frac{1}{2} \YTHETABF  \BLUE{\delta_{3}^{\ast}}\\
        -\YPSIBF \BLUE{\delta_{4}^{\ast}}-\frac{1}{2} \sqrt{\frac{3}{5}} \YTHETABF  
            \BLUE{\delta_{5}^{\ast}}+\frac{1}{2} \YTHETABF  \BLUE{\delta_{6}^{\ast}}& 
        \YPSIBF \BLUE{\delta_{1}^{\ast}}+\frac{1}{2} \sqrt{\frac{3}{5}} \YTHETABF 
            \BLUE{\delta_{2}^{\ast}}+\frac{1}{2} \YTHETABF  \BLUE{\delta_{3}^{\ast}}& 
        0 \\
    \end{pmatrix}. \label{eq:fermion-mass-R}
\end{align}
\endgroup

\begin{table}[htb]
\caption{The location of EW VEVs $\delta_{i}$ in terms of $G$-irreps for the cases of Table~\ref{tab:minimal-models}. The VEVs correspond to electrically-neutral components of weak doublets $(\mathbf{1},\mathbf{2},+1/2)$ in the $\mathbf{27}\oplus\mathbf{351'}$ part of the model's scalar sector. The trinification location and $R$-charge from $t^{3}_R$ uniquely specify the states (up to phase), and they also have well-defined charges $\psi$; their explicit definitions can be found in Appendix~\ref{app:state-definitions-SMdoublets}. The same basis of doublets cannot be well-adapted to both $3_C\,3_L\,3_R$ and $(10)'\,1_{\psi'}$, cf.~also Eq.~\eqref{eq:delta-relation}, hence the separation into table (a) and (b), respectively.   
\label{tab:SM-doublets}}
\begin{minipage}[t]{.70\linewidth}
\centering
\caption*{(a) Doublet locations in irreps of $3_C\,3_L\,3_R$,\ \  $6_{CL}\,2_{R}$,\ \  $6_{CR}\,2_{L}$,}
\begin{tabular}{lrrlllr@{$\;\sim\;$}l}
    \toprule
     Label & $1_\psi$ &$2R$ & $3_{C}\,3_L\,3_R$ & $6_{CL}\,2_{R}$ & $6_{CR}\,2_{L}$ &\multicolumn{2}{c}{$\mathrm{E}_{6}$ irrep}\\
     \midrule
     $\delta_{1}$  &$+2$& $+1$  & $(\mathbf{1},\mathbf{\bar{3}},\mathbf{3})^\ast$  & $(\mathbf{\bar{6}},\mathbf{2})^\ast$
        & $(\mathbf{\bar{6}},\mathbf{2})^\ast$ & $\mathbf{27}^\ast$ & $\PSI^\ast$\\
     $\delta_{2}$  &$+2$& $+1$  & $(\mathbf{1},\mathbf{\bar{3}},\mathbf{3})^\ast$  & $(\mathbf{84},\mathbf{2})$    
        & $(\mathbf{84},\mathbf{2})$  & $\mathbf{351'}$ & $\THETA$ \\
     $\delta_{3}$  &$+2$& $+1$  & $(\mathbf{1},\mathbf{\bar{6}},\mathbf{6})$       & $(\mathbf{84},\mathbf{2})$               
        & $(\mathbf{84},\mathbf{2})$  & $\mathbf{351'}$ & $\THETA$\\
     $\delta_{4}$  &$-1$& $+0$  & $(\mathbf{1},\mathbf{\bar{3}},\mathbf{3})^\ast$  & $(\mathbf{15},\mathbf{1})^\ast$ 
        & $(\mathbf{\bar{6}},\mathbf{2})^\ast$  & $\mathbf{27}^\ast$ & $\PSI^\ast$\\
     $\delta_{5}$  &$-1$& $+0$  & $(\mathbf{1},\mathbf{\bar{3}},\mathbf{3})^\ast$  & $(\mathbf{15},\mathbf{1})^\ast$  
        & $(\mathbf{84},\mathbf{2})$  & $\mathbf{351'}$ & $\THETA$\\
     $\delta_{6}$  &$-1$& $+0$  & $(\mathbf{1},\mathbf{\bar{6}},\mathbf{6})$       & $(\mathbf{\overline{21}},\mathbf{3})$  
        & $(\mathbf{84},\mathbf{2})$  & $\mathbf{351'}$ & $\THETA$\\
     $\delta_{7}$  &$-2$& $+1$  & $(\mathbf{1},\mathbf{\bar{3}},\mathbf{3})$       & $(\mathbf{\bar{6}},\mathbf{2})$    
        & $(\mathbf{\bar{6}},\mathbf{2})$  & $\mathbf{27}$ & $\PSI$\\
     $\delta_{8}$  &$-2$& $+1$  & $(\mathbf{1},\mathbf{\bar{3}},\mathbf{3})$       & $(\mathbf{84},\mathbf{2})^\ast$  
        & $(\mathbf{84},\mathbf{2})^\ast$  & $\mathbf{351'}^\ast$ & $\THETA^\ast$\\
     $\delta_{9}$  &$-2$& $+1$  & $(\mathbf{1},\mathbf{\bar{6}},\mathbf{6})^\ast$  & $(\mathbf{84},\mathbf{2})^\ast$  
        & $(\mathbf{84},\mathbf{2})^\ast$  & $\mathbf{351'}^\ast$ & $\THETA^\ast$\\
     $\delta_{10}$ &$+1$& $+2$  & $(\mathbf{1},\mathbf{\bar{6}},\mathbf{6})^\ast$  & $(\mathbf{\overline{21}},\mathbf{3})^\ast$  
        & $(\mathbf{84},\mathbf{2})^\ast$  & $\mathbf{351'}^\ast$ & $\THETA^\ast$\\
     $\delta_{11}$ &$-5$& $+0$  & $(\mathbf{1},\mathbf{\bar{6}},\mathbf{6})^\ast$  & $(\mathbf{\overline{105'}},\mathbf{1})^\ast$
        & $(\mathbf{84},\mathbf{2})^\ast$  & $\mathbf{351'}^\ast$ & $\THETA^\ast$\\
     \bottomrule
\end{tabular}
\end{minipage}\hspace{0.3cm}
\begin{minipage}[t]{.23\linewidth}
\centering
\caption*{(b) and $(10)'\,1_{\psi'}$.}
\begin{tabular}{lll}
    \toprule
    Label& $(10)'\,1_{\psi'}$ & $\mathrm{E}_{6}$\\
    \midrule
    $\delta_{1}$            & $(\mathbf{10},-2)^\ast$               & $\PSI^\ast$\\
    $\tilde{\delta}_{2}$    & $(\mathbf{10},-2)^\ast$               & $\THETA$ \\
    $\tilde{\delta}_{3}$    & $(\mathbf{126},+2)$                   & $\THETA$\\
    $\delta_{4}$            & $(\mathbf{10},-2)^\ast$               & $\PSI^\ast$\\
    $\tilde{\delta}_{5}$    & $(\mathbf{10},-2)^\ast$               & $\THETA$\\
    $\tilde{\delta_{6}}$    & $(\mathbf{126},+2)$                   & $\THETA$\\
    $\delta_{7}$            & $(\mathbf{16},+1)$                    & $\PSI$\\
    $\delta_{8}$            & $(\mathbf{\overline{144}},-1)^\ast$   & $\THETA^\ast$\\
    $\delta_{9}$            & $(\mathbf{\overline{144}},-1)^\ast$   & $\THETA^\ast$\\
    $\delta_{10}$           & $(\mathbf{\overline{144}},-1)^\ast$   & $\THETA^\ast$\\
    $\delta_{11}$           & $(\mathbf{\overline{144}},-1)^\ast$   & $\THETA^\ast$\\
    \bottomrule
\end{tabular}
\end{minipage}
\end{table}

\noindent
Let us briefly comment on the structure of the general result in Eqs.~\eqref{eq:fermion-mass-U}--\eqref{eq:fermion-mass-R}:
\begin{itemize}
    \item For better visual clarity we colored the intermediate scale VEVs as \RED{red}, while the EW scale VEVs are \BLUE{blue}. The pattern clearly shows that all fermions have either EW or intermediate scale masses, denoted by $\MZ$ and $\MI$ respectively. The former include states corresponding to SM fermions, while the latter are the right-handed neutrinos $\nu^{c}$ and $n$, as well as a vector-like pair of lepton doublets and down-type quarks. More specifically, up to leading order corrections in $\MZ/\MI$, the vector-like lepton doublet pair consists of $L'^{c}$  and a linear combination of $L$ and $L'$, while the down-type vector-like pair consists of $d'^{c}$ and a linear combination of $d$ and $d'$. This shows that in general, the SM leptons and down-type quarks are also an admixture of the states from the $\mathbf{16}$ and $\mathbf{10}$ of the standard $\SO(10)$. Equivalently stated, fermions of the two $\mathbf{\bar{5}}$ irreps of $\SU(5)$ in the fermionic $\mathbf{27}$ mix.
    \item Some of the $\delta_{i}$ in the mass matrices are conjugated. This is consistent with some sectors coupling to doublets with negative hypercharge $-1/2$: these are the usual sectors $D$ (down) and $E$ (charged leptons), as well as part of the neutrino sector $R$. Furthermore, each $\delta_{i}$ is always either conjugated or unconjugated throughout the mass expressions, in accordance with the Yukawa sector coupling the fermions to $\mathbf{27}$ and $\mathbf{351}'^\ast$. 
    \item Table~\ref{tab:SM-doublets} has to consider the doublet locations in $(10)'\,1_{\psi'}$ separately in case (b), since a new basis adapted to this subgroup's irreps is required. The new states $\tilde{\delta}_{i}$ are defined by
        \begin{align}
            \matrixp{\delta_{2}}{-\delta_{3}} &= \mathbf{P}\, \matrixp{\tilde{\delta}_{2}}{\tilde{\delta}_{3}}, &
            \matrixp{\delta_{5}}{\delta_{6}} &= \mathbf{P}\, \matrixp{\tilde{\delta}_{5}}{\tilde{\delta}_{6}},&
            \text{with}\quad
            \mathbf{P}&:= \frac{1}{4}\,\matrixp{\sqrt{6}&\sqrt{10}}{\sqrt{10}&-\sqrt{6}}.
            \label{eq:delta-relation}
        \end{align}
    \item The representation $\mathbf{650}$ also contains doublets $(\mathbf{1},\mathbf{2},\pm 1/2)$, namely $7$ copies, but we have not considered those. They play little role, since they are not present in the renormalizable Yukawa sector of Eq.~\eqref{eq:Yukawa-sector}. Furthermore, they have little effect on the admixture content of the Higgs in the breaking pattern of Eq.~\eqref{eq:breaking-pattern}, since there is no mixing between the $\mathbf{650}$ and $\mathbf{27}\oplus\mathbf{351'}$ parts in the first stage of breaking, see Section~\ref{sec:two-stage-breaking-and-masses}.
    In line with ESH we thus assume they are at the GUT scale and their involvement negligible.
\end{itemize}

We now proceed by considering the more specific $\text{ESH}+\mathbb{Z}_{2}^{\psi}$ scenario. With $\mathbb{Z}_{2}^{\psi}$ parity preserved, no irreps with even and odd spinorial parity can mix. This implies that the EW-scale SM fermions are entirely in the spinorial $\mathbf{16}$ of the standard $\SO(10)$, while the vector-like pairs $(\nu',\nu'^{c})$, $(e',e'^{c})$ and $(d',d'^{c})$ in the $\mathbf{10}$ are intermediate-scale mass eigenstates. Furthermore, only $\nu^{c}$ (and not $n$) is involved in the type I seesaw mechanism for neutrino mass generation.  

These features can be seen explicitly in Eq.~\eqref{eq:fermion-mass-U}--\eqref{eq:fermion-mass-R}. The spinorial SM-singlet VEVs vanish in accordance with Eq.~\eqref{eq:ansatz-spinorial-VEVs}. The spinorial ($\psi$-odd) and non-spinorial ($\psi$-even) doublets from Table~\ref{tab:SM-doublets} cannot mix. Since we must perform the Higgs fine-tuning in the non-spinorial sector of their mass matrix, the spinorial EW VEVs vanish:
\begin{align}
    \delta_{4}=\delta_{5}=\delta_{6}=\delta_{10}=\delta_{11}&=0. \label{eq:ansatz-spinorial-doublets}
\end{align}
With the spinorial ansatz no mixing occurs, the light fermion masses in the $D$ and $E$ sector simply correspond to the $(1,1)$-block in Eqs.~\eqref{eq:fermion-mass-D} and \eqref{eq:fermion-mass-E}. The light fermion mass matrices $\mathbf{m}$ for SM fermions are thus explicitly
\begin{align}
    \mathbf{m}_{U}&= -\YPSIBF \BLUE{\delta_{7}}+\tfrac{1}{\sqrt{15}}\YTHETABF \BLUE{\delta_{8}}, \label{eq:fermion-mass-light-U}\\
    \mathbf{m}_{D}&= \phantom{+}\YPSIBF \,\BLUE{\delta_{1}^{\ast}}-\tfrac{1}{\sqrt{15}}\YTHETABF  \BLUE{\delta_{2}^{\ast}}, \label{eq:fermion-mass-light-D}\\
    \mathbf{m}_{E}&= -\YPSIBF  \BLUE{\delta_{1}^{\ast}}-\tfrac{1}{2} \sqrt{\tfrac{3}{5}} \YTHETABF  
            \BLUE{\delta_{2}^{\ast}}+\tfrac{1}{2}\YTHETABF  \BLUE{\delta_{3}^{\ast}}, \label{eq:fermion-mass-light-E}\\
    \mathbf{m}_{N}&= -
        (\YPSIBF  \BLUE{\delta_{7}}+\tfrac{1}{2} \sqrt{\tfrac{3}{5}} \YTHETABF  \BLUE{\delta_{8}}- 
            \tfrac{1}{2}\YTHETABF  \BLUE{\delta_{9}})\;
        (\YTHETABF \RED{W_{3}}^{\ast})^{-1}\;
        (\YPSIBF  \BLUE{\delta_{7}}+\tfrac{1}{2} \sqrt{\tfrac{3}{5}} \YTHETABF  \BLUE{\delta_{8}}- 
            \tfrac{1}{2}\YTHETABF  \BLUE{\delta_{9}})^{T}, \label{eq:fermion-mass-light-N}
\end{align}
where we considered only the type I seesaw contribution (and not type II). Notice that due to $\mathbb{Z}_{2}^{\psi}$ only the SM-singlet $\nu^{c}$ is involved in the seesaw mechanism for light neutrinos, and the VEV $W_{3}$ must be non-vanishing already due to rank breaking, cf.~Section~\ref{sec:minimal-models-breaking}. Furthermore, there is an additional requirement that $V_{2}$ or $W_{2}$ must also be non-vanishing, so that the vector-like exotics in the $(2,2)$-blocks of $M_{D}$ and $M_{E}$ get masses of scale $M_{I}$.

The SM fermion masses of Eqs.~\eqref{eq:fermion-mass-light-U}--\eqref{eq:fermion-mass-light-N} contain only the EW VEVs $\delta_{1,2,3}$ and $\delta_{7,8,9}$, while the mass matrices $\YPSIBF$ and $\YTHETABF$ are symmetric. The mass matrices are thus similar the $\SO(10)$ GUT with $\mathbf{10}_{\mathbb{C}}\oplus\mathbf{126}$, PQ symmetry, and type~I seesaw, except that we have $6$ instead of $4$ EW VEVs. Since the $\SO(10)$ fit is known to work, see e.g.~\cite{Babu:1992ia,Bajc:2005zf,Joshipura:2011nn,Dueck:2013gca,Ohlsson:2019sja,Babu:2020tnf}, we are guaranteed to have a realistic Yukawa sector if all $\delta_{1,2,3,7,8,9}$ are present and independent. 

We are now prepared to supplement the considerations of second-stage breaking from Section~\ref{sec:minimal-models-breaking} with the requirements of the Yukawa sector, and apply them to determine the scalar content of each $G$-theory case in Table~\ref{tab:minimal-models} under $\text{ESH}+\mathbb{Z}_{2}^{\psi}$. In each case we require $G$-irreps so that $W_{3}$ and at least one of $V_{2}$ or $W_{2}$ are non-vanishing (second-stage symmetry breaking and masses of fermion vector-like exotics at $M_{I}$), along with the doublets $\delta_{1,2,3,7,8,9}$ (or at least some, as discussed in specific cases) for the fermion fit. The necessary information on their locations can be looked up in Tables~\ref{tab:vevs-27-and-351p} and \ref{tab:SM-doublets}, with the case-by-case considerations to get the minimal models as follows:
\begin{itemize}
    \item In the case of trinification $3_C\,3_L\,3_R$ with LR parity, the scalar $(\mathbf{1},\mathbf{\bar{6}},\mathbf{6})$ is necessary to have $W_{3}$, and one or two bitriplets $(\mathbf{1},\mathbf{\bar{3}},\mathbf{3})$ are necessary for $V_{2}$ and $W_{2}$. The $(\mathbf{1},\mathbf{\bar{3}},\mathbf{3})$ also automatically provides all the VEVs $\delta_{1,2,3,7,8,9}$. An effective trinification theory with $(\mathbf{1},\mathbf{\bar{3}},\mathbf{3})\oplus(\mathbf{1},\mathbf{\bar{6}},\mathbf{6})$ with only one bitriplet is however known not to be viable for the Yukawa sector~\cite{Babu:1985gi}, so we require two bitriplets, yielding the result for Case~1 in Table~\ref{tab:minimal-models}.
    \par
    One can understand the limitations of having only one bitriplet explicitly. The bitriplet has two $\psi$-even EW VEVs, which means that having one triplet instead of two imposes two constraints on the six VEVs $\delta_{1,2,3,7,8,9}$. The VEVs $\delta_{1}^\ast$ and $\delta_{7}$ are in the $\PSI$-bitriplet, while $\delta_{2}$ and $\delta_{8}^\ast$ are in the $\THETA$-bitriplet, with the mass matrix of first-stage breaking in the $(\PSI,\THETA^\ast)$-basis given in Eq.~\eqref{eq:masses-333-doublet133}. One mini fine-tuning from GUT to intermediate scale is required in that matrix to drag one bitriplet to the intermediate scale; if the eigenmode of this lighter state is $(\alpha,\beta)$, then 
        \begin{align}
            \frac{\delta_{1}^\ast}{\delta_{2}^{\ast}}=\frac{\delta_{7}}{\delta_{8}}=\frac{\alpha}{\beta},
        \end{align}
    implying that $\mathbf{m}_{U}$ and $\mathbf{m}_{D}$ in Eqs.~\eqref{eq:fermion-mass-light-U} and \eqref{eq:fermion-mass-light-D} are proportional, and thus there is no CKM mixing (at least up to order $\mathcal{O}(\MI/\MGUT)$, where $\MGUT$ is the GUT-scale), making the fermion fit unviable. The structural importance of the second bitriplet thus warrants its inclusion into the intermediate theory.
    \par
    The minimal set of scalar irreps for the case with $LR$ parity is already $LR$-symmetric, so no further amendments are required. If we instead consider cases with $CL$ or $CR$ parity, we simply need to complete the scalar spectrum of the $LR$ case to be consistent with that parity, yielding the results of Cases~2 and 3 in Table~\ref{tab:minimal-models}.
    \item 
        For $6_{CL}\,2_{R}$, we must necessarily take $(\mathbf{\overline{21}},\mathbf{3})$ for $W_{3}$, and one $(\mathbf{15},\mathbf{1})$ to have $V_{2}$ and $W_{2}$, with the ratio $V_{2}/W_{2}^{\ast}$ fixed by mini-tuning the mass matrix for $(\mathbf{15},\mathbf{1})$. In addition, the presence of $\delta_{1,7}$ requires $(\mathbf{\bar{6}},\mathbf{2})$, and the presence of $\delta_{2,3,8,9}$ requires $(\mathbf{84},\mathbf{2})$. Since $(\mathbf{\bar{6}},\mathbf{2})$ generates the Yukawa contributions proportional to $\YPSIBF$, and $(\mathbf{84},\mathbf{2})$ generates contributions proportional to $\YTHETABF$, the presence of both is required for a realistic Yukawa sector. The minimal case for $6_{CL}\,2_{R}$ thus consists altogether of $4$ irreps: $(\mathbf{15},\mathbf{1})$ $(\mathbf{\overline{21}},\mathbf{3})$, $(\mathbf{\bar{6}},\mathbf{2})$ and $(\mathbf{84},\mathbf{2})$, where $W_{5}$ is induced by corrections of order $\mathcal{O}(\MI/\MGUT)$, leading to a slightly lower mass of the SM-singlet fermions $n$ compared to $\nu^{c}$.
    \item 
        For $6_{CR}\,2_{L}$ we must necessarily take $(\mathbf{\overline{105'}},\mathbf{1})$ for $W_{3}$, and one $(\mathbf{15},\mathbf{1})$ for $V_{2}$ and $W_{2}$. That automatically generates all intermediate-scale VEVs $V_{2}$ and $W_{2,3,5}$. Furthermore, both $(\mathbf{\bar{6}},\mathbf{2})$ and $(\mathbf{84},\mathbf{2})$ are required for $\delta_{1,2,3,7,8,9}$, with the same argument as for the $6_{CL}\,2_{R}$ case above. 
    \item For the $(10)'\,1_{\psi'}$ intermediate theory, $(\mathbf{126},+2)$ is required for $W_{3}$, and either $(\mathbf{16},+1)$ for $V_{2}$ or $(\mathbf{\overline{144}},-1)$ for $W_{2}$. We shall attempt to build the minimal theory without $(\mathbf{\overline{144}},-1)$.
    \par
    The new basis of EW VEVs for $(10)'\,1_{\psi'}$ is given in panel (b) of Table~\ref{tab:SM-doublets}, and modifies Eqs.~\eqref{eq:fermion-mass-light-D} and \eqref{eq:fermion-mass-light-E} according to relation~\eqref{eq:delta-relation} into
        \begin{align}
            \mathbf{m}_{D}&= \YPSIBF \BLUE{\delta_{1}^{\ast}}-\tfrac{1}{2 \sqrt{10}} \YTHETABF \BLUE{\tilde{\delta}_{2}{}^{\ast}}-\tfrac{1}{2 \sqrt{6}} \YTHETABF \BLUE{\tilde{\delta}_{3}{}^{\ast}}, \\
    \mathbf{m}_{E}&= -\YPSIBF \BLUE{\delta_{1}^{\ast}}-\sqrt{\tfrac{2}{5}} \YTHETABF \BLUE{\tilde{\delta}_{2}{}^{\ast}}. 
        \end{align}
    The presence of $\delta_{1}$ and $\tilde{\delta}_{2}$ necessary for the fit of D and E sectors requires one $(\mathbf{10},-2)$ at the intermediate scale. This is achieved by mini-tuning the mass matrix in Eqs.~\eqref{eq:condition-doublets-101}, where the light eigenmode can be made to have an arbitrary direction, so the ratio $\delta_{1}/\tilde{\delta}_{2}$ is arbitrary. In addition, $\tilde{\delta}_{3}$ is already present in $(\mathbf{126},+2)$.
    \par
    Having $(\mathbf{16},+1)$ but omitting $(\mathbf{\overline{144}},-1)$ due to the earlier consideration of $G$-breaking VEVs also introduces $\delta_{7}$, but omits $\delta_{8,9}$. The omission of the latter two restricts the form of mass expressions in the U and N sectors; the neutrinos are helped by type II seesaw, however, and the VEVs $\delta_{8,9}$
    are induced by $\mathcal{O}(\MI/\MGUT)$ corrections. Although a dedicated fit would in principle be required, we assume that the above considerations are sufficient; unlike trinification, where structural reasons (CKM) required the presence of two bitriplets, the omission of $(\mathbf{\overline{144}},-1)$ in $(10)'\,1_{\psi'}$ has more of a numerical nature.
    \par 
    Altogether, we thus require the scalars $(\mathbf{16},+1)$, $(\mathbf{126},+2)$ and $(\mathbf{10},-2)$, with the limitation that $\delta_{7,8}$ are induced by corrections of order $\mathcal{O}(\MI/\MGUT)$.
\end{itemize}

\section{Unification analysis of minimal cases \label{sec:analysis-unification}}
\def\ZETA{\zeta}
\def\ETA{\eta}

We now turn to the basic phenomenological consideration of gauge coupling unification for the minimal cases of $G$-theories determined in Table~\ref{tab:minimal-models}. 

\subsection{Procedure \label{sec:analysis-unification-procedure}}

We consider 2-loop renormalization group equations (RGE) running with one-loop threshold corrections (TC), using standard RG techniques. Since this introduces the notation for some important quantities relevant for our analysis, we give a brief summary below:
\begin{itemize}
\item 
    The 2-loop RGE for gauge coupling running of a direct product of groups $\prod_{i} G_{i}$, assuming at most one $\UU$ factor (to insure no kinetic mixing, otherwise cf.~\cite{Fonseca:2013bua}) and ignoring the Yukawa contribution, are 
    \cite{Ellis:2015jwa,Bertolini:2009qj}
        \begin{align}
            \tfrac{d}{dt}\,\alpha^{-1}_{i}&= -\tfrac{1}{2\pi} \big(a_{i}+\tfrac{1}{4\pi} \sum_{j} b_{ij} \alpha_{j}\big),
            \label{eq:RG-running}
        \end{align}
    where the gauge couplings $g_{i}$ and the renormalization scale $\mu$ are encoded in
        \begin{align}
            \alpha_{i}^{-1} &:=4\pi/g_{i}^{2},& 
            t&=\log(\mu/\mathrm{GeV}),
        \end{align}
    while $a_{i}$ and $b_{ij}$ are the $1$- and $2$-loop beta coefficients that depend on the field content of the theory.
    Suppose we label the adjoint representation of the group factor $G_{i}$ by $\mathbf{G}_{i}$, write the scalar content as the direct sum of irreps $\bigoplus_{c} \mathbf{S}_{a}$, and the fermion content as $\bigoplus_{b} \mathbf{F}_{b}$. The $1$- and $2$-loop coefficients can then be computed by \cite{Jones:1981we}
        \begin{align}
	       a_{i}&=\sum_{a}\tfrac{1}{3}\,\ZETA(\mathbf{S}_{a})\, D_{i}(\mathbf{S}_{a}) + \sum_{b} \tfrac{2}{3}\, \ZETA(\mathbf{F}_{b})\, D_{i}(\mathbf{F}_{b})-\tfrac{11}{3} C_{i}(\mathbf{G}_{i}), \label{eq:RG-factor-a}\\
	       \begin{split}
	       b_{ij} &=\sum_{a} 4\,\ZETA(\mathbf{S}_{a})\, D_{i}(\mathbf{S}_{a})\,C_{j}(\mathbf{S}_{a})
                +\sum_{b} 2\,\ZETA(\mathbf{F}_{b})\,D_{i}(\mathbf{F}_{b})\,C_{j}(\mathbf{F}_{b})+\\
		          &\quad + \delta_{ij}\left[\sum_{a}\tfrac{2}{3}\,\ZETA(\mathbf{S}_{a})\,D_{i}(\mathbf{S}_{a})\;C_{i}(\mathbf{G}_{i})+\sum_{b} \tfrac{10}{3}\, \ZETA(\mathbf{F}_{b})\,D_{i}(\mathbf{F}_{b})\;C_{i}(\mathbf{G}_{i})-\tfrac{34}{3} C_{i}(\mathbf{G}_{i})^2\right],
	       \end{split} \label{eq:RG-factor-b}
        \end{align}
    where $\ZETA(\mathbf{S}_{a})=1/2$ or $1$ for a real or complex scalar irrep $\mathbf{S}_{a}$, while $\ZETA(\mathbf{F}_{b})=1$ or $2$ for Weyl or Dirac fermions $\mathbf{F}_{b}$. We use a compact notation for Dynkin indices $D_{i}$ and Casimir factors $C_{i}$, so that the subscript $i$ refers to these quantities with respect to group factor $G_{i}$. For the Dynkin index this means we consider each product irrep $\mathbf{R}=\prod_{i} \mathbf{R}_{i}$ of $G=\prod_i G_{i}$ in terms of irreps of $G_{i}$, i.e. we have
        \begin{align}
            D_{i}(\prod_{k}\mathbf{R}_{k})&= D_{i}(\mathbf{R}_{i})\cdot \prod_{k\neq i} \dim(\mathbf{R}_k),
        \end{align}
    while the sum $\sum_{a}\mathbf{T}_{a}\mathbf{T}_{a}$ for the Casimir operator $C_{i}$ is taken over the generators of $G_{i}$.  For numeric computation of these, we use \texttt{LieART 2.0}~\cite{Feger:2012bs,Feger:2019tvk}.
\item 
    Suppose we match at energy scale $\mu$ the high-energy EFT with gauge symmetry $\prod_{j} \tilde{G}_{j}$ to the low-energy EFT with gauge roup $\prod_{i} G_{i}$. Let the corresponding high- and low-energy gauge couplings be $\tilde{\alpha}^{-1}_{j}$ and $\alpha^{-1}_{i}$, respectively.\footnote{Note the consistent use of indices $i$ and $j$ for referring to factors of the low- and high-energy EFT groups, respectively.} Furthermore, let the representations integrated out at the threshold scale $\mu$ be labeled by $\bigoplus_{a}\mathbf{S}_{a}$, $\bigoplus_{b}\mathbf{F}_{a}$ and $\bigoplus_{c}\mathbf{V}_{c}$  for scalars, fermions and gauge bosons in terms of irreps of the low-energy symmetry $\prod_{i}G_{i}$, and the mass of any irrep $\mathbf{R}$ be denoted by $M_{\mathbf{R}}$. We then have the $1$-loop matching expression at $\mu$ in the $\overline{MS}$ scheme given by \cite{Weinberg:1980wa,Hall:1980kf,Ellis:2015jwa}
        \begin{align}
            \alpha^{-1}_{i}(\mu)&=\sum_{j}C_{ij} \,\tilde{\alpha}^{-1}_{j}(\mu)-\ETA_{i}(\mu), \label{eq:matching-condition}
        \end{align}
    where $C_{ij}$ depend on the embedding $\prod_{i}G_{i}\subset\prod_{j}\tilde{G}_{j}$, and the one-loop expression for the threshold effects $\ETA_{i}$ is given by
        \begin{align}
	       \ETA_{i}(\mu)&= \frac{1}{12\pi}
            \left[
            \sum_{c} D_{i}(\mathbf{V}_{c}) \left(1-21 \log\,\tfrac{M_{\mathbf{G}_c}}{\mu}\right)
	        +\sum_{a} 2\,\ZETA(\mathbf{S}_{a})\,D_{i}(\mathbf{S}_{a})\,\log\tfrac{M_{\mathbf{S}_{a}}}{\mu}
	        +\sum_{b} 4\,\ZETA(\mathbf{F}_{b})\,D_{i}(\mathbf{F}_{b})\,\log\tfrac{M_{\mathbf{F}_{b}}}{\mu}
            \right]. \label{eq:thresholds-definition}
    \end{align}
    The coefficients $\ZETA$ and Dynkin indices $D_{i}$ have already been introduced earlier for RGE running.
    Note that compared to literature we absorbed a numeric coefficient into the definiton of TC $\ETA_{i}$, so that it appears with coefficient $1$ in Eq.~\eqref{eq:matching-condition}. The expression in Eq.~\eqref{eq:thresholds-definition} assumes that the zero-mass would-be Goldstone modes are omitted from the sum over scalar irreps $\mathbf{S}_a$.  
\end{itemize}

We can integrate the above methods into a procedure suitable for our analysis. The more canonical approach using $2$-loop running would be top-down, since in general ambiguities may arise when matching theories in reverse (low to high). 
For all our cases of Table~\ref{tab:minimal-models}, however, all ambiguities can be uniquely resolved, and we shall rather consider the procedure bottom-up. The $3$ scales in the problem are the $Z$-boson scale $\MZ$, the intermediate scale $\MI$
and the unification scale $\MGUT$, hence the procedure will consists of $6$ steps alternating between matching and RG running.
Schematically the steps are defined by the diagram
    \begin{align}
        \overbrace{\MZ}^{1}\ \xrightarrow[\text{RGE in SM}]{\quad 2\quad }\ \overbrace{\MI}^{3} \ \xrightarrow[\text{RGE in }G]{\quad 4\quad}\ \overbrace{\MGUT}^{5}\ \xrightarrow[\text{RGE in }\mathrm{E}_{6}]{\quad 6\quad }\ \MPL,
        \label{eq:procedure-RG}
    \end{align}
where the running procedure terminates no higher than the (reduced) Planck scale $\MPL$. Assuming we have fixed values for threshold corrections, each of the steps involves the following:
\begin{enumerate}
    \item 
        At $\MZ$: set the values of the SM gauge couplings to their experimental values given by~\cite{Marciano:1991ix,Kuhn:1998ze,Martens:2010nm,Sturm:2013uka,ParticleDataGroup:2020ssz}
        \begin{align}
            \begin{split}
            \alpha^{-1}_{3}(M_Z) &= 8.550 \pm 0.065,\\
            \alpha^{-1}_{2}(M_Z) &=29.6261 \pm 0.0051,\\
            \alpha^{-1}_{1}(M_Z) &= 59.1054 \pm 0.0031,
            \end{split}\label{eq:SM-couplings-at-MZ}
	       \end{align}
        where the GUT normalization $ \alpha^{-1}_{1}=\tfrac{3}{5}\alpha^{-1}_{Y}$ is used for the $1_{Y}$ factor of the SM. We use central values as the boundary condition.
    \item 
        $\MZ\to\MI$: we use the RGE of Eq.~\eqref{eq:RG-running} with the SM coefficients (given in the order $3_C\,2_L\,1_{Y}$ and GUT normalization)
        \begin{align}
            a^{\text{SM}}_{i}&= (-7,-\tfrac{19}{6},\tfrac{41}{10}), &
            b^{\text{SM}}_{ij}&= 
                \begin{pmatrix}
                    -26 & \tfrac{9}{2} & \frac{11}{10} \\
                    12 & \tfrac{35}{6} & \frac{9}{10} \\
                    \tfrac{44}{5} & \tfrac{27}{10} & \tfrac{199}{50} \\
                \end{pmatrix}. 
                \label{eq:coefficients-ab-SM}
        \end{align}
        The termination condition for RG running that determines $\MI$ is given in Table~\ref{tab:matching-data-I} for each $G$-vacuum.
    \item At $M_I$: we match the SM gauge couplings to the $G$ couplings using Eq.~\eqref{eq:matching-condition}, where the matching coefficients $C_{ij}$ are found for each $G$-vacuum in Table~\ref{tab:matching-data-I}, while the values for the threshold corrections $\ETA_{i}=(\ETA_{3},\ETA_{2},\ETA_{1})$ are assumed to be given in advance.
    \item $\MI\to\MGUT$: we use the RGE of Eq.~\eqref{eq:RG-running} with coefficients $a$ and $b$ given for each $G$-theory in Table~\ref{tab:minimal-models}. These were computed using Eqs.~\eqref{eq:RG-factor-a}--\eqref{eq:RG-factor-b}.
    The termination condition for RG running that determines $\MGUT$ is given in Table~\ref{tab:matching-data-GUT} for each $G$-vacuum. 
    \item At $\MGUT$: we match the $G$ gauge couplings to the $\mathrm{E}_{6}$ unified coupling using Eq.~\eqref{eq:matching-condition}, where the matching coefficients $C_{ij}$ are found for each $G$-vacuum in Table~\ref{tab:matching-data-GUT}, while the GUT TC are assumed to be given in advance; we denote them by $\ETA'_{i}$.
    \item $\MGUT\to\MPL$: the RG running in $\mathrm{E}_{6}$ GUT proceeds according to Eq.~\eqref{eq:RG-running}, where the coefficients $a$ and $b$ are computed from Eqs.~\eqref{eq:RG-factor-a} and \eqref{eq:RG-factor-b} to be
        \begin{align}
            a^{\mathrm{E}_{6}}&=16, & 
            b^{\mathrm{E}_{6}}&=11956. \label{eq:ABcoefficients-E6}
        \end{align}
    Due to the large values of these coefficients, the unified coupling reaches the Landau pole before $\MPL$, as discussed later in Section~\ref{sec:analysis-unification-results}. 
\end{enumerate}

\def\ECO{\ETA_{0}}
\def\ECR{\ETA_{\text{cr}}}
\def\EFR{\ETA_{\text{fr}}}
\def\ECP{\ETA'_{0}}
\def\EGU{\ETA'_{\Delta}}

\noindent
While all necessary technical details of the procedure are given above, further discussion of some elements will enhance conceptual clarity. Our remarks are as follows:
\begin{itemize}
    \item  \textbf{Determining the matching scales}: 
        In all cases of intermediate $G$-theories in Table~\ref{tab:minimal-models}, there are always two different gauge coupling values, either because the group $G$ has two simple factors, or because it has three together with a discrete parity. 
        \par
        The matching condition of Eq.~\eqref{eq:matching-condition} thus connects at the scale $M_{I}$ three SM couplings to two (different valued) couplings of $G$. This set of expressions is overdetermined for bottom-up matching, so a consistent solution can be found for the two $G$-couplings only when the threshold corrections $\ETA_{i}$ and the SM couplings $\alpha_{i}^{-1}$ satisfy an additional constraint. This constraint is determined explicitly from Eq.~\eqref{eq:matching-condition} for each $G$-vacuum and is given in Table~\ref{tab:matching-data-I}. The scale when it is satisfied defines the intermediate scale $M_{I}$. For example, if threshold effects $\ETA_{i}$ vanish, the trinification scale in the case of $LR$ parity is defined where the weak and hypercharge couplings meet. 
        \par 
        Analogously at $\MGUT$, Eq.~\eqref{eq:matching-condition} connects two (different valued) couplings of $G$ to one unified $\mathrm{E}_{6}$ coupling. The system is again overdetermined, with explicit calculation giving the constraint in Table~\ref{tab:matching-data-GUT}. The scale where the constraint 
        is fulfilled is defined as the GUT scale. It may well happen that such a solution does not exist, i.e.~there may be no successful unification for a given set of threshold values. 
        \par 
        Note that the conditions for $\MI$ in Table~\ref{tab:matching-data-I} and $\MGUT$ in Table~\ref{tab:matching-data-GUT} are written with all couplings $\alpha^{-1}$ on the left-hand side of the equality and thresholds $\ETA$ on the right-hand side. Furthermore, the choice of overall sign is such that numerically the left-hand side of the $\MI$-constraint starts out positive in the RG running at $\MZ$, while the left-hand side of the $\MGUT$-constraint starts out positive at $\MI$ (assuming no TC at $\MI$).
    \item \textbf{Parametrizing the thresholds}:
        In all cases of intermediate $G$-theories in Table~\ref{tab:minimal-models},   there are always five different threshold corrections that unification depends on. In particular, there are three threshold corrections $(\ETA_{3},\ETA_{2},\ETA_{1})$ to the SM couplings at $\MI$, and then two more for the (different valued) couplings of $G$ at $\MGUT$. It is convenient to parametrize these thresholds so that their effect on unification for a given $G$-embedding is more transparent.
        \par
        Let's first consider the three threshold corrections to SM couplings. Notice that if two-loop running effects are negligible, i.e.~in the limit $b\to 0$ in Eq.~\eqref{eq:RG-running}, the RG solutions are straight lines and an equal shift in all three thresholds simply translates the running picture upwards or downwards with no change in angles. The two-loop effect in SM running is indeed very small, thus such an overall shift has little effect on $\MI$ or $\MGUT$; we parametrize it by $\ECO$. The scale $\MI$ is determined by the combination of thresholds given on the right-hand side of the condition for $\MI$ in Table~\ref{tab:matching-data-I}; the label $\ECR$ parametrizes this threshold combination ``critical'' for the position of the intermediate scale $\MI$. The third independent combination of thresholds has no influence on $\MI$ (it vanishes on the right-hand side of the $\MI$-constraint), but it does impact the difference of the two coupling values of $G$ at $\MI$ and thus ultimately the position of $\MGUT$; it can be chosen ``freely'' without impacting the position of $\MI$ and is labeled as $\EFR$. The parametrization of $(\ETA_{3},\ETA_{2},\ETA_{1})$ in terms of $\{\ECO,\ECR,\EFR\}$ for each $G$-vacuum is  given in Table~\ref{tab:matching-data-I}. As an example, consider trinification with $LR$ parity: $\ECR$ parametrizes the critical gap between the weak and hypercharge coupling just before they join together at $\MI$, while the gap to the color coupling $\EFR$ can be chosen freely with no impact on the position of $M_{I}$.
        \par Analogously, the two threshold corrections at $\MGUT$ can be parametrized with their overall shift $\ECP$ and their difference $\EGU$ (that comes out as the right-hand side of the constraint for $\MGUT$), see Table~\ref{tab:matching-data-GUT}. 
    \item \textbf{Scale dependence of threshold values}: 
        The entire RG procedure is uniquely defined once the five threshold corrections $\{\ECR,\EFR,\EGU,\ECO,\ECP\}$ are specified, with the latter two expected to have little effect on unification. These five threshold values are computed given their $G$-specific parametrizations in Tables~\ref{tab:matching-data-I} and \ref{tab:matching-data-GUT} from the original threshold corrections $\ETA_{i}$ at $\MI$ and $\MGUT$. These in turn are in principle computed from the expression in Eq.~\eqref{eq:thresholds-definition}. Their values thus ultimately depend on the particle spectrum, and thus on the parameters of the model, as well as on the matching scale $\mu$.  
        \par 
        In a top-down approach, the matching scale $\mu$ can be chosen freely anywhere around the mass scale of the associated particles being integrated out; changing the scale simply redistributes at a given order of perturbation theory the contribution between the RG running and threshold corrections. In our bottom up approach, the matching scales $\MI$ and $\MGUT$ are determined as part of the procedure, which however has the scale-dependent threshold values as input. 
        If the particle spectrum is given, then one can perform an iterative self-consistent calculation for determining $\mu$; this gives a valid result, but a change of matching scale would again induce only next-order perturbative corrections, and thus the arbitrariness of the matching scale is still present in the bottom-up approach at any fixed order.
        \par
        We use this arbitrariness of $\mu$ to our advantage, and set this scale to be equal to the mass $M_{\XI}$ of some particular irrep $\XI$. We make different choices for $\XI$ for different $G$-vacua; 
        we list the choices of SM irreps $\XI$ defining $\MI$ in Table~\ref{tab:matching-data-I}, while the $G$ irreps $\XI$ defining $\MGUT$ are listed in Table~\ref{tab:matching-data-GUT}. At $\MGUT$ the choice for $\XI$ is always the unique irrep of gauge bosons in the coset $\mathrm{E}_{6}/G$, while one of the vector-boson irreps from the coset $G/(3_C\,2_L\,1_{Y})$ was chosen for $\XI$ at $\MI$. The benefit of these choices is that the mass of vector bosons mediating proton decay is equal to the unification scale $\MGUT$ (or in some cases the intermediate scale $\MI$, as discussed later in Section~\ref{sec:analysis-proton}), making the phenomenological analysis unambiguous.
\end{itemize}

\begin{table}
\centering
\caption{Matching data for all cases of $G$-vacua in Table~\ref{tab:minimal-models} at the intermediate scale $\MI$. This data includes the matching coefficients $C_{ij}$ for Eq.~\eqref{eq:matching-condition}, the condition for determining when $\MI$ was reached during RG running, the parametrization of the intermediate threshold values $\ETA_{i}$ in terms of $\{\ECO,\ECR,\EFR\}$, and the irrep $\XI$ whose mass is set to the matching scale $\mu=\MI$ in Eq.~\eqref{eq:thresholds-definition}. The subscript $V$ in the irrep $\XI$ indicates vector bosons, and a subsequent number indicates which state is meant when there are multiple in a diagonal mass matrix, cf.~Appendix~\ref{app:masses-gauge}. Note regarding bases: the index $i$ runs over SM factors $3_C\,2_L\,1_Y$ and $j$ runs over the factors of $G$; we use the GUT normalization for $\UU$ factors, i.e.~$\alpha_{1}^{-1}=\tfrac{3}{5}\alpha_{Y}^{-1}$ for hypercharge $Y$ in SM and $\alpha_{1'}^{-1}=\tfrac{1}{24}\alpha_{\psi'}^{-1}$ for charge $\psi'$ in $G=(10)'\,1_{\psi'}$. \label{tab:matching-data-I}}
\begin{tabular}{llp{3.4cm}ll}
     \toprule
     $G$-vacuum& $C_{ij}$ & condition for $\MI$&$(\ETA_{3},\ETA_{2},\ETA_{1})$ at $\MI$& $\XI$ in $\MI=M_{\XI}$   \\
     \midrule
     $3_C\,3_L\,3_R\rtimes \mathbb{Z}_{2}^{LR}$& 
        $\left(\begin{smallmatrix}
            1&0&0\\
            0&1&0\\
            0&1/5&4/5\\
        \end{smallmatrix}\right)$&
        $\alpha^{-1}_{1}-\alpha^{-1}_{2}=\ETA_{2}-\ETA_{1}$&
        $(\ECO+\EFR,\ECO+\ECR,\ECO)$&
        $(\mathbf{1},\mathbf{2},+\tfrac{1}{2})_V $\\\addlinespace[6pt]
    $3_C\,3_L\,3_R\rtimes \mathbb{Z}_{2}^{CL}$& 
        $\left(\begin{smallmatrix}
            1&0&0\\
            0&1&0\\
            0&1/5&4/5\\
        \end{smallmatrix}\right)$&
        $\alpha^{-1}_{2}-\alpha^{-1}_{3}=\ETA_{3}-\ETA_{2}$&
        $(\ECO+\ECR,\ECO,\ECO+\EFR)$&
        $(\mathbf{1},\mathbf{2},+\tfrac{1}{2})_V $\\\addlinespace[6pt]
    $3_C\,3_L\,3_R\rtimes \mathbb{Z}_{2}^{CR}$& 
        $\left(\begin{smallmatrix}
            1&0&0\\
            0&1&0\\
            0&1/5&4/5\\
        \end{smallmatrix}\right)$&
        $\alpha_{1}^{-1}-\tfrac{1}{5}\alpha_{2}^{-1}-\tfrac{4}{5}\alpha_{3}^{-1}=-\ETA_{1}+\tfrac{1}{5}\ETA_{2}+\tfrac{4}{5}\ETA_{3}$&
        $(\ECO+\tfrac{1}{5}\EFR,\ECO-\tfrac{4}{5}\EFR,\ECO-\ECR)$&
        $(\mathbf{1},\mathbf{2},+\tfrac{1}{2})_V $\\\addlinespace[6pt]
    $6_{CL}\,2_{R}$& 
        $\left(\begin{smallmatrix}
            1&0\\
            1&0\\
            2/5&3/5\\
        \end{smallmatrix}\right)$&
        $\alpha^{-1}_{2}-\alpha^{-1}_{3}=\ETA_{3}-\ETA_{2}$&
        $(\ECO+\ECR,\ECO,\ECO+\EFR)$&
        $(\mathbf{3},\mathbf{2},+\tfrac{1}{6})_{V,2}$\\\addlinespace[6pt]
    $6_{CR}\,2_{L}$& 
        $\left(\begin{smallmatrix}
            1&0\\
            0&1\\
            1&0 \\
        \end{smallmatrix}\right)$&
        $\alpha^{-1}_{1}-\alpha^{-1}_{3}=\ETA_{3}-\ETA_{1}$&
        $(\ECO+\ECR,\ECO+\EFR,\ECO)$&
        $(\mathbf{3},\mathbf{1},+\tfrac{1}{3})_V$\\\addlinespace[6pt]
    $(10)'\,1_{\psi'}$& 
        $\left(\begin{smallmatrix}
            1&0\\
            1&0\\
            1/10&9/10 \\
        \end{smallmatrix}\right)$&
        $\alpha^{-1}_{2}-\alpha^{-1}_{3}=\ETA_{3}-\ETA_{2}$&
        $(\ECO+\ECR,\ECO,\ECO+\EFR)$&
        $(\mathbf{3},\mathbf{2},+\tfrac{1}{6})_{V,1}$\\
     \bottomrule
\end{tabular}
\end{table}

\begin{table}[htb]
\centering
\caption{Matching data for all cases of $G$-vacua in Table~\ref{tab:minimal-models} at the GUT scale $\MGUT$.
    This data includes the matching coefficients $C_{ij}$ for Eq.~\eqref{eq:matching-condition}, the condition for determining when unification scale $\MGUT$ was reached during RG running, the parametrization of the GUT threshold values $\ETA'_{i}$ in terms of $\{\ECP,\EGU\}$, and the irrep $\XI$ whose mass is set to the matching scale $\mu=\MGUT$ in Eq.~\eqref{eq:thresholds-definition}. The labels for couplings and thresholds in $G$ are based on Table~\ref{tab:minimal-models}. Note regarding bases: the index $i$ runs over the factors in $G$
    while $j$ refers only to the unified coupling; we use the GUT normalization $\alpha_{1'}^{-1}=\tfrac{1}{24}\alpha_{\psi'}^{-1}$ for charge $\psi'$ in $G=(10)'\,1_{\psi'}$. 
    \label{tab:matching-data-GUT}}
\begin{tabular}{lllll}
     \toprule
      $G$-vacuum& $C_{ij}$ & condition for $\MGUT$ & $\ETA'_{i}$ at $\MGUT$ & $\XI$ in $\MGUT=M_{\XI}$ \\
     \midrule
     $3_C\,3_L\,3_R\rtimes \mathbb{Z}_{2}^{LR}$&
        $\left(\begin{smallmatrix} 1\\ 1\\ 1\\ \end{smallmatrix}\right)$&
        $\alpha_L^{-1}-\alpha_C^{-1}=\ETA_{C}-\ETA_{L}$&
        $(\ECP+\EGU,\ECP,\ECP)$&
        $(\mathbf{3},\mathbf{\bar{3}},\mathbf{\bar{3}})_V $\\\addlinespace[6pt]
     $3_C\,3_L\,3_R\rtimes \mathbb{Z}_{2}^{CL}$&
        $\left(\begin{smallmatrix} 1\\ 1\\ 1\\ \end{smallmatrix}\right)$&
        $\alpha_C^{-1}-\alpha_R^{-1}=\ETA_{R}-\ETA_{C}$&
        $(\ECP,\ECP,\ECP+\EGU)$&
        $(\mathbf{3},\mathbf{\bar{3}},\mathbf{\bar{3}})_V $\\\addlinespace[6pt]
     $3_C\,3_L\,3_R\rtimes \mathbb{Z}_{2}^{CR}$&
        $\left(\begin{smallmatrix} 1\\ 1\\ 1\\ \end{smallmatrix}\right)$&
        $\alpha_L^{-1}-\alpha_C^{-1}=\ETA_{C}-\ETA_{L}$&
        $(\ECP+\EGU,\ECP,\ECP+\EGU)$&
        $(\mathbf{3},\mathbf{\bar{3}},\mathbf{\bar{3}})_V $\\\addlinespace[6pt]
     $6_{CL}\,2_{R}$&
        $\left(\begin{smallmatrix} 1\\ 1\\ \end{smallmatrix}\right)$&
        $\alpha_{CL}^{-1}-\alpha_{R}^{-1}=\ETA_{R}-\ETA_{CL}$&
        $(\ECP,\ECP+\EGU)$&
        $(\mathbf{20},\mathbf{2})_V $\\\addlinespace[6pt]
     $6_{CR}\,2_{L}$&
        $\left(\begin{smallmatrix} 1\\ 1\\ \end{smallmatrix}\right)$&
        $\alpha_{L}^{-1}-\alpha_{CR}^{-1}=\ETA_{CR}-\ETA_{L}$&
        $(\ECP+\EGU,\ECP)$&
        $(\mathbf{20},\mathbf{2})_V $\\\addlinespace[6pt]
     $(10)'\,1_{\psi'}$&
        $\left(\begin{smallmatrix} 1\\ 1\\ \end{smallmatrix}\right)$&
        $\alpha_{10'}^{-1}-\alpha_{1'}^{-1}=\ETA_{1'}-\ETA_{10'}$&
        $(\ECP,\ECP+\EGU)$&
        $(\mathbf{16},-3)_V $\\
     \bottomrule
\end{tabular}
\end{table}

\subsection{Results \label{sec:analysis-unification-results}}

\subsubsection{Position of intermediate scale \label{sec:analysis-unification-results-intermediateScale}}

\def\TGUT{\bar{t}_{U}}
\def\TZ{\bar{t}_{Z}}
\def\TPL{\bar{t}_{Pl}}
\def\TI{\bar{t}_{I}}
\def\XMIN{r_-}
\def\XMAX{r_+}

We start the unification analysis by determining the intermediate scale $\MI$ for all cases of $G$-vacua in Table~\ref{tab:minimal-models}.
As discussed in Section~\ref{sec:analysis-unification-procedure}, this only involves steps $1$ and $2$ in the schematic of Eq.~\eqref{eq:procedure-RG}, and the primary dependence is on the threshold parameter $\ECR$.

For further convenience, we label the base $10$ logarithm of the scale $\mu$ expressed in $\mathrm{GeV}$ by $\bar{t}$, i.e.
\begin{align}
    \bar{t}&:= \log_{10}(\mu/\mathrm{GeV}),
\end{align}
and correspondingly $\bar{t}_{X}=\log_{10}(M_{X}/\mathrm{GeV})$ for any scale $M_{X}$ of interest, such as $\MZ$, $\MI$, $\MGUT$ and $\MPL$.

\begin{figure}[htb]
    \centering
    \includegraphics[width=0.48\linewidth]{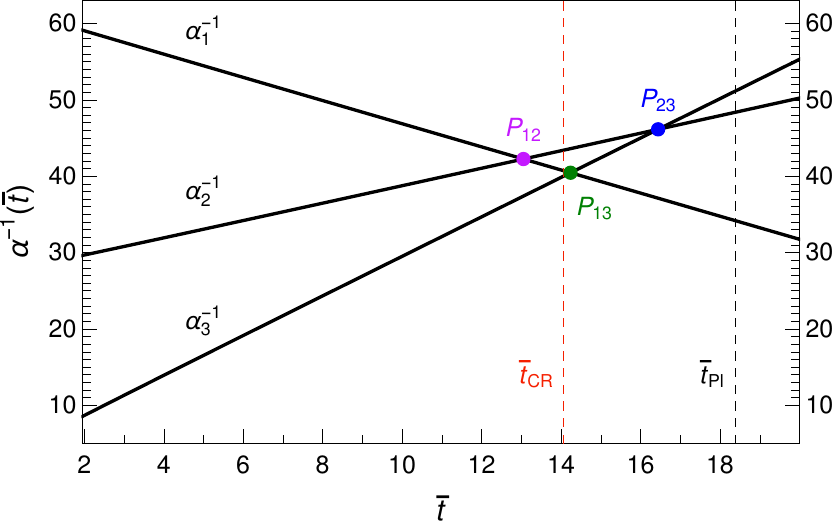}
    \HSPACEPLOT
    \includegraphics[width=0.48\linewidth]{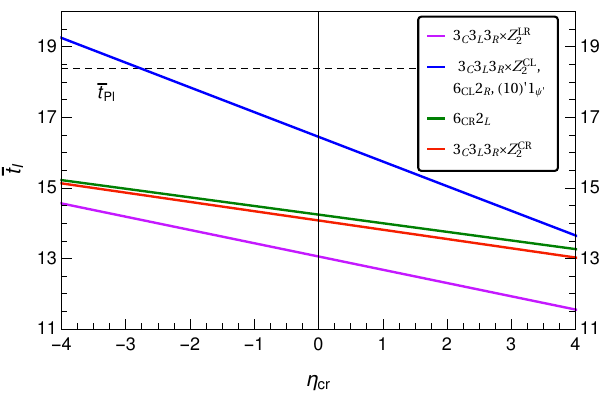}
    \caption{Left panel: the bottom-up two-loop running of SM gauge couplings, together with various intersection points defining the intermediate scale with no thresholds present. Right panel: the intermediate ($\log_{10}$)scale $\TI$ dependence on the threshold parameter $\ECR$,  as defined for all cases of $G$-vacuum in Table~\ref{tab:matching-data-I} . The color coding in both panels matches. \label{fig:RG-bottom-up-SM}}
\end{figure}

The analysis can be summarized in Figure~\ref{fig:RG-bottom-up-SM}. The left panel shows the bottom-up RG running of SM gauge couplings at two-loop order, where central values in Eq.~\eqref{eq:SM-couplings-at-MZ} are taken for the boundary condition at $\MZ$. The right panel shows the dependence of the intermediate ($\log_{10}$)scale $\TI$ on the threshold value $\ECR$ for all the scenarios of Table~\ref{tab:minimal-models}.

The intermediate scales in all scenarios are relatively close to the GUT scale, perhaps one or two orders of magnitude below where a viable unification scale is expected. This can give the masses of right-handed neutrinos in the correct ballpark for the seesaw mechanism (once the associated VEVs are multiplied with the Yukawa coupling in Eq.~\eqref{eq:fermion-mass-R}). 

We collect some side remarks below:
\begin{itemize}
\item 
    As is well known, two-loop effects have only a small effect in SM, as demonstrated by the straight-line appearance of the curves in the left panel. Furthermore, experimental errors in Eq.~\eqref{eq:SM-couplings-at-MZ} have a negligible effect on the positions of the intersection points; the largest change comes from $\delta\alpha_{3}^{-1}<0.1$, which is subdominant to the $\mathcal{O}(1)$ threshold effects $\ETA_{i}$ in Eq.~\eqref{eq:matching-condition} that are expected in our scenarios, as will be shown in Section~\ref{sec:analysis-unification-results-thresholdModeling}. We thus do not consider the experimental errors in any further analysis.
\item 
   The red vertical line $\bar{t}_{CR}$ in the left panel corresponds to the intermediate scale for trinification with $CR$ parity; it is not at an intersection of two lines, since it reflects a more complicated condition for $\MI$ in Table~\ref{tab:matching-data-I}. The scale $\bar{t}$ associated with each color in the left panel corresponds to the values at $\ECR=0$ in the right panel.
\item 
    In the right panel of Figure~\ref{fig:RG-bottom-up-SM}, the intermediate scale always drops with increasing $\ECR>0$, which is a consequence of our consistent choice of signs in the parametrization of thresholds defined in Table~\ref{tab:matching-data-I}. The rate of drop corresponds to the size of the intersection angle in the left panel, with the $CR$ trinification case very close to the intersection angle for point $P_{13}$ and the $6_{CR}\,2_L$ case.
\end{itemize}

\subsubsection{Unification in benchmark scenarios \label{sec:analysis-unification-results-benckmarkPoints}}

The analysis from Figure~\ref{fig:RG-bottom-up-SM} is now extended beyond just the first two steps of the procedure in Eq.~\eqref{eq:procedure-RG}. We consider the thresholds $\{\ECO,\ECR,\EFR\}$ from Table~\ref{tab:matching-data-I} and $\{\ECP,\EGU\}$ from Table~\ref{tab:matching-data-GUT} as input parameters, and study how unification depends on their values for all the minimal cases, cf.~Table~\ref{tab:minimal-models}. As was pointed out in Section~\ref{sec:analysis-unification-procedure}, the parametrizations of thresholds was chosen so that $\ECO$ and $\ECP$ have no effect (at one-loop) whether and at which scale 
couplings unify, so we set for now $\ECO=\ECP=0$ and effectively explore the $3D$ parameter space of $\{\ECR,\EFR,\EGU\}$. 

To this end, we consider for each case of $G$-vacuum two benchmark scenarios:

\begin{enumerate}
    \item \label{item:benchmark-scenario-1}
        We perform steps $1$--$4$ in the procedure of Eq.~\eqref{eq:procedure-RG} with no intermediate thresholds, i.e.~$\ECR=\EFR=0$. We keep RG running in the effective intermediate theory to high scales. The gap between the two couplings at any given scale $\bar{t}$ in the intermediate theory can be viewed as the required value for the threshold correction $\EGU$ for unification to occur at said scale. 
    \item \label{item:benchmark-scenario-2}
        Informed by the behavior of RG running in the first scenario, we choose for each $G$-vacuum specific values for $\ECR$ and $\EFR$ at the intermediate scale, such that unification happens with $\EGU=0$ and preferably at a scale $\MGUT > 10^{15.8}\,\mathrm{GeV}$ (benchmark for proton decay, as seen later in Section~\ref{sec:analysis-proton-general}). We perform steps $1$--$6$ in the procedure of Eq.~\eqref{eq:procedure-RG}, i.e.~we also show the RG running above $\MGUT$ of the unified coupling in the full $\mathrm{E}_{6}$ model.
\end{enumerate}

\begin{figure}[htb]
    \centering
    \includegraphics[width=0.48\linewidth]{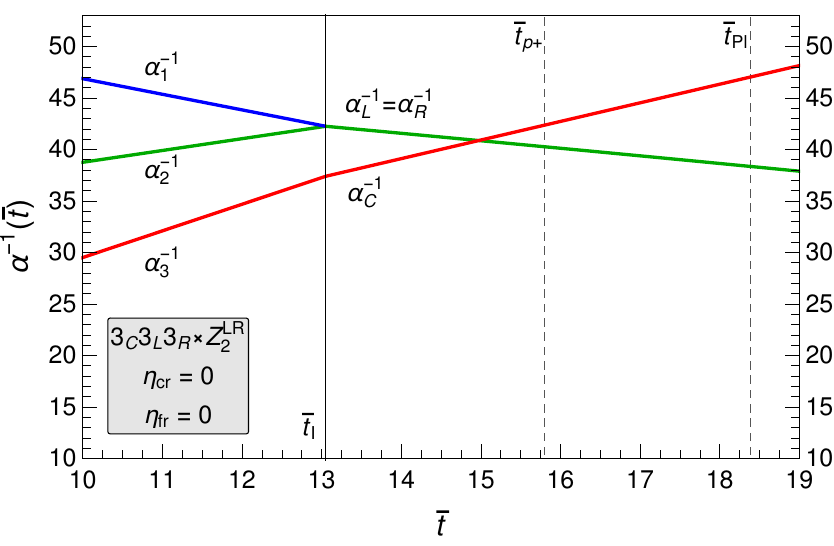}
    \HSPACEPLOT
    \includegraphics[width=0.48\linewidth]{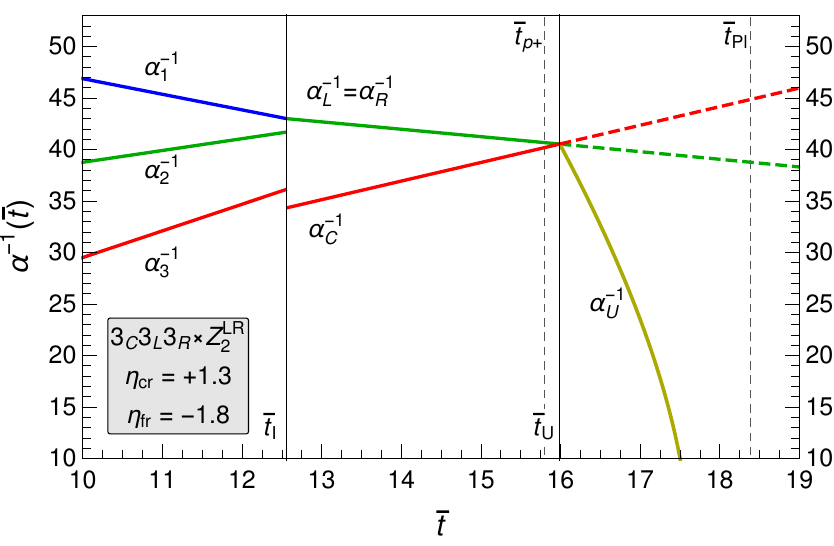}\\[6pt]
    \includegraphics[width=0.48\linewidth]{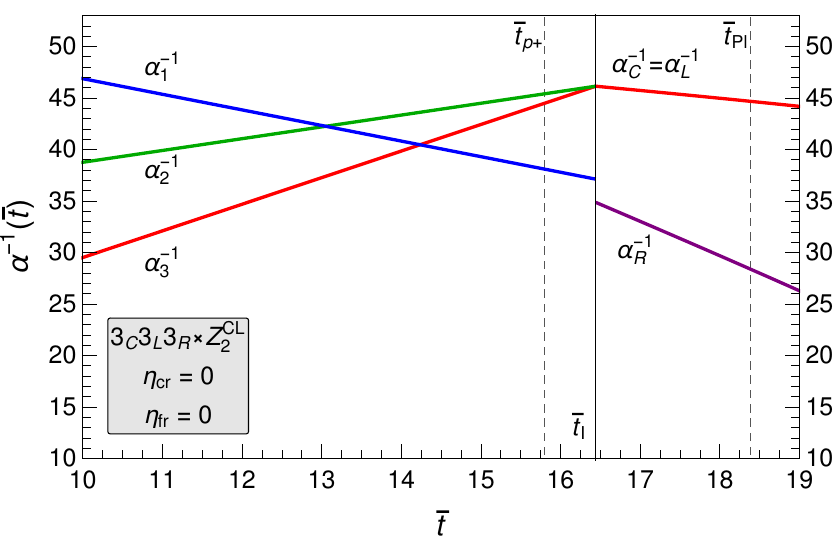}
    \HSPACEPLOT
    \includegraphics[width=0.48\linewidth]{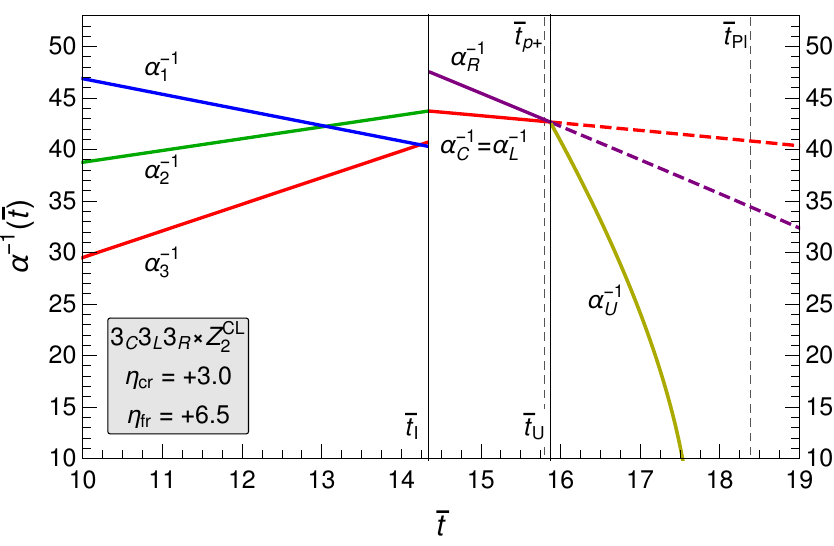}\\[6pt]
    \includegraphics[width=0.48\linewidth]{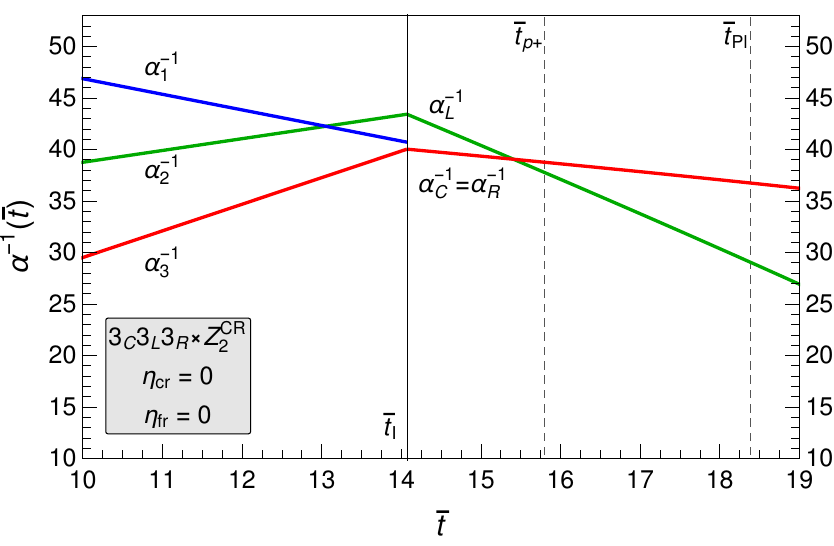}
    \HSPACEPLOT
    \includegraphics[width=0.48\linewidth]{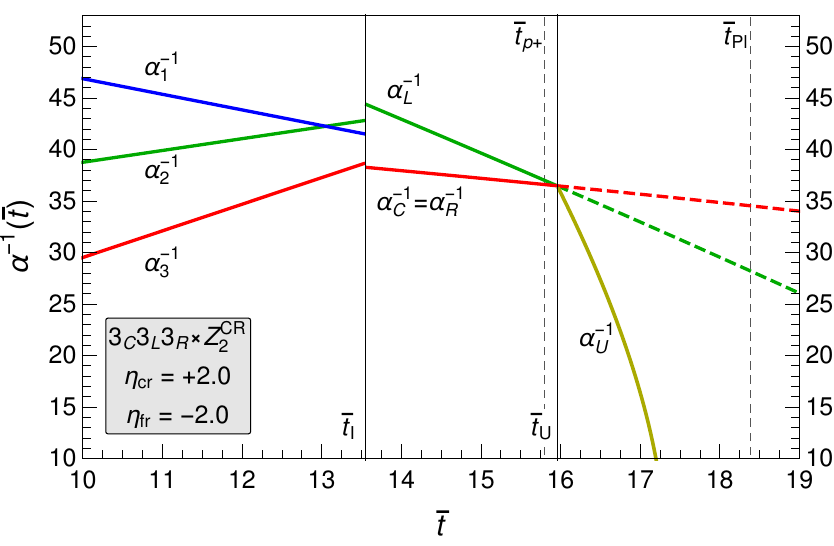}
    \caption{RG running of gauge couplings $\alpha_{i}^{-1}$ with ($\log_{10}$)scale $\bar{t}$ for all $\SU(3)_C\times\SU(3)_L\times\SU(3)_R$ trinification cases, cf.~Table~\ref{tab:minimal-models}. The three rows of panels correspond to a preserved $LR$, $CL$, or $CR$ parity, respectively. The left and right column of panels corresponds respectively to the benchmark scenarios~\ref{item:benchmark-scenario-1} and \ref{item:benchmark-scenario-2}, i.e.~to vanishing and non-vanishing intermediate threshold values $\ECR$ and $\EFR$. Their values in the right column were chosen so that unification occurs with $\EGU=0$ between the reference scales $\bar{t}_{p^{+}}$ and $\TPL$. 
    \label{fig:RG-benchmark-trinification}}
\end{figure}

\begin{figure}[htb]
    \centering
    \includegraphics[width=0.48\linewidth]{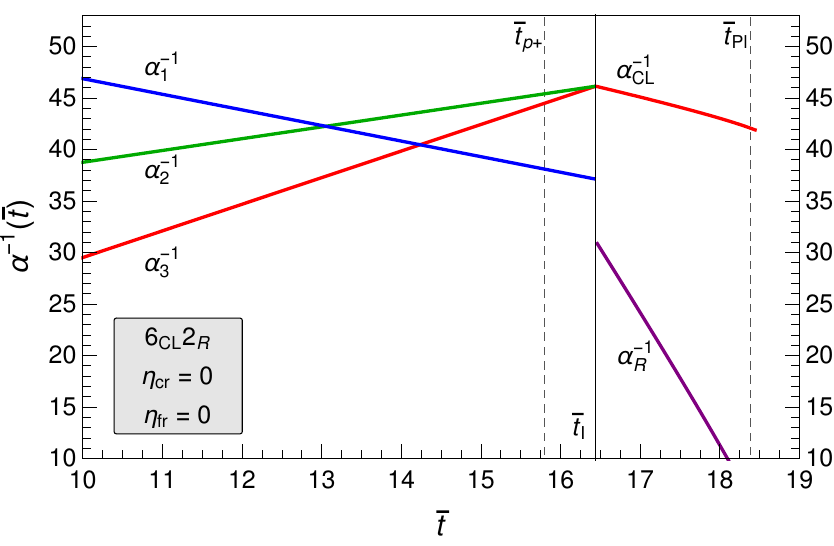}
    \HSPACEPLOT
    \includegraphics[width=0.48\linewidth]{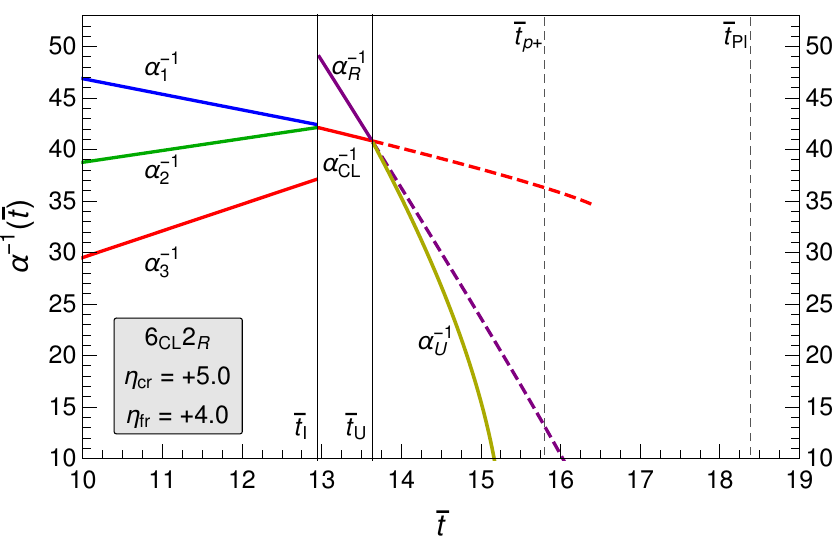}\\[6pt]
    \includegraphics[width=0.48\linewidth]{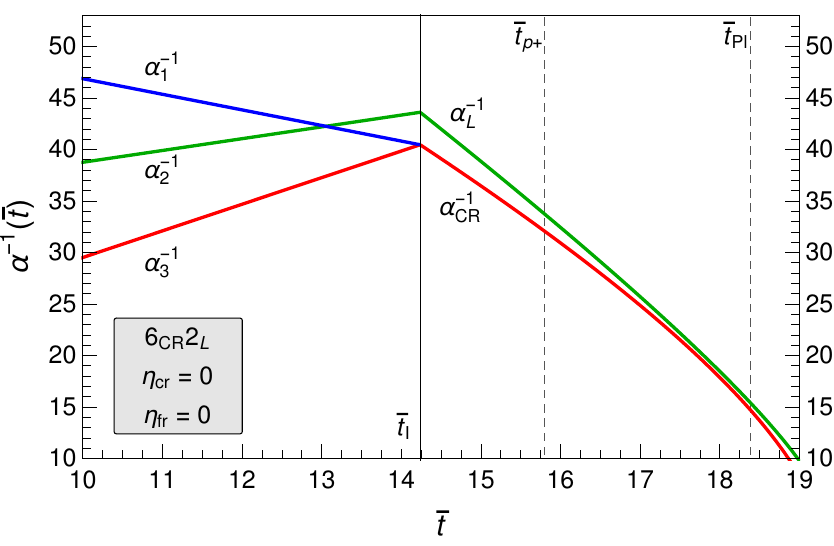}
    \HSPACEPLOT
    \includegraphics[width=0.48\linewidth]{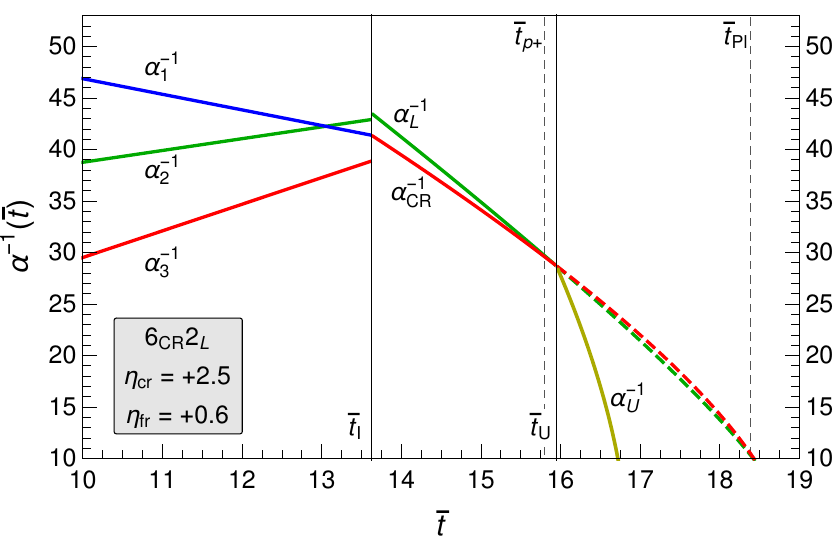}\\[6pt]
    \includegraphics[width=0.48\linewidth]{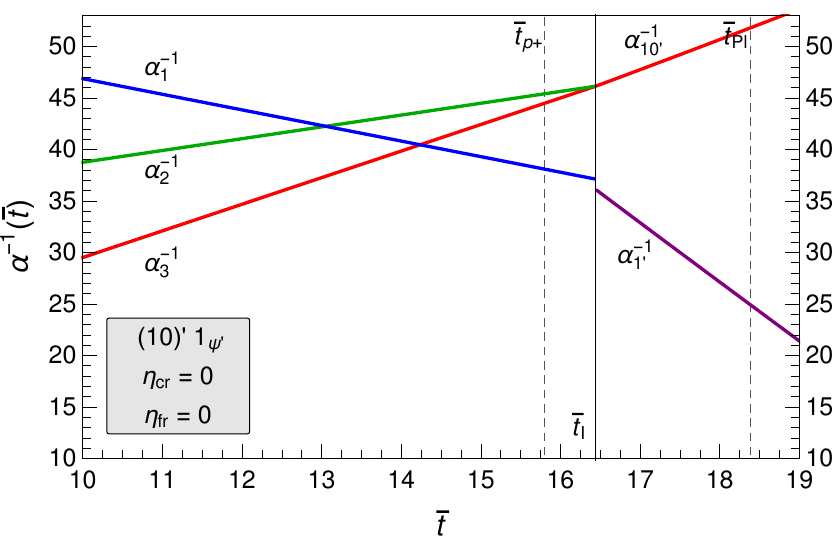}
    \HSPACEPLOT
    \includegraphics[width=0.48\linewidth]{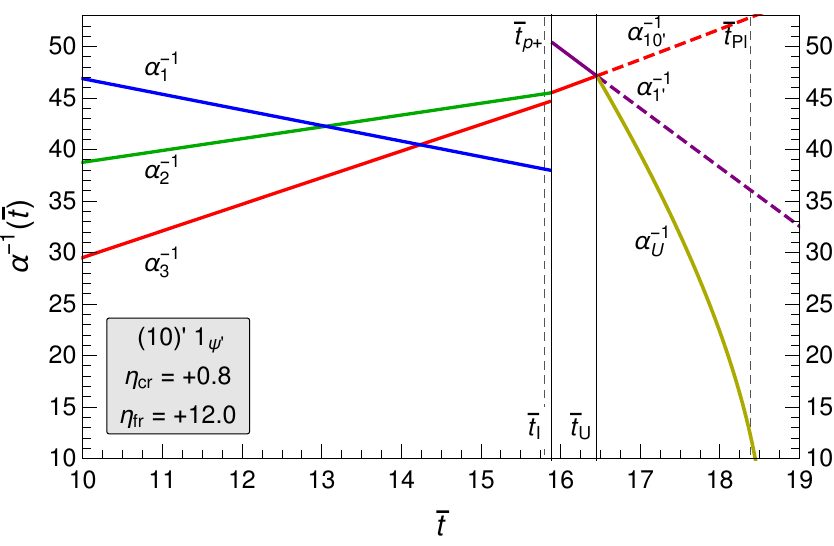}\\
    \caption{RG running of gauge couplings $\alpha_{i}^{-1}$ with ($\log_{10}$)scale $\bar{t}$ for all non-trinification cases, cf.~Table~\ref{tab:minimal-models}. The three rows of panels correspond to the minimal scenarios for the standard $\SU(6)_{CL}\times\SU(2)_R$, LR-flipped $\SU(6)_{CR}\times\SU(2)_L$ and flipped $\SO(10)'\times\UU_{\psi'}$. The left and right column of panels corresponds respectively to the benchmark scenarios~\ref{item:benchmark-scenario-1} and \ref{item:benchmark-scenario-2}, i.e.~to vanishing and non-vanishing intermediate threshold values $\ECR$ and $\EFR$. Their values in the right column were chosen so that unification occurs with $\EGU=0$, and if possible between the reference scales $\bar{t}_{p^{+}}$ and $\TPL$. 
    \label{fig:RG-benchmark-nonTrinification}}
\end{figure}

The results of the gauge coupling RG running  under these benchmark scenarios are shown in Figures~\ref{fig:RG-benchmark-trinification} and \ref{fig:RG-benchmark-nonTrinification}, where trinification cases are shown in rows of the former figure and non-trinification cases in the latter. The two benchmark scenarios without and with threshold corrections $\ECR$ and $\EFR$  
are shown as panels on the left- and right-hand side, respectively, as indicated in the bottom-left corner in each panel. We list below for quick reference all the features drawn in these plots:
\begin{itemize}
    \item All plots show the RG running only in the upper range $\bar{t}> 10$ relevant for analyzing unification. The intermediate and GUT matching scales $\TI$ and $\TGUT$ are shown as solid vertical lines, while the two dashed vertical lines correspond to two reference scales: the proton decay scale $\bar{t}_{p^{+}}= 15.8$ (an estimate for the experimental bound in $\mathrm{E}_{6}$, shown later in Section~\ref{sec:analysis-proton-general}) and the reduced Planck scale $\TPL= 18.38$.   
    \item We use consistent colors for curves in all the plots. The SM couplings $\alpha^{-1}_{3,2,1}$ are colored by red, green and blue. The intermediate couplings for factors containing the entire color group $3_C$ continue as red, while those containing the entire $2_L$ group continue as green, where the larger group has precedence if both $3_C$ and $2_L$ join together. For an intermediate coupling related to hypercharge and a more complicated matching via $C_{ij}$ from Table~\ref{tab:matching-data-I}, we color it purple. Notice that the complicated matching causes a discontinuity between all SM couplings and the purple line at $\TI$ even if all threshold corrections vanish. Finally, the running of the unified $\mathrm{E}_{6}$ coupling in the right-hand panels is drawn in yellow, while the extension above $\MGUT$ of RG running in the intermediate theory is drawn by dashed curves.
\end{itemize}

\begin{figure}[htb]
    \centering
    \includegraphics[width=0.55\linewidth]{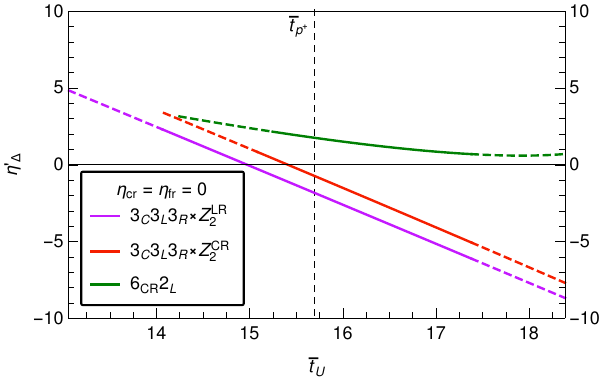}
    \caption{The required threshold corrections $\EGU$ for unification to happen at a given $\TGUT$ if no thresholds are present at $\TI$ ($\ECR=\EFR=0$). The three successful unification cases are distinguished by color, while a dashed curve in each case indicates a proximity of one order of magnitude or less to either $\TI$ or $\TPL$. The three unsuccessful cases have curves falling outside the shown range of $\EGU$. 
    \label{fig:plot-GUT-scale-and-thresholds}}
\end{figure}

\noindent
The running plots in Figures~\ref{fig:RG-benchmark-trinification} and \ref{fig:RG-benchmark-nonTrinification} 
prove very instructive and enable us to draw the following conclusions:
\begin{itemize}
    \item 
        There are three cases of $G$-theories where unification can happen with no threshold effects at the intermediate scale: $3_C\,3_L\,3_R\rtimes\mathbb{Z}_{2}^{LR}$, $3_C\,3_L\,3_R\rtimes\mathbb{Z}_{2}^{CR}$, and $6_{CR}\,2_L$, i.e.~the cases $1$, $3$ and $5$ from Table~\ref{tab:minimal-models}. For the first two, the intermediate couplings intersect and the threshold effects $\ECR$ and $\EFR$ in the right-hand panels merely help with repositioning the GUT scale. In the third case of the LR-flipped $6_{CR}\,2_L$, the intermediate couplings approach each other at a shallow angle rather than intersect, so a small GUT threshold correction $\EGU$ could alternatively be used instead of the intermediate thresholds $\ECR$ and $\EFR$. For all three successful cases of unification, the intermediate thresholds in the right-hand panels could be made even smaller if $\EGU\neq 0$ is brought into the picture. 
    \item 
        Three cases of $G$-theories do not unify with vanishing threshold corrections at $\TI$: $3_C\,3_L\,3_R\rtimes\mathbb{Z}_{2}^{CL}$, $6_{CL}\,2_{R}$, and $(10)'\,1_{\psi'}$, i.e.~the cases $2$, $4$ and $6$ from Table~\ref{tab:minimal-models}. The problem they encounter is that the two different values of the intermediate couplings start at $\TI$ in the wrong hierarchy for their one-loop beta coefficients $a$, so their curves diverge from then onward. This can in principle be fixed by assuming very large thresholds $\ECR$ and $\EFR$ that switch the relative hierarchy of the two couplings, as shown in the right-hand panels for these cases. Even allowing for this indulgence, the case $6_{CL}\,2_{R}$ still has trouble unifying above $\bar{t}_{p+}$.
    \item 
        The RG curves in Figures~\ref{fig:RG-benchmark-trinification} and \ref{fig:RG-benchmark-nonTrinification} deviate from the shape of a straight line when two-loop effects become important, an effect which is more pronounced in the lower regions of the graphs, i.e.~at larger coupling value $\alpha$, cf.~Eq.~\eqref{eq:RG-running}. Among the curves of intermediate $G$-theories, a slight effect of this sort can be noticed in the case $6_{CR}\,2_{L}$. The two-loop effect is visually always very pronounced, however, for the running in $\mathrm{E}_{6}$ (yellow curve) due to the large two-loop coefficient in Eq.~\eqref{eq:ABcoefficients-E6}. This raises questions of perturbativity, which we address in a dedicated discussion in Section~\ref{sec:analysis-unification-perturbativity}.
    \item 
        Under the assumption of vanishing intermediate-scale thresholds corrections, we can summarize the (non)unification information for all cases in one simple plot. For any given value of $\ECR$ and $\EFR$, we can always choose $\EGU$ such that unification happens at any desired scale above $\TI$.  We plot the required $\EGU$ as a function of the desired GUT scale $\TGUT$ for $\ECR=\EFR=0$ in Figure~\ref{fig:plot-GUT-scale-and-thresholds}. The three favored cases for unification are indicated by curves of different colors, while the curves for disfavored cases fall outside the plot range, i.e.~$|\EGU|>10$ for any suggested $\TGUT$. We see that the green curve is rather flat, indicating that the $6_{CR}\,2_L$ case prefers specific values for the threshold $\EGU$ and that the GUT scale $\TGUT$ is very sensitive to this parameter.
\end{itemize}

\subsubsection{Unification in minimally tuned scenarios \label{sec:analysis-unification-results-thresholdModeling}}

While we analyzed the (non)unification for all minimal cases from Table~\ref{tab:minimal-models} in Section~\ref{sec:analysis-unification-results-benckmarkPoints}, the threshold values there were considered simply as free parameters and the analysis considered benchmark cases. We now study what the expected size of the thresholds values $\ETA$ could be and what are the implications for unification.

The five thresholds of two-stage breaking, parametrized via $\{\ECO,\ECR,\EFR,\ECP,\EGU\}$ in Tables~\ref{tab:matching-data-I} and \ref{tab:matching-data-GUT}, are computed from the spectrum of the theory via Eqs.~\eqref{eq:thresholds-definition}. While the GUT-scale spectrum results are available in their entirety, the intermediate spectrum involves a prohibitively large number of SM irreps and parameters, as well as vector-like exotic fermions depending on Yukawa matrices and thus on the fermion fit. An analysis of this kind is beyond the scope of this paper.

Instead of computing the spectrum from the parameters in the scalar and Yukawa sectors, we instead model the masses themselves as free parameters. For the most part, the masses can indeed be considered largely independent, as we shall argue shortly. A sector-by-sector breakdown into irreps, which appear with their masses in the threshold corrections, is as follows:
\begin{itemize}
    \item 
        The scalar fields in the $\mathbf{650}$ are all heavy and their masses depend on the parameters of the part $V_{650}$ in Eq.~\eqref{eq:potential-all}. The breakdown into $G$-irreps is listed in Eqs.~\eqref{eq:decomposition-to333-650}, \eqref{eq:decomposition-to62-650} and \eqref{eq:decomposition-to101-650}. The masses have been shown to be independent~\cite{Babu:2023zsm}, except for two mass relations in the trinification case, one of which is redundant due to imposing a condition on the trinification singlet, 
        which does not contribute to threshold corrections. Note that $\mathbf{650}$ is a real representation, so in its decomposition the complex $G$-irreps $\mathbf{R}$ and $\mathbf{R}^{\ast}$ contain the same complex degrees of freedom, and are thus treated in threshold corrections accordingly.
    \item 
        The scalar fields in $\mathbf{27}\oplus\mathbf{351'}$ can be either at the GUT or intermediate scale. The 
        intermediate-scale fields depend on the $G$-vacuum and are listed in each case in Table~\ref{tab:minimal-models}, while the heavy fields are the remaining $G$-irreps in the decompositions of $\mathbf{27}$ and $\mathbf{351'}$ in Eqs.~\eqref{eq:decomposition-to333-27}--\eqref{eq:decomposition-to101-351p}.
        The masses of the heavy fields are expressed in terms of the parameters in $V_{\text{mix}}$; their explicit expressions are found in Appendix~\ref{app:masses-scalars} and are independent. 
        \par 
        Tuning conditions on the parameters $V_{\text{mix}}$ bring the desired $G$-irreps to the intermediate scale, in accordance with each case of $G$-vacuum. These $G$-irreps at scale $\MI$ on the other hand break down further to SM-irreps. We refrain from listing the large number of SM-irreps explicitly in the paper, but the result is straightforwardly computed using the public software \texttt{GroupMath}~\cite{Fonseca:2020vke} by using projection matrices of the embedding $G\supset 3_C\,2_L\,1_Y$ in Eqs.~\eqref{eq:projection-matrices-begin}--\eqref{eq:projection-matrices-end}  of Appendix~\ref{app:embeddings-projection-matrices}. The masses of the intermediate fields are characterized by the first appearance of parameters in the $V_{27,351'}$ part of the scalar potential, i.e.~the $3$ mass parameters $m'_{i}$ and $12$ quartic coefficients $\LAMBDAP_{j}$. We have not computed the mass expressions of these fields, and here we assume that taking them independent still gives a good enough approximation for threshold corrections.  
    \item 
        The fermions present around the intermediate scale are the exotics of Eqs.~\eqref{eq:exotic-fermions}, as discussed in detail when fermion masses were considered in Section~\ref{sec:minimal-models-Yukawa}. Furthermore, the SM-singlet fields do not contribute to the intermediate thresholds, and hence the contribution comes only from vector-like fermions. In particular, the contributions come from down-type quarks $(\mathbf{3},\mathbf{1},-1/3)$ and lepton doublets $(\mathbf{1},\mathbf{2},+1/2)$ of $3_C\,2_L\,1_Y$, each coming in three copies and treated as Dirac fermions. When spinorial parity is preserved, their masses are given by the $(2,2)$-block entry in Eqs.~\eqref{eq:fermion-mass-D} and \eqref{eq:fermion-mass-E}; their scale is thus set by the $\MI$-scale VEVs $V_{2}$ and $W_{2}$, as well as the Yukawa matrices $\YPSIBF$ and $\YTHETABF$. We assume in our threshold modeling that their dependence on Yukawa matrices causes a hierarchical suppression $10^{-4}:10^{-2}:1$ between families, but they can still be chosen with independent $\mathcal{O}(1)$ coefficients. 
    \item 
        The gauge bosons can be either at the GUT or intermediate scale. Their masses are given in Appendix~\ref{app:masses-gauge} and depend on the gauge coupling and VEVs. Individually these come with the largest coefficients into threshold corrections from any type of field; hence this is the only sector where we do not model their masses as independent, but instead implement in full any relations between their masses. Furthermore, note that one gauge boson irrep $\XI$ is chosen at each matching scale to set the scale $\mu$ of the threshold corrections, as discussed 
        in Section~\ref{sec:analysis-unification-procedure} and listed in Tables~\ref{tab:matching-data-I} and \ref{tab:matching-data-GUT}. The irrep $\XI$ thus does not contribute to the threshold corrections in our setup. Since there is only one (complex) irrep of gauge bosons in the coset $\mathrm{E}_{6}/G$ for all cases of $G$, there is no threshold correction from gauge bosons at the scale $\MGUT$. At $\MI$ there is more than one SM-irrep of heavy gauge bosons; we model them by taking the VEVs to be independent, while the gauge couplings are approximately all equal, and then they are normalized so that their masses are expressed in units of $M_{\XI}=\mu$. 
\end{itemize}
The only further curation the above lists of $G$-irreps at $\MGUT$ and SM-irreps at $\MI$ require is to remove the irreps that are would-be Goldstone modes at both scales, and to remove one doublet $(\mathbf{1},\mathbf{2},+1/2)$ representing the SM-Higgs from the list of states around $\MI$. For trinification cases, masses related by the preserved parity are taken as equal, which is reflected in threshold values respecting said parity, e.g.~if $\mathbb{Z}_{2}^{LR}$ is preserved then taking the spectrum to be $LR$ symmetric yields $\ETA_{R}=\ETA_{L}$. 

We model the masses of all states under the assumption of minimal fine-tuning, i.e., no state is much lower than it needs to be.\footnote{If an arbitrary number of states are fine-tuned to lower mass ranges, the generic GUT expectation when a large number of states is available is that it is always possible to unify the gauge couplings. This would manifest itself in our setup as very large threshold corrections. We are thus interested in scenarios with minimal fine-tunings, where thresholds have a predictable range.} The VEVs from which gauge boson masses are computed are randomly sampled within one order of magnitude; this gives the mass for $\XI$ and thus sets $\mu$, and the mass of scalars and fermions around $\mu$ is sampled as $10^r\mu$, where $r=\mathrm{rand}(-1,0)$ is sampled uniformly. The motivation behind taking an asymmetric interval is that scalar and fermion masses should be lighter than gauge bosons, since they depend on scalar-potential and Yukawa couplings rather than the larger gauge coupling. The aforementioned fermion hierarchy is imposed by further multiplying the fermion masses with suppression factors $(10^{-4},10^{-2},10^0)$ for the three families. 

The above procedure describes the random sampling of all irrep masses, from which the threshold corrections $\{\ECO,\ECR,\EFR,\ECP,\EGU\}$ can be computed for any given $G$-theory. The obtained threshold values can then be used to run the RG procedure in Eq.~\eqref{eq:procedure-RG} to see whether and where unification happens. 

The two-step procedure of computing threshold corrections and the associated unification can then be repeated multiple times, allowing to perform statistics on the results. We computed $10^{5}$ points independently for each case of $G$-theory in Table~\ref{tab:minimal-models}.\footnote{Note that each $G$-theory comes with different irreps for TC, so a single ``point'' of chosen masses cannot be used across multiple $G$-theory scenarios.} The resulting distributions of threshold values are essentially Gaussian, so they are well represented by computing the mean and standard deviation; the results are shown in Table~\ref{tab:results-statistics}. The last row of the table also specifies what percentage of points for each $G$-theory successfully unified.

\begin{table}[htb]
    \centering
    \caption{The modeled values for threshold corrections based on $10^{5}$ points in each case of $G$-theory from Table~\ref{tab:minimal-models}. Each entry in the table represents the average and standard deviation $\bar{\ETA}\pm\delta\ETA$ for a distribution $\ETA$. The generic labels $\ETA'_{1,2}$ denote the two different values for couplings of $G$ (in the order of its factors). The last row specifies the percentage of points --- independently generated for each $G$-theory --- in which unification was achieved below $\TPL$, but with no lower bound imposed. \label{tab:results-statistics}}
    \begin{tabular}{lrrrrrr}
        \toprule
        & 
            $3_C3_L3_R\rtimes \mathbb{Z}_{2}^{LR}$&
            $3_C3_L3_R\rtimes \mathbb{Z}_{2}^{CL}$&
            $3_C3_L3_R\rtimes \mathbb{Z}_{2}^{CR}$&
            $6_{CL}\,2_R$&
            $6_{CR}\,2_L$&
            $(10)'\,1_{\psi'}$ \\
        \midrule
            $\ETA_{3}$&$-1.8\pm 0.1$  & $-2.9\pm 0.3$  & $-2.9\pm 0.4$  & $-4.8\pm 0.6$  & $-5.6\pm 0.6$  & $-4.6\pm 1.1$ \\
            $\ETA_{2}$&$-2.8\pm 0.2$  & $-2.8\pm 0.2$  & $-3.9\pm 0.5$  & $-5.0\pm 0.5$  & $-4.6\pm 0.4$  & $-4.3\pm 1.3$ \\
            $\ETA_{1}$&$-2.1\pm 0.3$  & $-3.0\pm 0.4$  & $-2.3\pm 0.3$  & $-6.0\pm 1.3$  & $-4.4\pm 0.8$  & $-3.5\pm 0.3$ \\[4pt]
            $\ETA'_{1}$&$-9.4\pm 2.1$  & $-8.3\pm 2.0$  & $-8.3\pm 2.0$  & $-6.3\pm 1.9$  & $-5.1\pm 1.6$  & $-7.3\pm 2.0$ \\
            $\ETA'_{2}$&$-8.3\pm 2.0$  & $-7.2\pm 1.8$  & $-7.2\pm 1.8$  & $-4.0\pm 1.7$  & $-6.6\pm 2.3$  & $-7.8\pm 2.4$ \\[4pt]
            $\ECO$&$-2.1\pm 0.3$  & $-2.8\pm 0.2$  & $-3.1\pm 0.4$  & $-5.0\pm 0.5$  & $-4.4\pm 0.8$  & $-4.3\pm 1.3$ \\
            $\ECR$&$-0.7\pm 0.3$  & $-0.1\pm 0.3$  & $-0.8\pm 0.4$  & $0.2\pm 0.5$  & $-1.2\pm 0.6$  & $-0.3\pm 0.4$ \\
            $\EFR$&$0.3\pm 0.3$  & $-0.1\pm 0.4$  & $1.0\pm 0.3$  & $-1.0\pm 1.1$  & $-0.1\pm 0.7$  & $0.9\pm 1.1$ \\[4pt]
            $\ECP$&$-9.4\pm 2.1$  & $-7.2\pm 1.8$  & $-8.3\pm 2.0$  & $-4.0\pm 1.7$  & $-5.1\pm 1.6$  & $-7.8\pm 2.4$ \\
            $\EGU$&$1.1\pm 1.3$  & $-1.1\pm 1.3$  & $1.1\pm 1.3$  & $-2.3\pm 1.7$  & $-1.5\pm 1.7$  & $0.4\pm 1.9$ \\[4pt]
            $\%$ unify& $>99.9$ & $0.0$ & $99.1$ & $<0.1$ & $26.2$ & $<0.1$ \\ 
        \bottomrule
    \end{tabular}
\end{table}

Statistics from the modeled threshold values allows for a more comprehensive analysis of compatibility with unification, which requires first some preliminary considerations. For every case of $G$-theory, a given pair $(\ECR,\EFR)$ of TC values uniquely defines the matching scale $\MI$ and the intermediate-coupling values there up to an overall shift; then at any given scale $\mu$ in the interval\footnote{We take very mild unification constraints here, so we impose no additional buffer between $\MGUT$ and the scales $\MI$ or $\MPL$.} $\MI<\mu<\MPL$, unification can be achieved for some value of GUT thresholds $\EGU$, cf.~Table~\ref{tab:matching-data-GUT}; we label this function as $\EGU(\mu)$, while the modeled thresholds give a distribution with $\bar{\ETA}'_{\Delta}\pm\delta\EGU$, see Table~\ref{tab:results-statistics}. We can define for every case of $G$-vacuum a measure of incompatibility $\sigma_\Delta$ between the required and modeled value $\EGU$ to unify anywhere below the Planck scale: 
    \begin{align}
        \sigma_{\Delta}(\ECR,\EFR)&:= \mathrm{s_{\pm}}\cdot \mathrm{min}_{\mu\in (\MI(\ECR,\EFR),\MPL)}\left| \frac{\EGU(\mu)-\bar{\ETA}'_{\Delta}}{\delta\EGU}\right|, \label{eq:definitions-sigmadelta}
    \end{align}
where $s_{\pm}$ is the sign of the modulus argument in the minimum.
For example, if for some value of intermediate thresholds a value of $\sigma_\Delta=-3$ is obtained, that indicates any GUT threshold that unifies will be outside the $3$-$\sigma$ region of the modeled distribution, and the negative sign tells us that the required threshold is below the modeled values; such intermediate thresholds are then interpreted as $3$-$\sigma$ unlikely to successfully unify.

\begin{figure}[htb]
    \centering
    \includegraphics[width=0.31\linewidth]{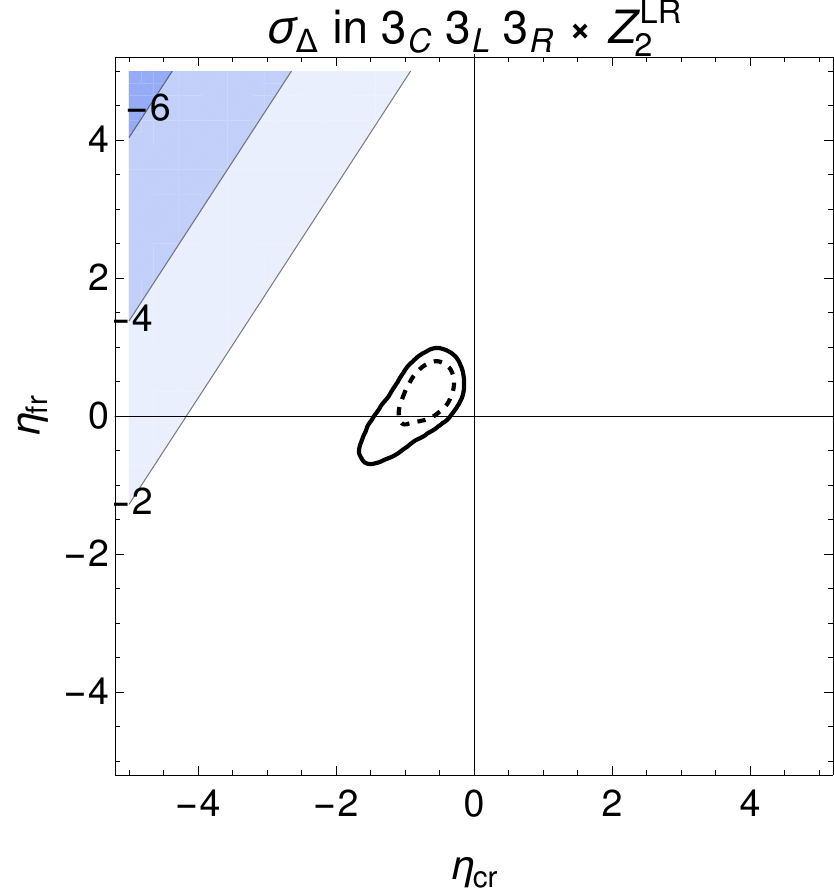}
    \includegraphics[width=0.31\linewidth]{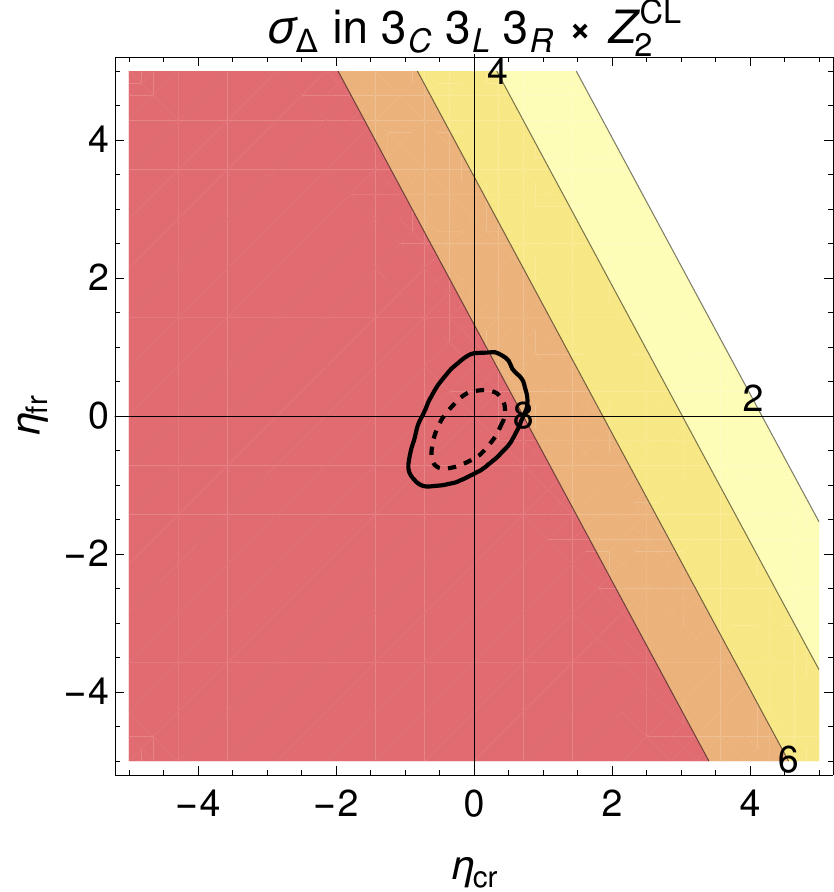}
    \includegraphics[width=0.31\linewidth]{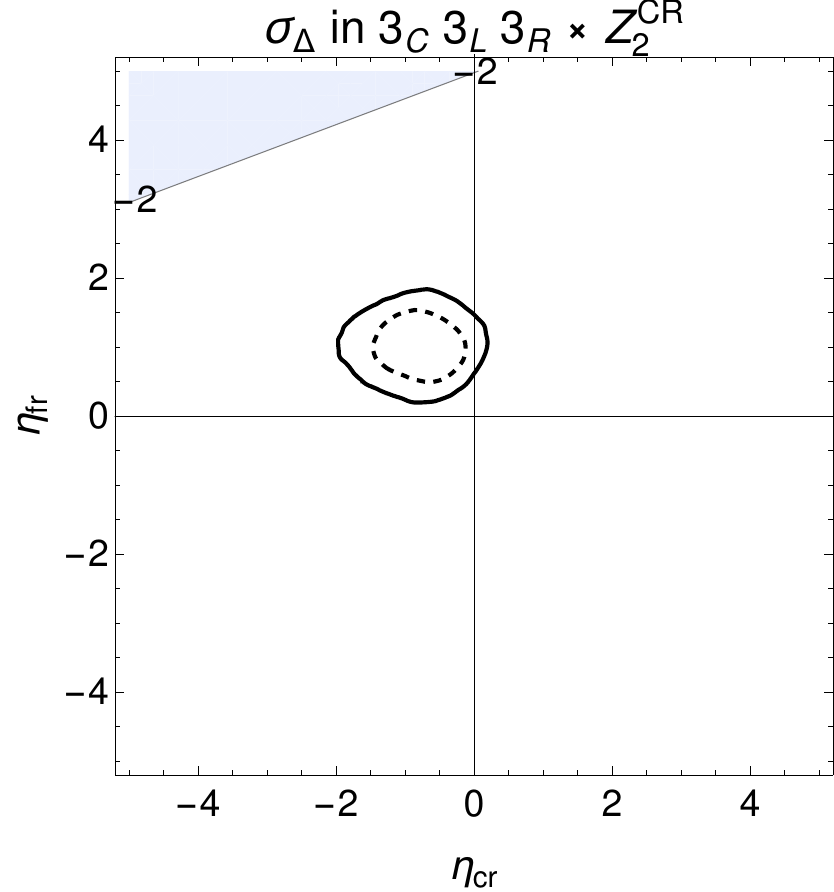}
    \includegraphics[height=0.32\linewidth]{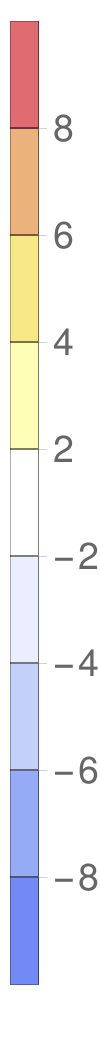}\\[4pt]
    \includegraphics[width=0.31\linewidth]{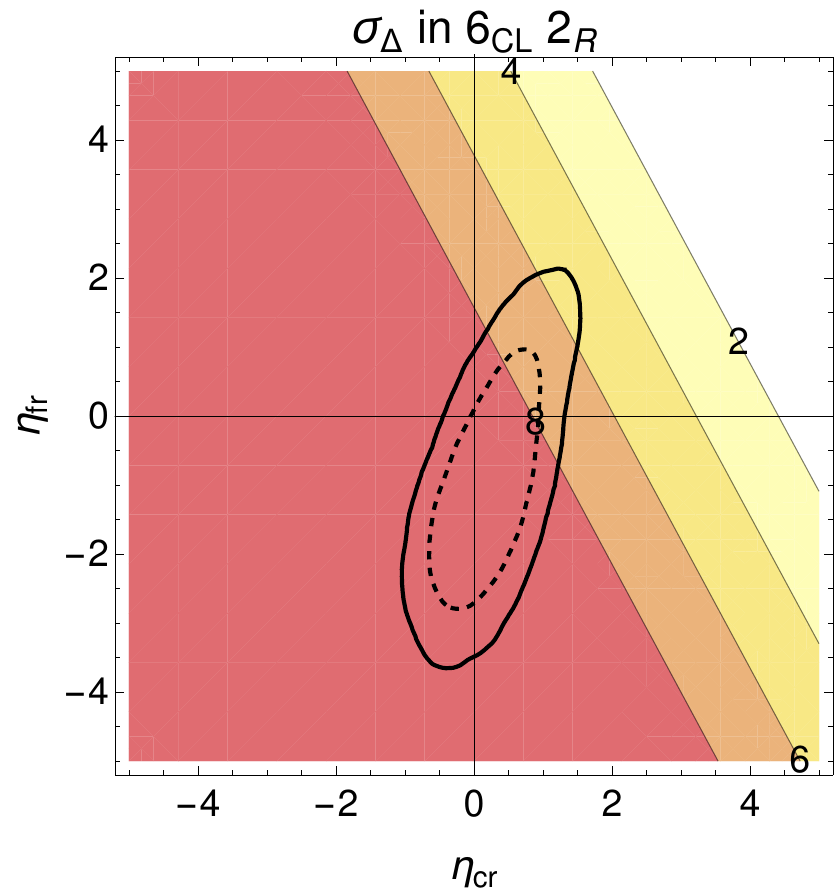}
    \includegraphics[width=0.31\linewidth]{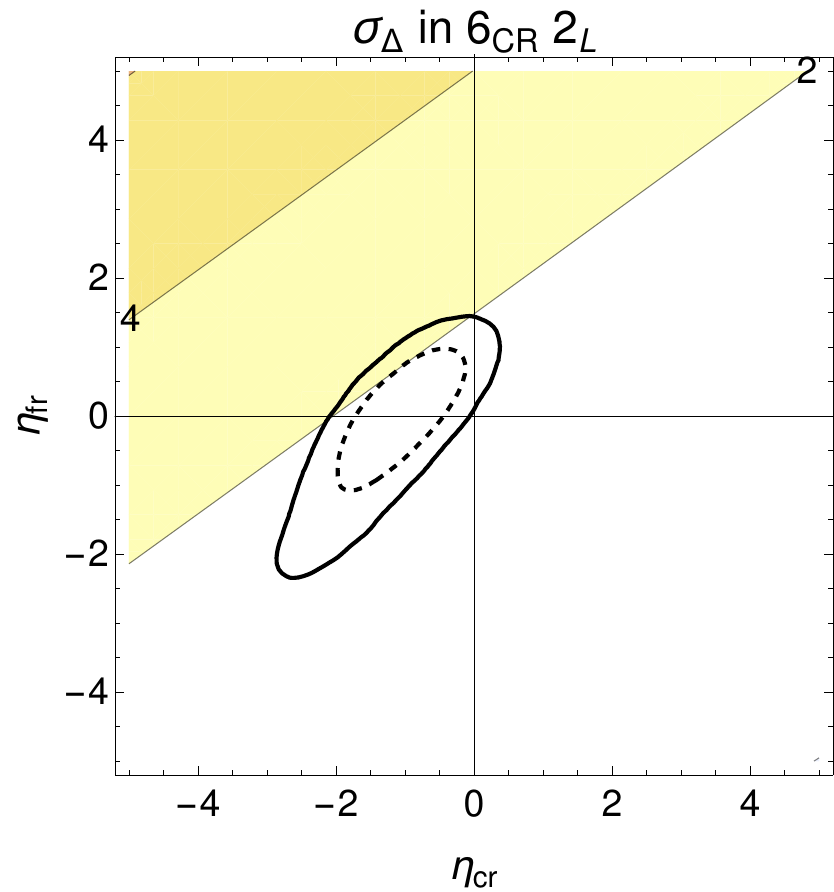}
    \includegraphics[width=0.31\linewidth]{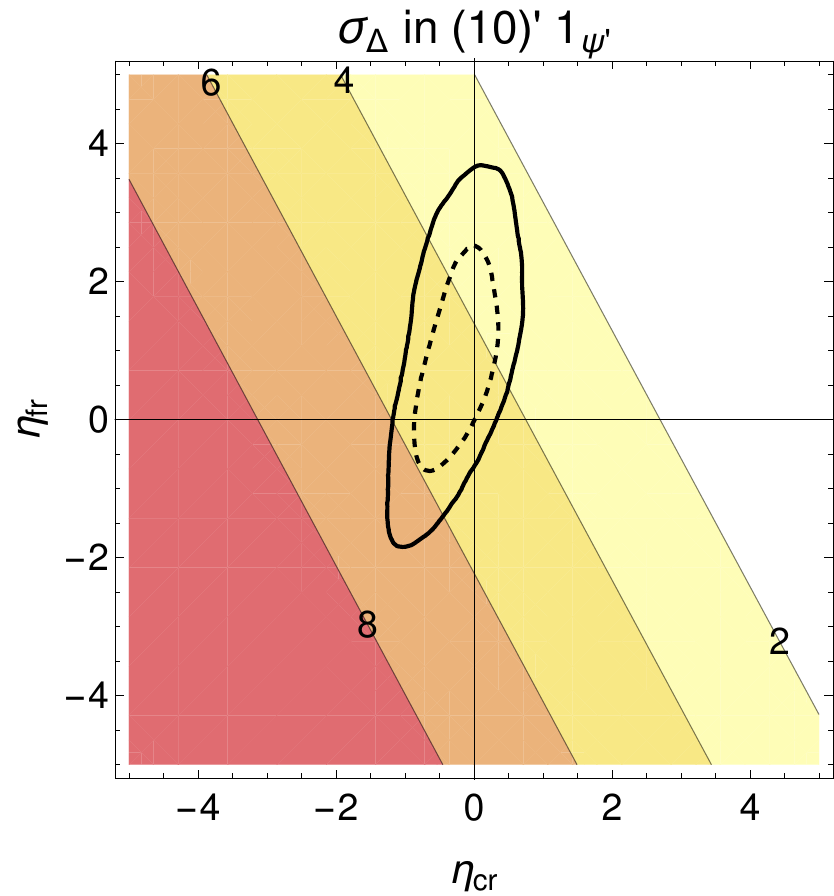}
    \includegraphics[height=0.32\linewidth]{graphics2/plot-contours-unification-legend1.pdf}
    \caption{Compatibility of threshold corrections under a minimally fine-tuned scenario with unification requirements. The plots show for each case of $G$-vacuum in Table~\ref{tab:minimal-models} a composite of two properties. First, colored regions indicate the value of $\sigma_{\Delta}(\ECR,\EFR)$, cf.~Eq.~\eqref{eq:definitions-sigmadelta}; every point in the $(\ECR,\EFR)$-plane uniquely defines the transition to the intermediate theory at $\MI$, and $\sigma_\Delta$ measures in units of $\delta\EGU$ the closest the gap between the intermediate-theory couplings in the region $(\MI,\MPL)$ come to the modeled distribution of the GUT threshold corrections characterized by $\bar{\ETA}'_{\Delta}\pm \delta\EGU$. Second, the dashed and solid black curve represent the $1$- and $2$-$\sigma$ highest probability region of where the modeled intermediate thresholds $(\ECR,\EFR)$ actually fall. Unification is generically achieved when the curves fall into the white region, while a large magnitude of $\sigma_\Delta$ indicates that the required GUT-threshold for unification is very different from where the modeled values fall. 
     \label{fig:unification-analysis}}
\end{figure}

The regions in the plane $(\ECR,\EFR)$ compatible with likely GUT thresholds, as measured by $\sigma_\Delta$, can then be compared with the modeled values of these same intermediate thresholds $(\ECR,\EFR)$. The results are shown in Figure~\ref{fig:unification-analysis}. The best compatibility is achieved when the $1$- and $2$-$\sigma$ highest probability regions for intermediate thresholds, shown as dashed and solid black lines, respectively, have a high overlap with regions of low $|\sigma_\Delta|$ colored in white. 

Having accumulated all the unification considerations in Table~\ref{tab:results-statistics} and Figure~\ref{fig:unification-analysis}, we can now study and comment on the results:
\begin{itemize}
    \item 
        The modeled size of TC values in the table shows that thresholds should not be neglected in the running, since they should be compared to the coupling values $\alpha_{i}^{-1}\approx 40$ in Eq.~\eqref{eq:matching-condition}. All TC in their original form, i.e.~the values $\ETA_{3,2,1}$ and $\ETA'_{1,2}$, are skewed to negative values due to the scalars and fermions having been modeled as lighter than gauge bosons. Once reparametrized to the physically relevant basis and computed for each point, cf.~Table~\ref{tab:matching-data-I} and \ref{tab:matching-data-GUT} for the relevant expressions, the bulk of their value is carried by the overall shifts $\ECO$ and $\ECP$, while the unification-impacting $\{\ECR,\EFR,\EGU\}$ come out as much smaller --- of order $\mathcal{O}(1)$ with a similarly-sized spread --- and can be of either sign. Another pattern seen from the table in the original basis is that the overall effect of thresholds can be distributed between those at $\MI$ and $\MGUT$, depending on how many scalars are present in the intermediate theory. Indeed, the $[6][2]$ cases for the $G$-theory have the largest scalar sector, mainly due to the necessary inclusion of $(\mathbf{84},\mathbf{2})$ as determined in Table~\ref{tab:minimal-models}, thus having the largest intermediate and the smallest GUT thresholds in terms of magnitude out of all cases.
    \item
        The $\%$ values of successfully unifying points in the table confirms our conclusions from studying benchmark scenarios in Section~\ref{sec:analysis-unification-results-benckmarkPoints}: only the three cases $3_C\,3_L\,3_R\rtimes\mathbb{Z}_{2}^{LR}$,
        $3_C\,3_L\,3_R\rtimes\mathbb{Z}_{2}^{CR}$ and $6_{CR}\,2_{L}$ lead to successful unification given the $\mathcal{O}(1)$ size of $\{\ECR,\EFR,\EGU\}$. The same result is better understood from the figure, where the modeled thresholds for the successful cases fall into the white region compatible with unification. In addition, for the case $6_{CR}\,2_{L}$ the ellipses partly overlap with the colored region, indicating unification is not always guaranteed, which is also reflected in the lower success rate of $26.2\,\%$ in the table. On the other hand, cases $2$, $4$ and $6$ have the points falling into the central colored region, while only an upper-right corner would be compatible with unification. That is the region where the right-hand panels for the benchmark scenarios in Figures~\ref{fig:RG-benchmark-trinification} and \ref{fig:RG-benchmark-nonTrinification} were located in. On a final note, we again stress that these results apply to minimally fine-tuned scenarios. 
    \item 
        We assumed the RG running between $\MZ$ and $\MI$ to be that of the SM,  and the first new states to appear around $\MI$. The inert doublet, however, is expected to be much lighter to reproduce the correct DM relic abundance~\cite{Griest:1989wd}, which we achieve by fine-tuning the mass matrix of $\psi$-odd doublets that otherwise live around $\MI$. This effect can be included in the RG running of gauge couplings by adding it to the threshold correction, in particular the irrep $(\mathbf{1},\mathbf{2},+1/2)$ of $3_C\,2_L\,1_Y$ gives the contribution
        \begin{align}
            (\Delta\ETA_{3},\Delta\ETA_{2},\Delta\ETA_{1})&=(0,\tfrac{1}{12\pi},\tfrac{1}{20\pi})\cdot\log(m/\MI)
        \end{align}
        to the SM TC (with GUT normalization of $\ETA_{1}$). For DM with mass $m=10^{3}\,\mathrm{GeV}$, and taking a large $\MI=10^{15}\,\mathrm{GeV}$, we obtain the shift $(\Delta\ETA_{3},\Delta\ETA_{2},\Delta\ETA_{1})\approx (0,-0.73,-0.44)$, which can then be further translated into $(\Delta\ECR,\Delta\EFR)$ for different $G$-vacua via Table~\ref{tab:matching-data-I}. This would manifest in Figure~\ref{fig:unification-analysis} as a shift of the $1$- and $2$-$\sigma$ ellipses in the modeled distributions of TC values. Since this shift is small (at most $0.73$ in any of the basis directions), it does not change the main conclusion on which cases successfully unify.  
\end{itemize}

\subsection{Remarks on perturbativity \label{sec:analysis-unification-perturbativity}}

Given the large number of degrees of freedom in our model, perturbativity is an important aspect to consider. The issue can arise in many different aspects, so we organize our discussion below accordingly:

\begin{itemize}
    \item
        When commenting on Figures~\ref{fig:RG-benchmark-trinification} and \ref{fig:RG-benchmark-nonTrinification} in Section~\ref{sec:analysis-unification-results-benckmarkPoints}, we observed a distinct curve to the RG running of the unified gauge coupling due to a strong two-loop effect.
        \par
        Indeed, assuming a coupling value $\alpha^{-1}_U=40$ at the GUT scale, and taking the values $a$ and $b$ from Eq.~\eqref{eq:ABcoefficients-E6} and inserting into \eqref{eq:RG-running}, the two-loop term slightly dominates over the one-loop term, raising perturbativity concerns.
        This can be addressed by computing the next-order coefficient in the perturbative expansion
            \begin{align}
                \tfrac{d}{dt}\alpha^{-1}&=-\tfrac{1}{2\pi}\Big(a+b\;(\tfrac{\alpha}{4\pi})+c\;(\tfrac{\alpha}{4\pi})^2+\ldots \Big) \label{eq:three-loop-expansion}
            \end{align}
        for the gauge coupling RGE, where only the gauge contributions are considered. Using three-loop expressions from~\cite{Pickering:2001aq}, we get $c=560730$, while $a$ and $b$ are given in Eq.~\eqref{eq:ABcoefficients-E6}, as stated earlier. The three-loop term in Eq.~\eqref{eq:three-loop-expansion} turns out to be almost an order of magnitude smaller than the first two, so the order-by-order contributions are not necessarily doomed to grow uncontrollably. 
        \par 
        In fact, we suspect the hierarchy is disrupted by the one-loop term being spuriously small. Its value $a=16$ in the model is for example somewhat small in magnitude compared to $a=-38$ when there are no scalars, see Eq.~\eqref{eq:Acoefficient-general}. Indeed, if one labels the RGE coefficients for a pure Yang-Mills $\mathrm{E}_{6}$ with no fermion or scalar content by $(a',b',c')$, explicit computation gives the ratios $(\tfrac{a}{a'},\tfrac{b}{b'},\tfrac{c}{c'})\approx (-0.36,-7.3,-6.1)$.
        Having a large fermion and scalar content, these ratios are expected much larger than $\mathcal{O}(1)$ in magnitude, so the smallness of the first ratio indicates that the gauge effect has not yet been fully overwhelmed in the one-loop coefficient. Although not a proof, this does give some confidence that the model is not intrinsically non-perturbative at the unification scale.
    \item
        Another perturbativity consideration is how quickly in the RG running the unified coupling reaches non-perturbative values. Around that scale $\bar{t}_{\text{pert}}$, effects of non-perturbative dynamics or new physics (such as quantum gravity) would become important, presumably saving the theory from destabilizing, but also 
        generating non-renormalizable operators suppressed by that scale. If the non-perturbative regime is reached too quickly, even the qualitative effects of the non-renormalizable operators cannot be neglected 
        at the GUT scale, and in that sense we may again lose perturbativity.
        \par
        Given the three-loop expansion of the running in Eq.~\eqref{eq:three-loop-expansion}, we show in Figure~\ref{fig:plot-perturbativity} how high above the GUT scale in powers of $10$ we lose perturbativity depending on the inverse coupling value $\alpha_{U}^{-1}$ at $\TGUT$. We see that a naive one-loop estimate would suggest perturbativity for $4$ or more orders of magnitude, all the way to the Planck scale, while the important two-loop contribution reduces that significantly to between $1$ and $2$ orders of magnitude for a starting value $\alpha_U^{-1}(\TGUT)\gtrsim 30$, which \textit{a posteriori} justifies neglecting non-renormalizable operators in our analysis. Crucially, this result also seems to stabilize at $3$-loop order. The horizontal dashed line in the plot indicates perturbativity for one order of magnitude above $\TGUT$, drawn primarily for psychological reasons rather than any special physical significance. 
    
    \begin{figure}[htb]
        \centering
        \includegraphics[width=0.55\linewidth]{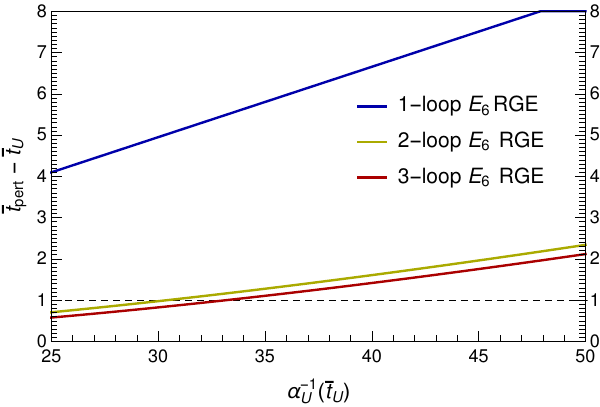}
            \caption{The scale of non-perturbativity (measured in orders of magnitude above the GUT scale) as a function of the unified coupling value $\alpha_{U}^{-1}$ at the GUT scale $\TGUT$. The scale $\bar{t}_{\text{pert}}$ is implicitly defined by $\alpha^{-1}_{U}(\bar{t}_{\text{pert}})=1$. The curves in the plot correspond to one-, two- and three-loop RG running.
        \label{fig:plot-perturbativity}}
    \end{figure}

    \item
        As recently pointed out in~\cite{Milagre:2024wcg} for $\mathrm{E}_{6}$ GUT models, another challenge to the model's perturbativity might also come from perturbative unitarity, see~\cite{Lee:1977yc,Lee:1977eg,Hally:2012pu} for famous examples and e.g.~\cite{Logan:2022uus} for a pedagogical overview. Following~\cite{Milagre:2024wcg}, the largest-magnitude eigenvalue for the partial-wave coefficient $a_{0}$ would in our model be
        \begin{align}
            (a_{0})^\text{max}= \frac{\alpha_U}{2\sqrt{78}}\sqrt{
            \frac{3\pi^2}{4}\,27\cdot C(\mathbf{27})^2
            + \sum_{\mathbf{R}} \ZETA(\mathbf{R})\,\dim(\mathbf{R})\,C(\mathbf{R})^{2}
            },
            \label{eq:perturbative-unitarity}
        \end{align}
        where $C(\mathbf{R})$ is the Casimir coefficient for the irrep $\mathbf{R}$, $\ZETA(\mathbf{R})$ is the reality/complexity coefficient defined below Eq.~\eqref{eq:RG-factor-b}, and the sum is taken over scalar irreps $\mathbf{R}\in\{\mathbf{27},\mathbf{351'},\mathbf{650}\}$. Taking the smallest coupling value $\alpha_{U}=40.5^{-1}$ attained in our exploration of unification from Section~\ref{sec:analysis-unification-results-thresholdModeling}, we obtain $(a_{0})^{\text{max}}=0.69$, which violates the unitarity bound given as $|\Re a_{0}|\le 0.5$.
        The expression in Eq.~\eqref{eq:perturbative-unitarity}, however, is based on tree level scattering in the limit of vanishing masses of particles. The massless limit is a good approximation at scales $\mu\gg \MGUT$, but we already expect the theory to be non-perturbative at $\mu\gsim 10\MGUT$,
        as alluded to in the previous point. The two issues may in fact be connected. At $\MGUT$, where the calculation aspires to be perturbative, the masses of the scattered particles can no longer be neglected and may rather easily shift the value
        $(a_{0})^{\text{max}}$ to below $0.5$. It is thus hard to say what a more accurate estimate of $(a_{0})^{\text{max}}$ at $\MGUT$ might be and how big of a shift in this quantity needs to be induced by loop corrections.
\end{itemize}

\noindent
Based on the the points above, the conclusion is that there are perturbativity challenges to our $\mathrm{E}_{6}$ model, but also some indications of successful resolution, at least in some regions of parameter space. A more elaborate perturbativity analysis beyond the scope of this paper would be required to fully assess this issue. 

Finally, we remark that the perturbativity situation is worse for any non-minimal $\mathrm{E}_{6}$ GUT alternative attempting to break through novel intermediate symmetries. Hence we consider our proposed model as the main potentially viable candidate in this class.

\section{Proton decay \label{sec:analysis-proton}}

\subsection{General considerations \label{sec:analysis-proton-general}}

As usual in GUTs, an important phenomenological prediction in our $\mathrm{E}_{6}$ model is proton decay (or more generally nucleon decay). We now study how proton decay manifests for each possible case of $G$-theory, cf.~Table~\ref{tab:minimal-models}. We start by considering the most general case of just $G$-vacua, and then progressively apply further assumptions, such as spinorial parity and the specifics of unification under the minimal scenarios as was determined in Section~\ref{sec:analysis-unification}.  

Our model is non-supersymmetric, so the leading contributions to proton decay are those from five $D=6$ operators in the SM language~\cite{Weinberg:1979sa,Wilczek:1979hc,Nath:2006ut}. These break baryon- ($B$) and lepton number ($L$), but preserve $B-L$. We focus on gauge-mediated proton decay as opposed to scalar-mediated, since the gauge coupling is much larger than the Yukawa coupling of the first family, and hence the former is expected to dominate. 

As is well known, only two out of five $D=6$ four-fermion operators in the SM effective theory are generated by gauge boson mediation~\cite{Wilczek:1979hc,Nath:2006ut}, where the gauge bosons must either transform as $(\mathbf{3},\mathbf{2},-5/6)$ or $(\mathbf{3},\mathbf{2},+1/6)$ under $3_C\,2_L\,1_Y$. There are three such irreps of gauge bosons in the adjoint $\mathbf{78}$ of $\mathrm{E}_{6}$. We list them and specify their subgroup origin in Table~\ref{tab:proton-gauge-bosons-definition}. The usual $\SU(5)$ leptoquark is denoted by $X$, the $\SO(10)$ leptoquark by $X'$, and the new leptoquark introduced in $\mathrm{E}_{6}$ by $X''$. We also see from the table that $X''$ has odd $\psi$-charge, and thus necessarily connects spinorial and non-spinorial fermions in the gauge current. If spinorial parity $\mathbb{Z}_{2}^{\psi}$ is preserved, there is no mixing between the fermion
 $\mathbf{16}$ and $\mathbf{10}$, and hence $X''$ does not generate four-fermion operators with only EW-scale outer states, making it irrelevant for proton decay considerations. 

\begin{table}[htb]
    \centering
    \caption{The inclusion of $B+L$ violating gauge bosons in $\mathrm{E}_{6}$ into irreps of the standard GUT-subgroup chain $3_C\,2_L\,1_Y\subset\SU(5)\subset\SO(10)\subset\mathrm{E}_6$, as well as their $\psi$-charges. \label{tab:proton-gauge-bosons-definition}}
    \begin{tabular}{lllllrl}
        \toprule
            label & $3_C\,2_L\,1_Y$ & $\SU(5)$ & $\SO(10)$ & $\mathrm{E}_{6}$ & $\psi$ & comment \\
        \midrule
            $X$&
                $\sim(\mathbf{3},\mathbf{2},-5/6)$&
                $\mathbf{24}$&$\mathbf{45}$&$\mathbf{78}$&$0$& 
                the $\SU(5)$ leptoquark\\
            $X'$&
                $\sim(\mathbf{3},\mathbf{2},+1/6)$&
                $\mathbf{10}$&$\mathbf{45}$&$\mathbf{78}$&$0$& 
                the $\SO(10)$ leptoquark\\
            $X''$&
                $\sim(\mathbf{3},\mathbf{2},+1/6)$&
                $\mathbf{10}$&$\mathbf{16}$&$\mathbf{78}$&$-3$& 
                the $\mathrm{E}_{6}$ leptoquark\\
        \bottomrule
    \end{tabular}
\end{table}

We identify the origin of the states $X$, $X'$ and $X''$ in terms of $G$-irreps in Table~\ref{tab:proton-gauge-bosons-Ginclusions} and specify whether they appear at the intermediate or GUT scale ($\MI$ or $\MGUT$). We assume the preservation of $\mathbb{Z}_{2}^{\psi}$ parity in determining the minimal cases of Table~\ref{tab:minimal-models}, so the $\mathrm{E}_{6}$ leptoquark $X''$ can be ignored. 

\begin{table}[htb]
    \centering
    \caption{The inclusion of $B+L$ gauge bosons into irreps of the relevant $G$-embeddings. We also list explicitly which of these gauge bosons are present at the intermediate scale $\MI$ or GUT scale $\MGUT$. \label{tab:proton-gauge-bosons-Ginclusions}}
    \begin{tabular}{llllll}
         \toprule
            $G$-embedding& $X$ irrep &$X'$ irrep & $X''$ irrep & at $\MI$ & at $\MGUT$ \\
         \midrule
            $3_C\,3_L\,3_R$&
                $(\mathbf{3},\mathbf{\bar{3}},\mathbf{\bar{3}})$ &
                $(\mathbf{3},\mathbf{\bar{3}},\mathbf{\bar{3}})$&
                $(\mathbf{3},\mathbf{\bar{3}},\mathbf{\bar{3}})$&
                ---&$X,X',X''$\\
            $6_{CL}\,2_{R}$&
                $(\mathbf{20},\mathbf{2})$&
                $(\mathbf{20},\mathbf{2})$&
                $(\mathbf{35},\mathbf{1})$&
                $X''$&$X,X'$\\
            $6_{CR}\,2_L$&
                $(\mathbf{20},\mathbf{2})$&
                $(\mathbf{20},\mathbf{2})$&
                $(\mathbf{20},\mathbf{2})$&
                ---&$X,X',X''$\\
            $(10)'\,1_{\psi}'$&
                $(\mathbf{16},-3)$&
                $(\mathbf{45},0)$&
                $(\mathbf{45},0)$&
                $X',X''$&$X$\\
         \bottomrule
    \end{tabular}
\end{table}

Preserved spinorial parity makes proton decay in $\mathrm{E}_{6}$ proceed through exactly the same diagrams as in the standard $\mathrm{SO}(10)$ via $X$ and $X'$ leptoquarks, and imposes the fermion fit to be $\SO(10)$-like as well. For this type of $\SO(10)$ GUT, the dominant proton decay channels from gauge mediation are known to be $p^{+}\to\pi^{0}e^{+}$ and $p^{+}\to\pi^{+}\bar{\nu}$ (sum over all neutrino flavors), and the decay rate expressions for these two channels can be approximated by taking factors $1$ for the flavor structure~\cite{Bajc:2008dc,Babu:2015bna}.
This yields the expressions~\cite{Nath:2006ut,Babu:2015bna}
\begin{align}
    \Gamma(p^{+}\to\pi^{0}e^{+})&= \frac{\pi}{4} \frac{m_{p}\alpha_{U}^{2}}{f_{\pi}^{2}}\; |\alpha_{H}|^2 R_{\mathrm{L}}^{2} (1+F+D)^2 \left(
    A_{SR}^{2}\left(\frac{1}{M_{X}^2}+\frac{1}{M_{X'}^{2}}\right)^2+ \frac{4A_{SL}^{2}}{M_{X}^4}
    \right), \label{eq:proton-decay-1}\\
    \Gamma(p^{+}\to\pi^{+}\bar{\nu}) &=\frac{\pi}{2} \frac{m_{p}\alpha_{U}^{2}}{f_{\pi}^{2}}\; |\alpha_{H}|^2 R_{\mathrm{L}}^{2} (1+F+D)^2 \left(
    A_{SR}^{2}\left(\frac{1}{M_{X}^2}+\frac{1}{M_{X'}^{2}}\right)^2 \right). \label{eq:proton-decay-2}
\end{align}
Above $f_{\pi}=130.2\,\mathrm{MeV}$ is the pion decay constant (PDG2022, 72.14), $m_p=938.27\,\mathrm{GeV}$ (PDG2022) is the mass of the proton, $D=0.804$ and $F=0.463$ are parameters in the chiral Lagrangian~\cite{Claudson:1981gh,Cabibbo:2003cu}, $\alpha_H=-0.01257\,\mathrm{GeV}^3$ is the hadronic matrix element~\cite{Yoo:2021gql},  $R_L=1.35$ is the long-distance factor of renormalizing the four-fermion operators for our strong coupling value~\cite{Nihei:1994tx}, $A_{SL}$ and $A_{SR}$ are the short distance factors, $M_{X}$ and $M_{X'}$ are the masses of the two gauge bosons, while $\alpha_{U}$ is the value of the unified coupling at the GUT scale.
We specify the procedure for determining $A_{SL}$ and $A_{SR}$ in Appendix~\ref{app:proton-decay-gamma-factors}, but one can take as a rough approximation
\begin{align}
    A_{SL}\approx A_{SR}&\approx 2. \label{eq:short-range-approximation}
\end{align}
The current experimental bounds and future experimental reach for the two decay modes $p^{+}\to\pi^{0}e^{+}$ and $p^{+}\to\pi^{+}\bar{\nu}$ are summarized in Table~\ref{tab:proton-decay-experimental}.

\begin{table}[htb]
    \centering
    \caption{Experimental bounds for the lifetime of the proton $\tau_{p^{+}}/\mathcal{B}$ for different processes. The Super-K limits come from~\cite{Super-Kamiokande:2020wjk,Super-Kamiokande:2013rwg}, while the Hyper-K estimate is based on~\cite{Hyper-Kamiokande:2018ofw} assuming $20$ years of data with a $185\,\mathrm{kt}$ fiducial volume. \label{tab:proton-decay-experimental}}
    \begin{tabular}{lll}
         \toprule
         process& Super-K & Hyper-K \\
         \midrule
         $p^{+}\to \pi^{0}e^{+}$& $2.4\cdot 10^{34}\,\mathrm{y}$ & $1.4\cdot 10^{35}\,\mathrm{y}$\\
         $p^{+}\to\pi^{+}\bar{\nu}$&$3.9\cdot 10^{32}\,\mathrm{y}$& ---\\
         \bottomrule
    \end{tabular}
\end{table}

Further estimation of proton decay based on Eqs.~\eqref{eq:proton-decay-1}--\eqref{eq:proton-decay-2} proceeds along one of two paths, depending on $G$:
\begin{itemize}
    \item \underline{All $G$-vacua except $(10)'\,1_{\psi'}$:} 
        We see from Table~\ref{tab:proton-gauge-bosons-Ginclusions} that both $X$ and $X'$ are in the same coset irrep outside of the $G$-adjoint. In fact, they are located in the representation $\XI$ used in Table~\ref{tab:matching-data-GUT} that defines the GUT scale in our convention, i.e.~$M_{X}=M_{X'}=\MGUT$.\footnote{This is unlike $\SO(10)$ GUT, where the ratio $M_{X}/M_{X'}$ can take a continuous value (for a rich enough scalar sector), since the two gauge bosons are in different $\SO(10)$-irreps.} The decay rates can then be written as
            \begin{align}
                \Gamma(p^{+}\to\pi^{0}e^{+})&= \pi\;\frac{m_{p}\alpha_{U}^{2}}{f_{\pi}^{2}}\; |\alpha_{H}|^2 R_{\mathrm{L}}^{2} (1+F+D)^2 \;
                    \frac{A_{SR}^{2}+A_{SL}^{2}}{\MGUT^4}, \label{eq:proton-decay-1b}\\
                \Gamma(p^{+}\to\pi^{+}\bar{\nu})&=\pi\; \frac{m_{p}\alpha_{U}^{2}}{f_{\pi}^{2}}\; |\alpha_{H}|^2 R_{\mathrm{L}}^{2} (1+F+D)^2 \;
                    \frac{2A_{SR}^{2}}{\MGUT^4}. \label{eq:proton-decay-2b}
            \end{align}
        Inserting the numerical values, including the approximation of Eq.~\eqref{eq:short-range-approximation}, we obtain for the proton lifetime $\tau_{p^{+}}$ (based on one decay channel) in years
            \begin{align}
                \tau_{p^{+}}/\mathcal{Br}(p^{+}\to\pi^{0}e^{+})&\approx 2.57\cdot 10^{34}\,\mathrm{y}\;\left(\frac{\alpha_{U}^{-1}}{40}\right)^2\;\left(\frac{\MGUT}{10^{15.8}\,\mathrm{GeV}}\right)^4,
                \label{eq:proton-decay-numeric-approximation-1}
            \end{align}
        where the branching ratios of the two main decay modes are roughly equal:
            \begin{align}
               \mathcal{B}(p^{+}\to\pi^{0}e^{+}) &\approx \mathcal{B}(p^{+}\to\pi^{+}\bar{\nu}). \label{eq:proton-branching-relation}
            \end{align}
        The numeric value for the lifetime in Eq.~\eqref{eq:proton-decay-numeric-approximation-1} is close to the experimental bound in Table~\ref{tab:proton-decay-experimental}, so we consider \hbox{$\MGUT=10^{15.8}\,\mathrm{GeV}$} a rough reference value for the lower bound on unification (we referred to it as such earlier in Section~\ref{sec:analysis-unification}).
        \par
        We can relate the decay rate estimates to those of $\SU(5)$, where one applies the expressions Eqs.~\eqref{eq:proton-decay-1}--\eqref{eq:proton-decay-2} with $M_{X'}\to\infty$. Approximating the short-distance factors again via Eq.~\eqref{eq:short-range-approximation}, assuming the same flavor structure from the fermion fit in $\SU(5)$ and $\mathrm{E}_{6}$, and the same numeric values $\alpha_{U}$ and $\MGUT$ in both theories, the $\mathrm{E}_{6}$ enhancements in the decay rates are
            \begin{align}
            \Gamma_{\mathrm{E}_{6}}(p^{+}\to\pi^{0}e^{+}) & \approx  \tfrac{8}{5}\;\Gamma_{\SU(5)}(p^{+}\to\pi^{0}e^{+}),\\
            \Gamma_{\mathrm{E}_{6}}(p^{+}\to\pi^{+}\bar{\nu}) & \approx 4\;\Gamma_{\SU(5)}(p^{+}\to\pi^{+}\bar{\nu}).
            \end{align} 
            
    \item \underline{$G=(10)'\,1_{\psi'}$:} 
        From Table~\ref{tab:proton-gauge-bosons-Ginclusions}, together with the choice of irrep $\XI$ defining the matching scales, cf.~Tables~\ref{tab:matching-data-I} and \ref{tab:matching-data-GUT}, we obtain $M_{X'}=\MI$ and $M_{X}=\MGUT$. With the two scales hierarchical, proton decay is dominated by the $X'$ contribution, enforcing strong lower bounds on the scale $\MI$ instead of $\MGUT$. The decay rates with $M_{X'}\ll M_{X}$ become
            \begin{align}
                \Gamma(p^{+}\to\pi^{0}e^{+})&= \frac{\pi}{4}\;\frac{m_{p}\alpha_{10'}^{2}}{f_{\pi}^{2}}\; |\alpha_{H}|^2 R_{\mathrm{L}}^{2} (1+F+D)^2 \;
                    \frac{A_{SR}^{2}}{\MI^4}, \label{eq:proton-decay-1c}\\
                \Gamma(p^{+}\to\pi^{+}\bar{\nu})&=\frac{\pi}{2}\; \frac{m_{p}\alpha_{10'}^{2}}{f_{\pi}^{2}}\; |\alpha_{H}|^2 R_{\mathrm{L}}^{2} (1+F+D)^2 \;
                    \frac{A_{SR}^{2}}{\MI^4}. \label{eq:proton-decay-2c}
            \end{align}
        where $\alpha_{10'}$ is the gauge coupling of $\SO(10)'$ at scale $\MI$. The branching ratio relation now becomes 
            \begin{align}
                \mathcal{B}(p^{+}\to\pi^{0}e^{+}) & = \tfrac{1}{2}\; \mathcal{B}(p^{+}\to\pi^{+}\bar{\nu}),
            \end{align}
        and the numerical value for proton lifetime is
            \begin{align}
                \tau_{p^{+}}/\mathcal{Br}(p^{+}\to\pi^{0}e^{+})&\approx 6.86\cdot 10^{37}\,\mathrm{y}\;\left(\frac{\alpha_{10'}^{-1}}{46.1}\right)^2\;\left(\frac{\MI}{10^{16.4}\,\mathrm{GeV}}\right)^4
            \end{align}
        where the benchmark values for $\alpha_{10'}^{-1}$ and $\MI$ were chosen from the scenario with no intermediate thresholds, i.e.~point $P_{23}$ in the left-hand panel of Figure~\ref{fig:RG-bottom-up-SM}. Despite proton decay dominantly depending on $\MI$, the lifetime is large due to $\MI$ being large in this scenario. Also, we see from the right-hand panel of the same figure that $\mathcal{O}(1)$ values for $\ECR$ keep the scale $\MI$ sufficiently high, although unification issues in $(10)'\,1_{\psi}$ from Section~\ref{sec:analysis-unification-results-thresholdModeling} might apply.
            \par
        Finally, the comparison with $\SU(5)$ GUT with the approximation of Eq.~\eqref{eq:short-range-approximation} yields 
            \begin{align}
                \Gamma_{\mathrm{E}_{6}}(p^{+}\to\pi^{0}e^{+}) & \approx  \tfrac{1}{5}\;\Gamma_{\SU(5)}(p^{+}\to\pi^{0}e^{+})
                    \;\left(\alpha^{\mathrm{E}_{6}}_{10'}/\alpha_U^{\SU(5)}\right)^{2}
                    \;\left(\MGUT^{\SU5)}/\MI^{\mathrm{E}_{6}}\right)^{4},\\
                \Gamma_{\mathrm{E}_{6}}(p^{+}\to\pi^{+}\bar{\nu}) & \approx \Gamma_{\SU(5)}(p^{+}\to\pi^{+}\bar{\nu})
                    \;\left(\alpha^{\mathrm{E}_{6}}_{10'}/\alpha_U^{\SU(5)}\right)^{2}
                    \;\left(\MGUT^{\SU5)}/\MI^{\mathrm{E}_{6}}\right)^{4},
            \end{align}
        obtaining a suppression factor $1/5$ in the $\pi^{0}e^{+}$ channel. These approximations assume the flavor structure from the fermion fit to be the same in $\SU(5)$ and $\mathrm{E}_{6}$.
\end{itemize}

\subsection{Results for minimally tuned scenarios \label{sec:analysis-proton-numeric}}

We saw in  the unification analysis of Section~\ref{sec:analysis-unification} that only three out of six cases 
in Table~\ref{tab:minimal-models} can successfully unify under the assumption of minimal fine-tuning: the trinifications $3_C\,3_L\,3_R$ cases with $LR$ and $CR$ parity, as well as $6_{CR}\,2_L$. Under the modeling of thresholds corrections of Section~\ref{sec:analysis-unification-results-thresholdModeling},
these cases provided a set of points with successful unification, which can be analyzed with respect to their associated intermediate and GUT scales, as well as proton decay.

For each of the three viable cases, we take the subset of $10^{5}$ simulated points for which unification is successful, and in addition the GUT scale $\MGUT$ is more than one order of magnitude below the (reduced) Planck scale $\MPL$ and the scale of perturbativity associated with non-renormalizable operators, as defined in Figure~\ref{fig:plot-perturbativity}. The reason for taking this \textit{perturbative subset} is to consider only points for which non-perturbative or gravity contributions are potentially under control, albeit the cut-off of one order of magnitude is chosen arbitrarily. The percentage of points from the original $10^5$ that achieve this is given in Table~\ref{tab:proton-A-coefficients}, which can be compared with the less demanding criterion in Table~\ref{tab:results-statistics}.

\begin{table}[htb]
\centering
\caption{The mean and standard deviation for the distribution of the values of short-distance factors $A_{SR}$ and $A_{SL}$, computed according to Appendix~\ref{app:proton-decay-gamma-factors}, for the three $G$-theories compatible with unification (assuming minimal fine-tuning). The last row specifies the percentage of points in the \text{perturbative subset}, see main text. \label{tab:proton-A-coefficients}}
\begin{tabular}{lllr}
    \toprule
    $G$-vacuum&$A_{SR}$&$A_{SL}$&$\%$ subset\\
    \midrule
    $3_C\,3_L\,3_R\rtimes \mathbb{Z}_{2}^{LR}$&$2.27\pm 0.06$ & $2.39\pm 0.06$& $>99.9$\\
    $3_C\,3_L\,3_R\rtimes \mathbb{Z}_{2}^{CR}$&$2.31\pm 0.06$ & $2.44\pm 0.07$& $99.1$\\
    $6_{CR}\,2_{L}$&$2.38\pm 0.12$ & $2.52\pm 0.12$& $15.7$\\
    \bottomrule
\end{tabular}
\end{table}

\begin{figure}[htb]
    \centering
    \includegraphics[width=0.49\linewidth]{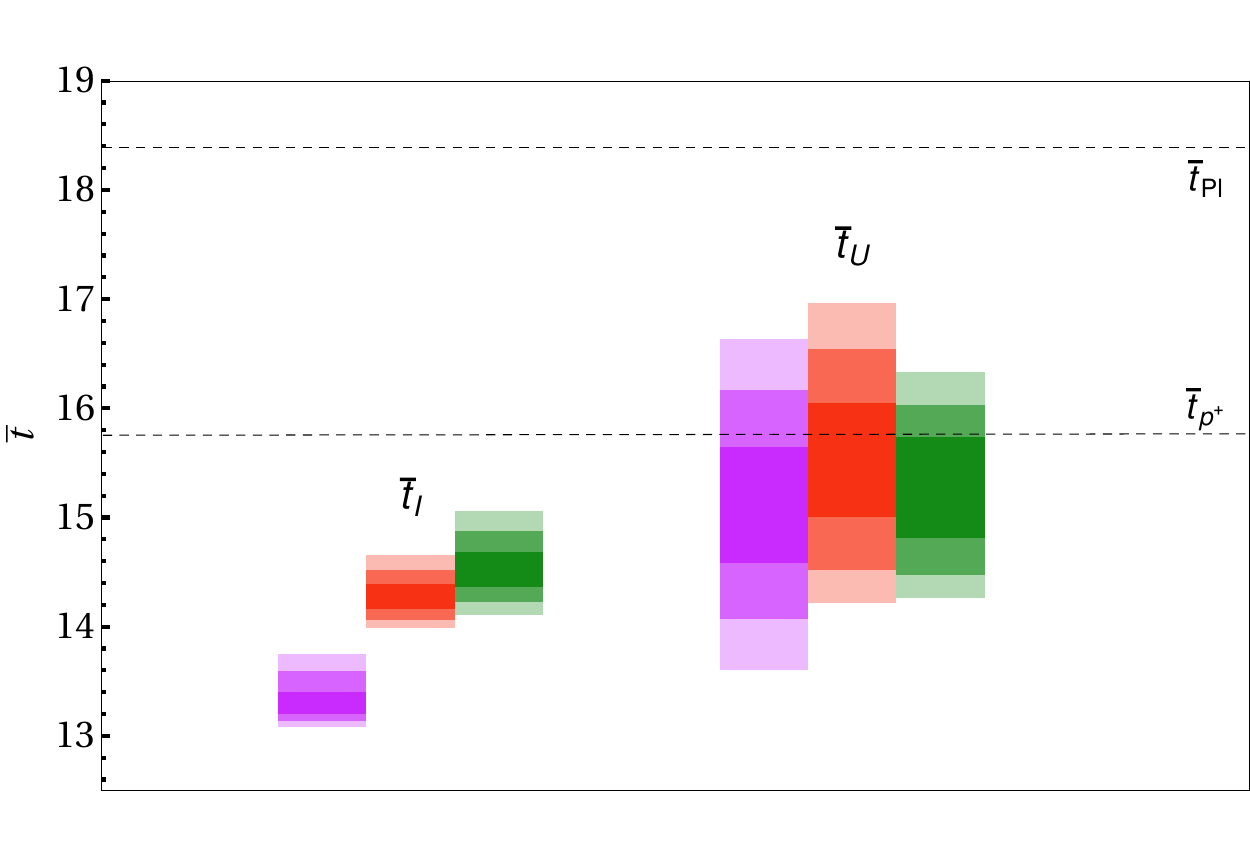}
    \includegraphics[width=0.49\linewidth]{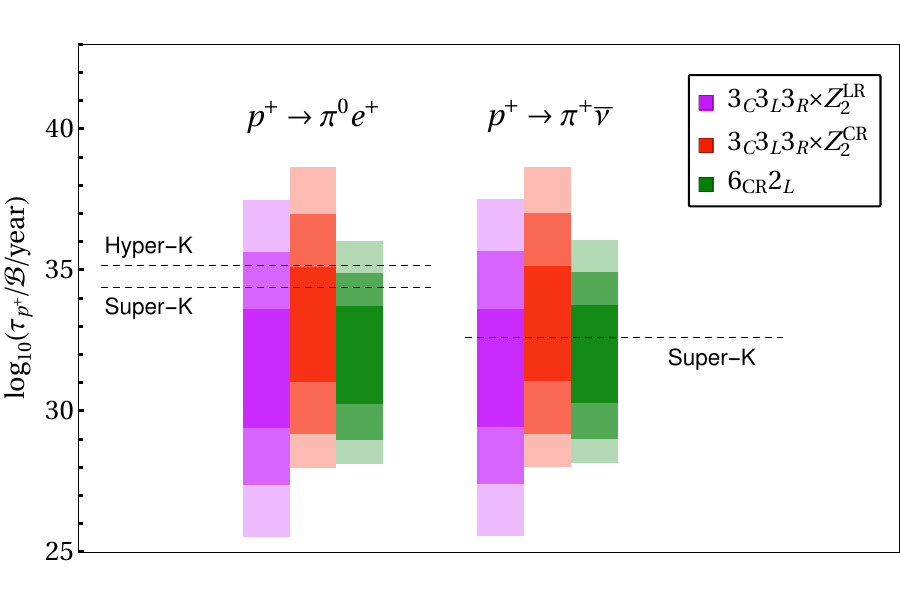}
    \caption{Left panel: the $1$-, $2$- and $3$-$\sigma$ highest probability density intervals for the intermediate and GUT scales $\TI$ and $\TGUT$ in the \text{perturbative subset}, cf.~main text, when modeling threshold corrections according to Section~\ref{sec:analysis-unification-results-thresholdModeling}, i.e.~under the minimal fine-tuning assumption. Dashed lines indicate the benchmark scale $\bar{t}_{p^{+}}$ for proton decay from Eq.~\eqref{eq:proton-decay-numeric-approximation-1} and the reduced Planck scale $\TPL$.
    Right panel: the $1$-, $2$- and $3$-$\sigma$ HPD intervals of proton lifetime estimates under the two dominant channels $p^{+}\to \pi^{0}e^{+}$ and $p^{+}\to\pi^{+}\bar{\nu}$ for the same perturbative subset as the left panel. The dashed lines correspond to the present and future bounds, cf.~Table~\ref{tab:proton-decay-experimental}. The color code is the same for both panels.
    \label{fig:hpd-scales-and-proton}}
\end{figure}

The highest probability density (HPD) intervals for the ($\log_{10}$)scales $\TI$ and $\TGUT$ under the modeled threshold corrections are shown in the left-hand panel of Figure~\ref{fig:hpd-scales-and-proton}, where $1$-, $2$- and $3$-$\sigma$ regions are shown with decreasing opacity. In the right-hand panel, the HPD intervals show the predictions for proton decay lifetime (with statistics performed on $\log_{10}(\tau_{p^{+}}/\mathcal{B}/\mathrm{y})$ values) for the two main decay channels. These values were computed using Eqs.~\eqref{eq:proton-decay-1b} and \eqref{eq:proton-decay-2b}; the short-range factors $A_{SR}$ and $A_{SL}$ were computed for each point individually using the procedure described in Appendix~\ref{app:proton-decay-gamma-factors}, with the mean and standard deviation of their distribution given in Table~\ref{tab:proton-A-coefficients} for each case of $G$-theory.

Regarding the results from Figure~\ref{fig:hpd-scales-and-proton}, we reiterate the assumptions made for the proton decay estimates:
\begin{itemize}
    \item The intermediate theory is one of the cases $1$, $3$ and $5$ from Table~\ref{tab:minimal-models}, which were determined in the $\text{ESH}+\mathbb{Z}_{2}^{\psi}$ scenario (extended survival hypothesis with spinorial parity).
    \item Minimal fine-tuning of states was assumed in the modeling of threshold corrections (scalar states scattered across one order of magnitude below their associated matching scale).
    \item Fermion fits from the analogous $\SO(10)$ setup 
    shows that the main decay modes can be approximated by taking a factor $1$ for the flavor structure.
\end{itemize}
The results thus represent merely rough estimates, but they suffice for drawing two important conclusions: 
\begin{enumerate}
    \item 
        All of the three viable unification cases $3_C\,3_L\,3_R\rtimes \mathbb{Z}_{2}^{LR}$, $3_C\,3_L\,3_R\rtimes \mathbb{Z}_{2}^{CR}$ and $6_{CR}\,2_L$ produce points compatible with present day bounds in their $2$-$\sigma$ HDP intervals, including in the future Hyper-K region. The trinification case with $CR$ parity (red bars) has slightly longer proton lifetimes; less of its points are currently excluded, but also making future proton decay detection in this scenario harder.   
    \item 
        The predicted ranges for the proton lifetime span many orders of magnitude, and hence are likely to disappoint the experimentalist. They are in our case, however, a straightforward consequence of the mild uncertainties in the threshold correction values. Given the large number of fields in the model, the conclusion is that unless the particle spectrum is well predicted,\footnote{Constraints might arise, for example, from a perturbativity analysis as performed in~\cite{Jarkovska:2023zwv}.} the absolute proton decay prediction cannot be under control, while the relative rates in our scenario are more predictive than in a typical $\SO(10)$ GUT, cf.~Eq.~\eqref{eq:proton-branching-relation}.
\end{enumerate}

\section{Conclusions and outlook \label{sec:conclusions}}

The broader idea studied in this paper is how a realistic non-supersymmetric $\mathrm{E}_{6}$ GUT can break to the Standard Model through a novel intermediate symmetry (that does not completely unify by itself). The novel symmetries considered are trinification $SU(3)_C\times\SU(3)_L\times\SU(3)_R$, the standard $\SU(6)_{CL}\times\SU(2)_{R}$, the LR-flipped $\SU(6)_{CR}\times\SU(2)_{L}$, and the flipped $\SO(10)'\times\UU_{\psi'}$, cf.~Table~\ref{tab:embeddings} for definitions. 

The minimal realistic theory for all these cases, assuming a renormalizable Yukawa sector with three fermion families of $\mathbf{27}_F$, turns out to have the scalar-sector content $\mathbf{650}\oplus \mathbf{351'}\oplus\mathbf{27}$, where the first irrep takes care of the first-stage breaking into the novel intermediate symmetry, and the latter two irreps take care of the subsequent breaking into the SM with a Higgs doublet and a realistic Yukawa sector. Which intermediate symmetry manifests depends on the parameters in the part of the scalar potential involving only the irrep $\mathbf{650}$ --- the $V_{650}$ --- as investigated in~\cite{Babu:2023zsm}. 

We investigated the full model in detail for all intermediate cases of interest under the assumption of the extended survival hypothesis and a choice of vacuum which preserves spinorial parity. The latter disentangles the fermion mixing between the $\SO(10)$-irreps $\mathbf{16}$ and $\mathbf{10}$ in the $\mathbf{27}$, and also provides a DM candidate, which we envision to be an inert doublet scalar. Interestingly, the case of $\mathrm{E}_{6}$ with preserved $\mathbb{Z}_{2}^{\psi}$ leads to Eqs.~\eqref{eq:fermion-mass-light-U}--\eqref{eq:fermion-mass-light-N}
and offers $6$ EW VEVs compared to $4$ in $\SO(10)$ with $\mathbf{10}_{\mathbb{C}}\oplus\mathbf{126}$, implying a fit with less tension.

Under the assumption of minimal fine-tuning, we showed that gauge coupling unification can be successful in only three cases of novel symmetries: trinification $SU(3)_C\times\SU(3)_L\times\SU(3)_R$ that preserves either $LR$ or $CR$ parity, and the LR-flipped $\SU(6)_{CR}\times\SU(2)_L$. The large number of fields with unknown masses in the spectrum imply a considerable uncertainty in the unification scale, so the predicted range for proton lifetime spans many orders of magnitude. Nevertheless, all three successful cases of unification show regions compatible with current experimental bounds from Super-K, including an overlap with the reach of Hyper-K. 

This paper ultimately presents considerable clarification and limitations on obtaining one of the novel intermediate symmetries from $\mathrm{E}_{6}$ GUT.
We see the possibility to extend the results of this paper in two ways:
\begin{enumerate}
    \item The choice of vacuum with spinorial parity is appealing for providing a DM candidate, but it also greatly simplifies the analysis. Withdrawing this choice and allowing for mixing between the chiral and vector-like fermions in $\mathbf{27}_F$ broadens the predictions. An investigation into fermion fits, together with proton decay mediated by the $\mathrm{E}_{6}$ leptoquark, would however make the analysis intrinsically $\mathrm{E}_{6}$-like, and unlike any in $\SO(10)$ GUT. 
    \item
    We argued that the considered $\mathrm{E}_{6}$ model is essentially unique due to the inherent non-perturbativity of any non-minimal setup. As discussed in Section~\ref{sec:analysis-unification-perturbativity}, the proposed model itself is expected to be challenged by perturbativity. A dedicated perturbativity analysis might clarify
    the situation: finding small viable regions of parameter space may narrow the uncertainties in the proton decay estimates, but on the other hand may question the perturbative predictions of the model altogether if no such regions are found. 
\end{enumerate}

\section*{Acknowledgments}

The work of KSB is supported in part by the U.S.~Department of Energy under grant number
DE-SC0016013. BB acknowledges the financial support from the Slovenian Research Agency (research core funding No. P1-0035 and in part research grant J1-4389). VS acknowledges financial support from the Charles University Research Center Grant No.~UNCE 24/SCI/016. VS would like to thank Renato Fonseca for a discussion on the use of \texttt{GroupMath} software. 

\appendix


\section{State definitions \label{app:state-definitions}}

This appendix provides the explicit definitions of various states that are used in the paper. The basis for the fundamental representation $\mathbf{27}$ of $\mathrm{E}_{6}$ used for specifying the states is the one with well-defined quantum numbers under trinification, and the labels used are those of fermion fields in Eq.~\eqref{eq:particles-in-27}. We are using fermionic labels for scalar states, so we shall use the symbol $\simeq$ and interpret the relation as ``transforms as''. This is the same convention as used in~\cite{Babu:2023zsm}.

We label this basis of $\mathbf{27}$ generically as $\mathbf{e}_{i}$, where $i$ goes from 1 to 27, and the conjugate basis of the $\mathbf{27}^\ast$ as $\mathbf{e}^{i\ast}$. States in any other irreducible representation of $\mathrm{E}_{6}$ can then be written in a basis extended by tensoriality; given the index notation in Eq.~\eqref{eq:irrep-labels}, the basis for writing states in the $\mathbf{351}'$ is $\mathbf{e}_i\mathbf{e}_{j}\equiv \tfrac{1}{2}(\mathbf{e}_{i}\otimes \mathbf{e}_{j}+\mathbf{e}_{j}\otimes \mathbf{e}_{i})$, while the basis for writing states in the $\mathbf{650}$ is $\mathbf{e}_{i}\mathbf{e}^{j\ast}\equiv\mathbf{e}_{i}\otimes\mathbf{e}^{j\ast}$.

A note on the normalization convention: all the states are written in the normalization where the quadratic invariant of any irrep gives a coefficient $1/2$ for a real degree of freedom or $1$ for a complex degree of freedom. Specifically, if the field $\phi$ is in $\mathbf{27}$, $\mathbf{351}'$, or $\mathbf{650}$, then in the irrep notation of Eq.~\eqref{eq:irrep-labels} we respectively have 
\begin{align}
    \PSI^{i}\PSI_{i}^{\ast}=\ZETA |\phi|^{2} +\ldots, \\
    \THETA^{ij}\THETA_{ij}^{\ast}=\ZETA |\phi|^{2} +\ldots, \\
    \mathrm{Tr}(\mathbf{X}^{2})=\ZETA |\phi|^{2} +\ldots,
\end{align}
where $\ZETA=1/2$ or $1$ if $\phi$ is real or complex. This leaves the sign or phase to be set by our explicit definition. 

Finally, all states we define are color singlets, so all colored indices are implicitly assumed to be summed over, e.g.~$uu^{\ast}$ is understood as $\sum_{\alpha=1}^{3}u_{\alpha}(u_{\alpha})^{\ast}$.

\subsection{Definitions for $G$-singlets under different embeddings \label{app:state-definitions-G-singlets}}

\def\PaU{uu^{\ast}}
\def\PaD{dd^{\ast}}
\def\PaDP{d'd'^{\ast}}
\def\PaNPC{\nu'^{c}\nu'^{c\ast}}
\def\PaEP{e'e'^{\ast}}
\def\PaE{ee^{\ast}}
\def\PaEPC{e'^c e'^{c\ast}}
\def\PaNP{\nu'\nu'^{\ast}}
\def\PaN{\nu\nu^{\ast}}
\def\PaEC{e^{c}e^{c\ast}}
\def\PaNC{\nu^{c}\nu^{c\ast}}
\def\PaS{nn^{\ast}}
\def\PaUC{u^{c}u^{c\ast}}
\def\PaDC{d^{c}d^{c\ast}}
\def\PaDPC{d'^{c}d'^{c\ast}}

We give the definitions of the $G$-singlet in the irrep $\mathbf{650}$ of $\mathbf{E}_{6}$ for all cases of $G$-vacua generated from Table~\ref{tab:embeddings}. There is always one such $G$-singlet state, except for the case of trinification, where two singlets are reduced to one by imposing one of three discrete parities. 

\begin{table}
\begin{center}
\caption{The coefficient $\alpha_{i}$ for the $G$-singlet defined in Eq.~\eqref{eq:G-singlet-definition}, as well as the discrete parities this state preserves. \label{tab:G-singlets-for-embeddings}}
\begin{tabular}{lrccc}
     \toprule
        $G$-vacuum& $\alpha_{i}$ for $\tilde{s}_{G}$&$P_{LR}$ & $P_{CL}$ & $P_{CR}$ \\
     \midrule
        $3_C\,3_L\,3_R \rtimes\mathbb{Z}_{2}^{LR}$ &
            $\tfrac{1}{6\sqrt{3}}(+1,+1,+1,+1,+1,-2,-2,-2,-2,-2,-2)$
            & $\checkmark$& --- &---\\
        $3_C\,3_L\,3_R \rtimes\mathbb{Z}_{2}^{CL}$ &
            $\tfrac{1}{6\sqrt{3}}({-2,-2,+1,+1,+1,+1,+1,+1,+1,+1,+1})$
            & --- & $\checkmark$&---\\
        $3_C\,3_L\,3_R \rtimes\mathbb{Z}_{2}^{CR}$ &
            $\tfrac{1}{6\sqrt{3}}(+1,+1,-2,-2,-2,+1,+1,+1,+1,+1,+1)$
            & ---&--- & $\checkmark$\\
        $6_{CL}\,2_{R}$ &
            $\tfrac{1}{6\sqrt{30}}(+4,+4,-5,-5,+4,+4,-5,-5,-5,-5,+4)$
            & ---&$\checkmark$&---\\
        $6_{CL}\,2_{R'}$ &
            $\tfrac{1}{6\sqrt{30}}(+4,+4,+4,-5,-5,-5,-5,+4,+4,-5,-5)$
            & ---&$\checkmark$&---\\
        $6_{CR}\,2_{L}$ &
            $\tfrac{1}{6\sqrt{30}}(-5,+4,+4,+4,+4,-5,-5,-5,+4,+4,+4)$
            & ---&---&$\checkmark$\\
        $(10)\,1_{\psi}$ &
            $\tfrac{1}{12\sqrt{30}}(+5,-4,+5,+5,-4,+5,-4,-4,+5,+5,-40)$
            & $\checkmark$&---&---\\
        $(10)'\,1_{\psi'}$ &
            $\tfrac{1}{12\sqrt{30}}(+5,-4,-4,+5,+5,-4,-4,+5,-40,+5,+5)$
            & ---&---&---\\
     \bottomrule 
\end{tabular}
\end{center}
\end{table}

The generic definition of such a singlet $\tilde{s}_{G}$ is
\begin{align}
    \tilde{s}_G&\simeq \alpha_{1}(\PaU+\PaD)+\alpha_{2}\,\PaDP
        +\alpha_{3} \,\PaUC +\alpha_{4} \,\PaDC  +\alpha_{5} \,\PaDPC 
        +\alpha_{6} (\PaN+\PaE) +\nonumber \\ 
        &\quad +\alpha_{7} (\PaNP+\PaEP) +\alpha_{8} (\PaNPC+\PaEPC) 
        +\alpha_{9} \,\PaEC + \alpha_{10} \,\PaNC + \alpha_{11} \,\PaS, \label{eq:G-singlet-definition}
\end{align}
where the coefficients $\alpha_i$ for different cases of $G$ are given in Table~\ref{tab:G-singlets-for-embeddings}. The table also specifies which of the discrete parities $LR$, $CL$, and $CR$ is preserved by these
singlets, as determined by using their explicit definitions from Appendix~B in \cite{Babu:2023zsm}.

\subsection{Definitions for SM-singlets \label{app:state-definitions-SMsinglets}} 

We list here all SM-singlet states in the scalar irreps of our model. 

The definitions of the SM-singlet states from $\mathbf{27}\oplus\mathbf{351}'$ given in Table~\ref{tab:vevs-27-and-351p} are
\begingroup
\allowdisplaybreaks
\begin{align}
    V_{1}&\simeq \nu^c,\\
    V_{2}&\simeq n,\\
    W_{1}&\simeq \tfrac{2}{\sqrt{15}}\,d'_{i} d^{c}_{i}-\sqrt{\tfrac{3}{5}}\, (\nu\nu'^c+ee'^{c}),\\
    W_{2}&\simeq \tfrac{2}{\sqrt{15}}\,d'_{i} d'^{c}_{i}-\sqrt{\tfrac{3}{5}}\,(\nu'\nu'^c+e'e'^{c}),\\
    W_{3}&\simeq \nu^c \nu^c,\\
    W_{4}&\simeq \sqrt{2}\,\nu^c n,\\
    W_{5}&\simeq nn.
\end{align}
\endgroup

\def\PaU{uu^{\ast}}
\def\PaD{dd^{\ast}}
\def\PaDP{d'd'^{\ast}}
\def\PaNPC{\nu'^{c}\nu'^{c\ast}}
\def\PaEP{e'e'^{\ast}}
\def\PaE{ee^{\ast}}
\def\PaEPC{e'^c e'^{c\ast}}
\def\PaNP{\nu'\nu'^{\ast}}
\def\PaN{\nu\nu^{\ast}}
\def\PaEC{e^{c}e^{c\ast}}
\def\PaNC{\nu^{c}\nu^{c\ast}}
\def\PaS{nn^{\ast}}
\def\PaUC{u^{c}u^{c\ast}}
\def\PaDC{d^{c}d^{c\ast}}
\def\PaDPC{d'^{c}d'^{c\ast}}

The definitions of the SM-singlet states in the irrep $\mathbf{650}$ were already given in our previous work~\cite{Babu:2023zsm}. Nevertheless we reiterate them below for completeness: 
\begingroup
\allowdisplaybreaks
\begin{align}
s&\simeq \tfrac{1}{6\sqrt{3}}\big(
 (\PaU +\PaD +\PaDP) \nonumber\\ 
 &\qquad -2(\PaNPC +\PaEP +\PaE
  +\PaEPC +\PaNP +\PaN
  +\PaEC  +\PaNC +\PaS) \nonumber\\
 &\qquad +(\PaUC +\PaDC +\PaDPC) \big), \label{eq:650-SM-singlets-begin}\\
a&\simeq \tfrac{1}{6} \big(
(\PaU+\PaD+\PaDP)-(\PaUC+\PaDC+\PaDPC) \big),
\\
x_1&\simeq \tfrac{1}{6\sqrt{2}} \big(
2(\PaUC+\PaNPC+\PaEPC +\PaEC) \nonumber\\
 &\qquad -(\PaDC +\PaDPC +\PaEP +\PaE +\PaNP +\PaN +\PaNC +\PaS)
\big),\\
x_2&\simeq \tfrac{1}{2\sqrt{6}} \big(
(\PaDC + \PaEP + \PaNP + \PaNC)
- (\PaDPC + \PaE + \PaN + \PaS)
\big),\\
X_3&\simeq \tfrac{1}{2\sqrt{3}}(
e' e^{\ast} + \nu' \nu^{\ast} + \nu^{c} n^{\ast} + d'^{c} d^{c\ast}
),\\
y_1&\simeq \tfrac{1}{6\sqrt{2}} \big(
(2\PaNPC -\PaEP -\PaE)
  +(2 \PaEPC -\PaNP -\PaN)
  -2(2\PaEC  - \PaNC -\PaS)
\big),\\
y_2&\simeq \tfrac{1}{2\sqrt{6}}\big(
 -(\PaEP -\PaE)
  -(\PaNP -\PaN)
  +2(\PaNC -\PaS)
\big),\\
Y_3&\simeq \tfrac{1}{2\sqrt{3}} ( e' e^{\ast} + \nu' \nu^{\ast}-2 \nu^{c} n^{\ast}),\\
z&\simeq \tfrac{1}{6\sqrt{2}} \big(
(\PaU +\PaD + \PaNPC +\PaEP +\PaE
  +\PaEPC +\PaNP +\PaN) \nonumber\\
 &\qquad -2 (\PaDP + \PaEC  +\PaNC +\PaS)
\big). \label{eq:650-SM-singlets-end}
\end{align}
\endgroup

Note that the state $s$ corresponds to $\tilde{s}_{G}$ of Eq.~\eqref{eq:G-singlet-definition} with $G=3_C\,3_L\,3_R\rtimes\mathbb{Z}_{2}^{LR}$ in Table~\ref{tab:G-singlets-for-embeddings}. The above SM-singlets of the $\mathbf{650}$ are written in a basis that is adapted to trinification irreps. The $1_{R'}\subset 2_{R'}\subset 3_C\,3_L\,3_R$ origin of the SM-singlets is given in Table~\ref{tab:SM-singlets-in-650} on the left-hand side, and the same states adapted to the $(10)\,1_{\psi}$ basis and their $4_C\,2_L\,2_R\subset (10)\,1_{\psi}$ origin on the right-hand side. The latter case is important for identifying the $\psi$-charges of the states, and thus which VEVs can preserve spinorial parity $\mathbb{Z}_{2}^{\psi}$ in Section~\ref{sec:minimal-models-breaking}. We do not provide a basis adapted to any other $G$, since this is not pertinent for the purposes of this paper.

\begin{table}[htb]
\centering
\caption{The basis of SM-singlet states from Eq.~\eqref{eq:650-SM-singlets-begin}--\eqref{eq:650-SM-singlets-end}
adapted to trinification irreps (left) and $(10)\,1_{\psi}$-irreps (right). \label{tab:SM-singlets-in-650}}
\scalebox{0.96}{
\begin{tabular}[t]{llll}
    \toprule
        state&
        $R'$&
        $2_{R'}$&
        $3_C\,3_L\,3_R$\\
    \midrule
        $s$&$0$&$\mathbf{1}$&$(\mathbf{1},\mathbf{1},\mathbf{1})$\\
        $a$&$0$&$\mathbf{1}$&$(\mathbf{1},\mathbf{1},\mathbf{1})$\\
        $z$&$0$&$\mathbf{1}$&$(\mathbf{1},\mathbf{8},\mathbf{1})$\\
        $x_1$&$0$&$\mathbf{1}$&$(\mathbf{1},\mathbf{1},\mathbf{8})$\\
        $x_2$&$0$&$\mathbf{3}$&$(\mathbf{1},\mathbf{1},\mathbf{8})$\\
        $X_3$&$1$&$\mathbf{3}$&$(\mathbf{1},\mathbf{1},\mathbf{8})$\\
        $y_1$&$0$&$\mathbf{1}$&$(\mathbf{1},\mathbf{8},\mathbf{8})$\\
        $y_2$&$0$&$\mathbf{3}$&$(\mathbf{1},\mathbf{8},\mathbf{8})$\\
        $Y_{3}$&$1$&$\mathbf{3}$&$(\mathbf{1},\mathbf{8},\mathbf{8})$\\
     \bottomrule
\end{tabular}
\hspace{0.3cm}
\begin{tabular}[t]{lll}
    \toprule
        state & 
        $4_C\,2_{L}\,2_{R}$ &
        $(10)\,1_{\psi}$\\
    \midrule
        $\tfrac{1}{\sqrt{10}}s+\sqrt{\tfrac{3}{20}}z+\tfrac{\sqrt{3}}{4\sqrt{5}}x_{1}+\tfrac{3}{4\sqrt{5}}x_{2}-\tfrac{\sqrt{3}}{2\sqrt{5}}y_1+\tfrac{3}{2\sqrt{5}}y_2$&
        $(\mathbf{1},\mathbf{1},\mathbf{1})$&
        $(\mathbf{1},0)$\\[8pt]
        $\sqrt{\tfrac{3}{8}}x_1-\tfrac{1}{\sqrt{8}}x_2+\sqrt{\tfrac{3}{8}}y_1+\tfrac{1}{\sqrt{8}}y_2$&
        $(\mathbf{1},\mathbf{1},\mathbf{3})$&
        $(\mathbf{45},0)$\\[8pt]
        $-\tfrac{1}{\sqrt{2}}a+\tfrac{1}{2}z-\tfrac{1}{4}x_1-\tfrac{\sqrt{3}}{4}x_2$&
        $(\mathbf{15},\mathbf{1},\mathbf{1})$&
        $(\mathbf{45},0)$\\[8pt]
        $\sqrt{\tfrac{2}{5}}s-\tfrac{2}{\sqrt{15}}z-\tfrac{1}{\sqrt{15}}x_1-\tfrac{1}{\sqrt{5}}x_2-\tfrac{1}{2\sqrt{15}}y_1+\tfrac{1}{2\sqrt{5}}y_2$&
        $(\mathbf{1},\mathbf{1},\mathbf{1})$&
        $(\mathbf{54},0)$\\[8pt]
        $-\tfrac{1}{\sqrt{2}}a-\tfrac{1}{2}z+\tfrac{1}{4}x_1+\tfrac{\sqrt{3}}{4}x_2$&
        $(\mathbf{1},\mathbf{1},\mathbf{1})$&
        $(\mathbf{210},0)$\\[8pt]
        $\tfrac{1}{\sqrt{2}}s+\tfrac{1}{2\sqrt{3}}z+\tfrac{1}{4\sqrt{3}}x_1+\tfrac{1}{4}x_2+\tfrac{1}{2\sqrt{3}}y_1-\tfrac{1}{2}y_2$&
        $(\mathbf{15},\mathbf{1},\mathbf{1})$&
        $(\mathbf{210},0)$\\[8pt]
        $-\sqrt{\tfrac{3}{8}}x_1+\tfrac{1}{\sqrt{8}}x_2+\sqrt{\tfrac{3}{8}}y_1+\tfrac{1}{\sqrt{8}}y_2$&
        $(\mathbf{15},\mathbf{1},\mathbf{3})$&
        $(\mathbf{210},0)$\\[8pt]
        $\tfrac{1}{\sqrt{5}}X_3-\tfrac{2}{\sqrt{5}}Y_3$&
        $(\mathbf{\bar{4}},\mathbf{1},\mathbf{2})$&
        $(\mathbf{16},-3)$\\[8pt]
        $\tfrac{2}{\sqrt{5}}X_3+\tfrac{1}{\sqrt{5}}Y_3$&
        $(\mathbf{\bar{4}},\mathbf{1},\mathbf{2})$&
        $(\mathbf{144},-3)$\\
    \bottomrule
\end{tabular}
}
\end{table}

\subsection{Definitions for SM-doublets \label{app:state-definitions-SMdoublets}}

The definition for the EW VEVs of the EM-neutral components $\delta_{i}$ of the SM-doublets $(\mathbf{1},\mathbf{2},+1/2)$ from Table~\ref{tab:SM-doublets} are the following:
\begingroup
\allowdisplaybreaks
\begin{align}
    \delta_{1} & \simeq -\nu'^{\ast}, &
    \delta_{2} & \simeq \tfrac{1}{\sqrt{15}}(-2 d d^{c}+3n\nu'^c-3ee^c), &
    \delta_{3} & \simeq n\nu'^{c}+ee^c,\\
    \delta_{4} & \simeq -\nu^{\ast}, &
    \delta_{5} & \simeq \tfrac{1}{\sqrt{15}}(2dd'^{c}-3\nu^{c}\nu'^{c}+3e'e^{c}), &
    \delta_{6} & \simeq \nu^{c}\nu'^{c}+e'e^{c},\\
    \delta_{7} & \simeq \nu'^{c}, &
    \delta_{8} & \simeq \tfrac{1}{\sqrt{15}}(2uu^c-3\nu' n+3\nu\nu^{c})^\ast, &
    \delta_{9} & \simeq -\nu'^\ast n^\ast - \nu^\ast\nu^{c\ast},\\
    \delta_{10} & \simeq -\sqrt{2} \nu'^\ast\nu^{c\ast}, &
    \delta_{11} & \simeq -\sqrt{2} \nu^\ast n^\ast. &&&
\end{align}
\endgroup

\section{Mass matrices \label{app:masses}}

\subsection{Scalar masses after first-stage breaking \label{app:masses-scalars}}
\def\MWIDTH{15cm}
\def\SS{\tilde{s}}

This Appendix is supplementary material for Section~\ref{sec:two-stage-breaking-and-masses}. We give here the mass expressions for the scalars in $\mathbf{27}\oplus\mathbf{351'}$ after the first breaking stage in Eq.~\eqref{eq:breaking-pattern} for all novel symmetries $G$ in Eq.~\eqref{eq:G-choices}. These expressions depend on the VEV $\langle\mathbf{650}\rangle$ (labeled as $s$ or $\tilde{s}$), the pure mass terms $M^{2}_{\PSI}$ and $M^{2}_\THETA$, and on the double-primed parameters of $V_{\text{mix}}$ in Eq.~\eqref{eq:potential-mix}. Note that the value of $s$ or $\tilde{s}$ is independent of the parameters in $V_{\text{mix}}$;\footnote{For reasons of continuity with previous work, the label $s$ is used for the specific singlet under $3_C\,3_L\,3_R\rtimes\mathbb{Z}_{2}^{LR}$, while the general label $\tilde{s}$ is used for a singlet under any $G$-vacuum.} specifically, it is determined from $V_{650}$, and is thus related to the non-primed parameters $M_{X}$, $m_{i}$ and $\LAMBDA_{i}$ via the stationarity condition, cf.~\cite{Babu:2023zsm} for further details. The value of the VEV does, however, depend on the group $G$.

The mass expressions for scalars in the $\mathbf{650}$ are already given in~\cite{Babu:2023zsm}; we do not repeat them here and those fields do not mix with the ones in $\mathbf{27}\oplus\mathbf{351'}$ during first-stage breaking.

We separate the results into three subsections, based on the choice of intermediate symmetry $G$ from Eq.~\eqref{eq:G-choices}. We do not distinguish different $G$-embeddings in this section, since all embeddings for a given group $G$ are $\mathrm{E}_{6}$-equivalent, consult Section~\ref{sec:embeddings-intermediate}, 
and thus the mass expressions for all embeddings of a given group are the same. The dependence on the embedding is implicit in which SM irreps are present in a given irrep of $G$.

Finally, a comment on our convention for bases of mass matrices: most $G$-irreps are present in the scalar sector in a single copy, so the choice of basis is unambiguous (up to overall sign or phase); in cases where a mass matrix for a $G$-irrep $\mathbf{R}$ is $2\times 2$, the first row/column always refers to $\mathbf{R}$ from $\mathbf{27}$ and the second row/column to $\mathbf{R}$ from $\mathbf{351'}$.

\subsubsection{Scalar masses for $G=\SU(3)\times\SU(3)\times\SU(3)$}

First, we consider trinification with LR parity $3_C\,3_L\,3_R\ltimes \mathbb{Z}_{2}^{LR}$. The masses of the fields from Eqs.~\eqref{eq:decomposition-to333-27} and \eqref{eq:decomposition-to333-351p} are 
\begingroup
\allowdisplaybreaks
\begin{align}
M^{2}_{(\mathbf{3},\mathbf{3},\mathbf{1})}&=
    \resizebox{\MWIDTH}{!}{$
    \begin{pmatrix}
        M^2_{\PSI}
        + \tfrac{\MPP_{1} s}{6 \sqrt{3}}
        + \tfrac{s^2}{108} (54 \LAMBDAPP_{11}+\LAMBDAPP_{12}+19 \LAMBDAPP_{13}-8 \LAMBDAPP_{14})
    &
        -\tfrac{\MPP^{\ast}_{3} s}{2 \sqrt{5}}
        + \tfrac{s^2}{12 \sqrt{15}} \left(10 \LAMBDAPP^{\ast}_{10}+\LAMBDAPP^{\ast}_{7}-2 \LAMBDAPP^{\ast}_{8}-\LAMBDAPP^{\ast}_{9}\right)
    \\
        -\tfrac{\MPP_{3} s}{2 \sqrt{5}}
        + \tfrac{s^2}{12 \sqrt{15}} \left(10 \LAMBDAPP_{10}+\LAMBDAPP_{7}-2 \LAMBDAPP_{8}-\LAMBDAPP_{9} \right)
    &
        M^2_{\THETA}
        + \tfrac{\MPP_{2} s}{15 \sqrt{3}} 
        + \tfrac{s^{2}}{540} (270 \LAMBDAPP_{15}-\LAMBDAPP_{16}+8 \LAMBDAPP_{17}+27 \LAMBDAPP_{18}+98 \LAMBDAPP_{19}-199 \LAMBDAPP_{20}-10 \LAMBDAPP_{21})
    \\
    \end{pmatrix}
    $}, \label{eq:masses-333-begin}\\[6pt]
M^2_{(\mathbf{1},\mathbf{\bar{3}},\mathbf{3})}&=
    \resizebox{\MWIDTH}{!}{$
    \begin{pmatrix}
        M^2_{\PSI}
        -\tfrac{\MPP_{1} s}{3 \sqrt{3}}
        +\tfrac{s^2}{54} (27 \LAMBDAPP_{11}+2 \LAMBDAPP_{12}+11 \LAMBDAPP_{13}+11\LAMBDAPP_{14})
    &
        -\tfrac{\MPP^{\ast}_{3} s}{\sqrt{5}}
        +\tfrac{s^2}{6 \sqrt{15}} \left(10 \LAMBDAPP^{\ast}_{10}+\LAMBDAPP^{\ast}_{7}+\LAMBDAPP^{\ast}_{8}+2 \LAMBDAPP^{\ast}_{9}\right)
    \\
        -\tfrac{\MPP_{3} s}{\sqrt{5}}
        +\tfrac{s^2}{6 \sqrt{15}} (10 \LAMBDAPP_{10}+\LAMBDAPP_{7}+\LAMBDAPP_{8}+2 \LAMBDAPP_{9})
    &
        M^2_{\THETA}
        -\tfrac{2 \MPP_{2} s}{15 \sqrt{3}}
        +\tfrac{s^2}{270} (135 \LAMBDAPP_{15}+7 \LAMBDAPP_{16}+7 \LAMBDAPP_{17}+54 \LAMBDAPP_{18}+52 \LAMBDAPP_{19}-20 \LAMBDAPP_{20}+25 \LAMBDAPP_{21})
    \\
    \end{pmatrix}
    $}, \label{eq:masses-333-doublet133}\\[6pt]
M^{2}_{(\mathbf{6},\mathbf{6},\mathbf{1})}&=
    M^2_{\THETA}
    + \tfrac{\MPP_{2} s}{6 \sqrt{3}}
    + \tfrac{s^2}{108} (54 \LAMBDAPP_{15}+\LAMBDAPP_{16}+\LAMBDAPP_{17}+19 \LAMBDAPP_{19}+19 \LAMBDAPP_{20}-8 \LAMBDAPP_{21}),\\[6pt]
M^{2}_{(\mathbf{1},\mathbf{\bar{6}},\mathbf{6})} &=
    M^2_{\THETA}
    -\tfrac{\MPP_{2}s}{3 \sqrt{3}}
    +\tfrac{s^2}{54} (27 \LAMBDAPP_{15}+2 \LAMBDAPP_{16}+2 \LAMBDAPP_{17}+11 \LAMBDAPP_{19}+11 \LAMBDAPP_{20}+11 \LAMBDAPP_{21}),\\[6pt]
M^{2}_{(\mathbf{3},\mathbf{3},\mathbf{8})} &=
    M^2_{\THETA}
    -\tfrac{\MPP_{2}s}{12 \sqrt{3}}
    +\tfrac{s^2}{216} (108 \LAMBDAPP_{15}-4 \LAMBDAPP_{16}+5 \LAMBDAPP_{17}+41 \LAMBDAPP_{19}+14 \LAMBDAPP_{20}+14 \LAMBDAPP_{21}),\\[6pt]
M^{2}_{(\mathbf{8},\mathbf{\bar{3}},\mathbf{3})}&=
    M^2_{\THETA}
    +\tfrac{\MPP_{2}s}{6 \sqrt{3}}
    +\tfrac{s^2}{108} (54 \LAMBDAPP_{15}+\LAMBDAPP_{16}+\LAMBDAPP_{17}+19 \LAMBDAPP_{19}+\LAMBDAPP_{20}-8 \LAMBDAPP_{21}),
\end{align}
\begin{align}
    M^{2}_{(\mathbf{\bar{3}},\mathbf{1},\mathbf{\bar{3}})}
        &= M^{2}_{(\mathbf{3},\mathbf{3},\mathbf{1})},\\
    M^{2}_{(\mathbf{6},\mathbf{6},\mathbf{1})}
        &=M^{2}_{(\mathbf{\bar{6}},\mathbf{1},\mathbf{\bar{6}})},\\
    M^{2}_{(\mathbf{3},\mathbf{3},\mathbf{8})}
        &=M^{2}_{(\mathbf{\bar{3}},\mathbf{8},\mathbf{\bar{3}})}. \label{eq:masses-333-end}
\end{align}
\endgroup

This spectrum can be transformed to a trinification vacuum with a different preserved parity by reshuffling the labels of representations. In particular, the three parities themselves transform the representation $(\mathbf{R}_{C},\mathbf{R}_{L},\mathbf{R}_{R})$ of $3_C\,3_L\,3_R$ in the following way (cf.~Appendix~C in Ref.~\cite{Babu:2023zsm}):
\begin{align}
    P_{LR}:&\quad (\mathbf{R}_{C},\mathbf{R}_{L},\mathbf{R}_{R}) \mapsto (\mathbf{R}_{C}^\ast,\mathbf{R}_{R}^\ast,\mathbf{R}_{L}^\ast),\\
    P_{CL}:&\quad (\mathbf{R}_{C},\mathbf{R}_{L},\mathbf{R}_{R}) \mapsto (\mathbf{R}_{L},\mathbf{R}_{C},\mathbf{R}_{R}^\ast),\\
    P_{CR}:&\quad (\mathbf{R}_{C},\mathbf{R}_{L},\mathbf{R}_{R}) \mapsto (\mathbf{R}_{R},\mathbf{R}_{L}^\ast,\mathbf{R}_{C}).
\end{align}
The expressions of Eqs.~\eqref{eq:masses-333-begin}--\eqref{eq:masses-333-end} clearly correspond 
to a $LR$-symmetric spectrum. The expressions for the $CL$-symmetric vacuum are obtained by applying $P_{CR}$ on the irrep labels, while those for a $CR$-symmetric spectrum are obtained by applying $P_{CL}$.

\subsubsection{Scalar masses for $G=\SU(6)\times\SU(2)$}

The masses of the fields from Eqs.~\eqref{eq:decomposition-to62-27} and \eqref{eq:decomposition-to62-351p}
are the following:
\begingroup
\allowdisplaybreaks
\begin{align}
M^{2}_{(\mathbf{15},\mathbf{1})} &= 
    \resizebox{\MWIDTH}{!}{$
    \begin{pmatrix}
        M^2_{\PSI}
        + \tfrac{\MPP_{1}\SS}{3} \sqrt{\tfrac{2}{15}}
        + \tfrac{\SS^{2}}{270} (135 \LAMBDAPP_{11}+4 \LAMBDAPP_{12}+49 \LAMBDAPP_{13}+49\LAMBDAPP_{14})
    &
        -\tfrac{3 \MPP^{\ast}_{3}\SS}{5 \sqrt{2}}
        +\tfrac{\SS^{2}}{20 \sqrt{15}} (-44 \LAMBDAPP^{\ast}_{10}+\LAMBDAPP^{\ast}_{7}+\LAMBDAPP^{\ast}_{8}-4 \LAMBDAPP^{\ast}_{9})
    \\
        -\tfrac{3 \MPP_{3}\SS}{5 \sqrt{2}}
        +\tfrac{\SS^{2}}{20 \sqrt{15}}(-44 \LAMBDAPP_{10}+\LAMBDAPP_{7}+\LAMBDAPP_{8}-4 \LAMBDAPP_{9})
    &
        M^2_{\THETA}
        -\tfrac{7 \MPP_{2}\SS}{30 \sqrt{30}}
        +\tfrac{\SS^{2}}{5400} (2700 \LAMBDAPP_{15}+107 \LAMBDAPP_{16}+107 \LAMBDAPP_{17}+972 \LAMBDAPP_{18}+1007 \LAMBDAPP_{19}-2881 \LAMBDAPP_{20}-208 \LAMBDAPP_{21})
    \end{pmatrix}
    $},\\[6pt]
M^{2}_{(\mathbf{\bar{6}},\mathbf{2})} &=
    M^2_{\PSI}
    -\tfrac{\MPP_{1}\SS}{6} \sqrt{\tfrac{5}{6}}
    +\tfrac{\SS^{2}}{216} (108 \LAMBDAPP_{11}+5 \LAMBDAPP_{12}+41 \LAMBDAPP_{13}-40 \LAMBDAPP_{14}),\\[6pt]
M^{2}_{(\mathbf{\overline{21}},\mathbf{3})} &=
    M^2_{\THETA}
    -\tfrac{\MPP_{2}\SS}{6} \sqrt{\tfrac{5}{6}} 
    +\tfrac{\SS^{2}}{216} (108 \LAMBDAPP_{15}+5 \LAMBDAPP_{16}+5 \LAMBDAPP_{17}+41 \LAMBDAPP_{19}+41 \LAMBDAPP_{20}-40 \LAMBDAPP_{21}),\\[6pt]
M^{2}_{(\mathbf{84},\mathbf{2})} &=
    M^2_{\THETA}
    -\tfrac{\MPP_{2}\SS}{12 \sqrt{30}}
    +\tfrac{\SS^2}{2160} (1080 \LAMBDAPP_{15}-40 \LAMBDAPP_{16}+41 \LAMBDAPP_{17}+401 \LAMBDAPP_{19}-4 \LAMBDAPP_{20}-4 \LAMBDAPP_{21}),\\[6pt]
M^{2}_{(\mathbf{\overline{105'}},\mathbf{1})} &=
    M^2_{\THETA}
    +\tfrac{\MPP_{2}\SS}{3} \sqrt{\tfrac{2}{15}}
    +\tfrac{\SS^2}{270} (135 \LAMBDAPP_{15}+4 \LAMBDAPP_{16}+4 \LAMBDAPP_{17}+49 \LAMBDAPP_{19}+49 \LAMBDAPP_{20}+49 \LAMBDAPP_{21}).
\end{align}
\endgroup

\subsubsection{Scalar masses for $G=\SO(10)\times\UU$}

The masses of the fields from Eqs.~\eqref{eq:decomposition-to101-27} and \eqref{eq:decomposition-to101-351p} are the following:

\begin{align}
M^{2}_{(\mathbf{1},+4)} &=
    M^2_{\PSI}
    -\tfrac{\MPP_{1}\SS}{3} \sqrt{\tfrac{10}{3}}
    +\tfrac{\SS^{2}}{54}(27\LAMBDAPP_{11}+20 \LAMBDAPP_{12}+2\LAMBDAPP_{13}+2\LAMBDAPP_{14}),\\[6pt]
M^{2}_{(\mathbf{1},+8)} &=
    M^2_{\THETA}
    -\tfrac{\MPP_{2}\SS}{3} \sqrt{\tfrac{10}{3}}
    +\tfrac{\SS^2}{54} (27\LAMBDAPP_{15}+20 \LAMBDAPP_{16}+20 \LAMBDAPP_{17}+2\LAMBDAPP_{19}+2\LAMBDAPP_{20}+2\LAMBDAPP_{21}),\\[6pt]
M^{2}_{(\mathbf{10},-2)} &=
    \resizebox{\MWIDTH}{!}{$
    \begin{pmatrix}
        M^2_{\PSI}
        -\tfrac{\MPP_{1}\SS}{3 \sqrt{30}}
        +\tfrac{\SS^2}{540} (270 \LAMBDAPP_{11}+2 \LAMBDAPP_{12}+227 \LAMBDAPP_{13}+65 \LAMBDAPP_{14})
    &
        -\tfrac{3 \sqrt{3} \MPP^{\ast}_{3}\SS}{10}
        +\tfrac{\SS^2}{40 \sqrt{10}} (20 \LAMBDAPP^{\ast}_{10}+29 \LAMBDAPP^{\ast}_{7}+5 \LAMBDAPP^{\ast}_{8}+4 \LAMBDAPP^{\ast}_{9})
    \\
        -\tfrac{3 \sqrt{3} \MPP_{3}\SS}{10}
        +\tfrac{\SS^2}{40 \sqrt{10}} (20 \LAMBDAPP_{10}+29 \LAMBDAPP_{7}+5 \LAMBDAPP_{8}+4 \LAMBDAPP_{9})
    &
        M^2_{\THETA}
        -\tfrac{83 \MPP_{2}\SS}{60 \sqrt{30}}
        +\tfrac{\SS^2}{21600} (10800 \LAMBDAPP_{15}+665 \LAMBDAPP_{16}+3257 \LAMBDAPP_{17}+5832 \LAMBDAPP_{18}+4157 \LAMBDAPP_{19}-1027 \LAMBDAPP_{20}+1160 \LAMBDAPP_{21})
    \\
    \end{pmatrix}
    $}, \label{eq:condition-doublets-101}\\[6pt]
M^{2}_{(\mathbf{16},+1)} &=
    M^2_{\PSI}
    +\tfrac{\MPP_{1} \SS}{12} \sqrt{\tfrac{5}{6}}
    +\tfrac{\SS^2}{864} (432 \LAMBDAPP_{11}+5 \LAMBDAPP_{12}+41 \LAMBDAPP_{13}-40 \LAMBDAPP_{14}),\\[6pt]
M^{2}_{(\mathbf{16},+5)} &=
    M^2_{\THETA}
    -\tfrac{7\MPP_{2}\SS}{24} \sqrt{\tfrac{5}{6}}
    +\tfrac{\SS^{2}}{1728} (864 \LAMBDAPP_{15}-80 \LAMBDAPP_{16}+325 \LAMBDAPP_{17}+73 \LAMBDAPP_{19}-8 \LAMBDAPP_{20}-8 \LAMBDAPP_{21}),\\[6pt]
M^{2}_{(\mathbf{54},-4)} &=
    M^2_{\THETA}
    -\tfrac{\MPP_{2}\SS}{3 \sqrt{30}}
    +\tfrac{\SS^2}{540} (270 \LAMBDAPP_{15}+2 \LAMBDAPP_{16}+2 \LAMBDAPP_{17}+227 \LAMBDAPP_{19}+227 \LAMBDAPP_{20}+65 \LAMBDAPP_{21}),\\[6pt]
M^{2}_{(\mathbf{126},+2)} &=
    M^2_{\THETA}
    + \tfrac{\MPP_{2}\SS}{12} \sqrt{\tfrac{5}{6}} 
    + \tfrac{\SS^2}{864} (432 \LAMBDAPP_{15}+5 \LAMBDAPP_{16}+5 \LAMBDAPP_{17}+41 \LAMBDAPP_{19}+41 \LAMBDAPP_{20}-40 \LAMBDAPP_{21}),\\[6pt]
M^{2}_{(\mathbf{\overline{144}},-1)} &=
    M^2_{\THETA}
    +\tfrac{\MPP_{2}\SS}{24 \sqrt{30}}
    +\tfrac{\SS^2}{8640} (4320 \LAMBDAPP_{15}-40 \LAMBDAPP_{16}+41 \LAMBDAPP_{17}+2021 \LAMBDAPP_{19}-328 \LAMBDAPP_{20}+320 \LAMBDAPP_{21}).
\end{align}

\subsection{Gauge boson masses \label{app:masses-gauge}}

We provide below the gauge boson masses of all SM-irreps after the two stages of symmetry breaking from Eq.~\eqref{eq:breaking-pattern}. The general expressions are too complicated to write in full, so we rather provide an abbreviated version where only the VEVs considered in the $\text{ESH}+\mathbb{Z}_{2}^{\psi}$ scenario are engaged: the $\mathrm{E}_{6}$-breaking VEV from the irrep $\mathbf{650}$ in first-stage breaking (which is $G$-specific), and the $G$-breaking non-spinorial VEVs $V_{2}$ and $W_{2,3,5}$ from $\mathbf{27}\oplus\mathbf{351'}$ in second-stage breaking, cf.~Table~\ref{tab:vevs-27-and-351p}.

Note that we shall not treat the second-stage breaking as part of an effective theory, where the heavy GUT gauge bosons are integrated out and the couplings then run down in the intermediate theory to scale $\MI$. Instead, the calculation treats all the SM-singlet VEVs as engaged already at the GUT scale, but hierarchical values between them are assumed in accordance with two-stage breaking. This approximate one-scale procedure is sufficient for confirming that the choices of non-zero VEVs from Section~\ref{sec:minimal-models-breaking} indeed break all the way to the SM.

The mass of each SM-irrep of vector bosons $\mathbf{R}_V$ is given in two parts via
\begin{align}
    M^{2}_{\mathbf{R}_V}&=g^{2}\,(M^{2}_{\mathbf{R}_V;G}+M^{2}_{\mathbf{R}_V;I}), \label{eq:gauge-boson-masses-twopart}
\end{align}
where $g$ is the unified $\mathrm{E}_{6}$ gauge coupling. The contributions $M^{2}_{\mathbf{R}_V;G}$ are dependent on the $G$-embedding, contain the GUT-scale VEV $\tilde{s}$, and are listed in Table~\ref{tab:masses-gauge-G}. The contributions $M^{2}_{\mathbf{R}_V;I}$ on the other hand contain intermediate-scale VEVs from the second-stage breaking, do not depend on the $G$-embedding, and are listed below:
\begin{align}
    M^{2}_{(\mathbf{1},\mathbf{1},0)_V;I} &= 
        \scalebox{0.72}{$
        \begin{pmatrix}
            5 |\RED{W_{3}}|^{2} & \sqrt{\tfrac{5}{3}} |\RED{W_{3}}|^2 & 0 & 0 \\
            \sqrt{\frac{5}{3}} |\RED{W_{3}}|^{2} & \tfrac{4}{3}|\RED{V_{2}}|^{2}+\tfrac{4}{3} |\RED{W_{2}}|^{2} +\tfrac{1}{3}|\RED{W_{3}}|^2+\tfrac{16}{3} |\RED{W_{5}}|^{2} & 0 & 0 \\
            0 & 0 & \frac{1}{2}|\RED{V_{2}}|^{2}+\tfrac{1}{2}|\RED{W_{2}}|^{2}+|\RED{W_{3}}|^{2}+|\RED{W_{5}}|^{2} & 2 \RED{W_{3}} \RED{W_{5}}^{\ast} \\
            0 & 0 & 2 \RED{W_{3}}^{\ast} \RED{W_{5}} & \frac{1}{2}|\RED{V_{2}}|^{2}+\tfrac{1}{2}|\RED{W_{2}}|^{2}+|\RED{W_{3}}|^{2}+|\RED{W_{5}}|^{2} \\
        \end{pmatrix}
        $},\\[6pt]
    M^{2}_{(\mathbf{1},\mathbf{1},+1)_V;I} &= 
        \begin{pmatrix}
            |\RED{W_{3}}|^2 & 0 \\
            0 & \tfrac{1}{2} |\RED{V_{2}}|^{2}+ \tfrac{1}{2} |\RED{W_{2}}|^{2}+|\RED{W_{5}}|^{2}
        \end{pmatrix},\\[6pt]
    M^{2}_{(\mathbf{1},\mathbf{2},+1/2)_V;I} &= 
        \tfrac{1}{2} |\RED{V_{2}}|^{2} + \tfrac{1}{2} |\RED{W_{2}}|^{2}+ |\RED{W_{3}}|^{2}+ |\RED{W_{5}}|^{2},\\[6pt]
    M^{2}_{(\mathbf{3},\mathbf{2},-5/6)_V;I} &= 
        \tfrac{5}{6} |\RED{W_{2}}|^{2},\\[6pt]
    M^{2}_{(\mathbf{3},\mathbf{2},+1/6)_V;I} &= 
        \begin{pmatrix}
            \tfrac{5}{6} |\RED{W_{2}}|^{2} +|\RED{W_{3}}|^{2} & 0 \\
            0 & \tfrac{1}{2} |\RED{V_{2}}|^{2} + \tfrac{1}{2} |\RED{W_{2}}|^{2}+|\RED{W_{5}}|^{2} \\
        \end{pmatrix},\\[6pt]
    M^{2}_{(\mathbf{3},\mathbf{1},+1/3)_V;I} &= 
        \tfrac{1}{2} |\RED{V_{2}}|^{2} +\tfrac{1}{2} |\RED{W_{2}}|^{2} + |\RED{W_{3}}|^{2} + |\RED{W_{5}}|^{2},\\[6pt]
    M^{2}_{(\mathbf{3},\mathbf{1},-2/3)_V;I} &= 
        \begin{pmatrix}
            |\RED{W_{3}}|^{2} & 0 \\
            0 &  \tfrac{1}{2} |\RED{V_{2}}|^{2} +\tfrac{1}{2} |\RED{W_{2}}|^{2} + |\RED{W_{5}}|^{2} \\
        \end{pmatrix}.
\end{align}
The intermediate-scale VEVs are colored red as in Section~\ref{sec:minimal-models-Yukawa}. We omitted writing the 
SM gauge bosons $(\mathbf{8},\mathbf{1},0)$, $(\mathbf{1},\mathbf{3},0)$ and $(\mathbf{1},\mathbf{1},0)$, which are all massless before EW symmetry breaking. Together with the list of irreps in Table~\ref{tab:masses-gauge-G} they form the adjoint $\mathbf{78}$ of $\mathrm{E}_{6}$, as can be cross-checked by counting real degrees of freedom. 

\def\DIAG{\mathrm{diag}}

\begin{table}[htb]
\centering
\caption{The $G$-dependent parts of gauge boson masses in Eq.~\eqref{eq:gauge-boson-masses-twopart}. The VEV $\tilde{s}$ depends on the parameters in $V_{650}$ and is also $G$-dependent, see~\cite{Babu:2023zsm} for details. The masses depend only on the $G$-embedding and not on any choice of perserved parity, i.e.~not on the $G$-vacuum.\label{tab:masses-gauge-G}}
\begin{tabular}{lllll}
    \toprule
    &$3_C\,3_L\,3_R$&$6_{CL}\,2_{R}$&$6_{CR}\,2_{L}$&$(10)'\,1_{\psi'}$ \\
    \midrule
    $M^{2}_{(\mathbf{1},\mathbf{1},0)_V;G}$&
        $\DIAG(0,0,0,0)$&
        $\tfrac{9}{20}\tilde{s}^{2}\,\DIAG(0,0,1,1)$&
        $\DIAG(0,0,0,0)$&
        $\DIAG(0,0,0,0)$\\
    $M^{2}_{(\mathbf{1},\mathbf{1},+1)_V;G}$&
        $\DIAG(0,0)$&
        $\tfrac{9}{20}\tilde{s}^{2}\,\DIAG(0,1)$&
        $\DIAG(0,0)$&
        $\tfrac{9}{16}\tilde{s}^{2}\,\DIAG(1,1)$\\
    $M^{2}_{(\mathbf{1},\mathbf{2},+1/2)_V;G}$&
        $0$&
        $0$&
        $\tfrac{9}{20}\tilde{s}^{2}$&
        $\tfrac{9}{16}\tilde{s}^{2}$\\
    $M^{2}_{(\mathbf{3},\mathbf{2},-5/6)_V;G}$&
        $\tfrac{1}{3}\tilde{s}^{2}$&
        $\tfrac{9}{20}\tilde{s}^{2}$&
        $\tfrac{9}{20}\tilde{s}^{2}$&
        $\tfrac{9}{16}\tilde{s}^{2}$\\
    $M^{2}_{(\mathbf{3},\mathbf{2},+1/6)_V;G}$&
        $\tfrac{1}{3}\tilde{s}^{2}\,\DIAG(1,1)$&
        $\tfrac{9}{20}\tilde{s}^{2}\,\DIAG(1,0)$&
        $\tfrac{9}{20}\tilde{s}^{2}\,\DIAG(1,1)$&
        $\DIAG(0,0)$\\
    $M^{2}_{(\mathbf{3},\mathbf{1},+1/3)_V;G}$&
        $\tfrac{1}{3}\tilde{s}^{2}$&
        $\tfrac{9}{20}\tilde{s}^{2}$&
        $0$&
        $0$\\
    $M^{2}_{(\mathbf{3},\mathbf{1},-2/3)_V;G}$&
        $\tfrac{1}{3}\tilde{s}^{2}\,\DIAG(1,1)$&
        $\tfrac{9}{20}\tilde{s}^{2}\,\DIAG(1,0)$&
        $\DIAG(0,0)$&
        $\tfrac{9}{16}\tilde{s}^{2}\,\DIAG(1,1)$\\
    \bottomrule
\end{tabular}
\end{table}

As for the convention in which the matrices are written, the only ambiguity is for cases where the matrix is $2\times 2$ or larger; we organize the basis in those cases according to the transformation properties under $\SO(10)\times\UU_{\psi}$. For SM-irreps $(\mathbf{1},\mathbf{1},+1)_V$, $(\mathbf{3},\mathbf{2},+1/6)_V$ or $(\mathbf{3},\mathbf{1},-2/3)_V$, the mass matrices are $2\times 2$, and we always take the first state of the basis to be located in $(\mathbf{45},0)$ of $(10)\,1_{\psi}$, while the second state is from the spinorial $(\mathbf{16},-3)$ or the conjugate. The off-diagonal terms are zero, since non-spinorial and spinorial fields do not mix if $\mathbb{Z}_{2}^{\psi}$ is preserved. The basis for the $4\times 4$ mass matrix of SM-singlets $(\mathbf{1},\mathbf{1},0)_V$ is written in terms of generators $\{ \tfrac{1}{2\sqrt{10}}t_{\chi},\tfrac{1}{2\sqrt{6}}t_{\psi},\tfrac{1}{\sqrt{2}}t_{R}^{67+},\tfrac{1}{\sqrt{2}}t_{R}^{67-}\}$, where the coefficients provide proper normalization of the states, $t_{\chi}$ and $t_{\psi}$ are defined in Eqs.~\eqref{eq:generator-chi} and \eqref{eq:generator-psi}, while $t_{R}^{67\pm}:=t_{R}^{6}\pm i t_{R}^{7}$ when using the notation of Eq~\eqref{eq:e6-adjoint}.
Note that this basis is complex, so the $4\times 4$ matrix is Hermitian; it's block structure is again due to spinorial parity.


\section{Embeddings \label{app:embeddings}}

\subsection{Properties of subalgebra diagrams \label{app:subalgebra-diagram}}

Subalgebra diagrams were defined in Figure~\ref{figure:subgroup-diagram-definition} as a visual aid to regular embeddings of subgroups in $\mathrm{E}_{6}$.\footnote{For regular embeddings, the generators of subalgebras can be chosen to be a subset of all generators, provided these are written in a basis with well-defined quantum numbers under the generators of the Cartan subalgebra. This allows for the subalgebra diagram to be drawn unambiguously.} These diagrams were then used in the results of Table~\ref{tab:embeddings} and discussed more broadly throughout Section~\ref{sec:embeddings-intermediate}.

In this appendix, we elaborate on how subalgebra diagrams make certain subgroup properties visually manifest:
\begin{itemize}
    \item In the L and R panels of the diagram, cf.~Figure~\ref{figure:subgroup-diagram-definition}, the 3rd column corresponds to diagonal generators, while the first 2 columns correspond to generators which form raising and lowering operators. In particular, the operators in the 1st, 2nd and 3rd column rotate respectively between the $1$-$2$, $1$-$3$ and $2$-$3$ entries of a triplet.
    \item For subalgebras of rank $6$, the 3rd columns of the L and R panel must be colored.
    \item In the last panel, each row is a $\SU(3)_R$ triplet, and each column is a $\SU(3)_L$ triplet.
    \item Since $\SU(2)_L$ is always part of an intermediate symmetry $G$, the first row in the L panel will always be colored. Furthermore, in the last panel the first two rows form $\SU(2)_L$ doublets, so the coloring of the 1st row will have the same pattern as the 2nd row. 
    \item The raising/lowering generators of $\SU(2)_R$ and $\SU(2)_{R}'$ are in the 1st and 3rd row of the R panel, respectively. If $2_R\subset G$, then the 1st and 2nd column of the last panel must be colored in the same pattern. If $2_{R'}\subset G$, the 2nd and 3rd column of the last panel are colored the same. 
    \item  Whether a discrete parity maps $G$ into itself is also visually manifest:
        \begin{center}
        \begin{tabular}{p{3.5cm}@{$\quad\Leftrightarrow\quad$}p{10cm}}
            $G$ is $LR$-symmetric. & 
            The L and R panels are colored in the same pattern, and the last panel is colored symmetrically with respect to its up-left to down-right diagonal.\\ 
            $G$ is $CL$-symmetric. & The entire L panel is colored and the last panel has all rows colored in the same pattern.\\
            $G$ is $CR$-symmetric. & The entire R panel is colored and the last panel has all columns colored in the same pattern.\\
        \end{tabular}
        \end{center}
        These rules allow us to easily determine the discrete part $D$ that preserves $G$ from Table~\ref{tab:embeddings}. 
    \item Subalgebra diagrams help with the intuition behind the naming convention of Table~\ref{tab:embeddings}, in particular with the adjectives ``standard'', ``flipped'' and ``LR-flipped''. ``Flipping'' exchanges $\SU(2)_{R}$ rotations in the ``standard'' embedding for those of $\SU(2)_{R'}$; visually this manifests in exchanging the 1st and 3rd row of the R panel in the diagram, as well as exchanging the 1st and 3rd column in the third panel. ``LR-flipping'' 
    applies the parity $P_{LR}$ on the generators of the ``standard'' case, visually manifesting in exchanging the L and R panels, and reflecting the third panel across the up-left to down-right diagonal. Clearly, performing the same type of flipping twice has no effect.
    \par
    Let us briefly consider what happens to the ``standard'' cases of $G$ from Eq.~\eqref{eq:G-choices}. In the case of $G=[3][3][3]$, ``flipping'' and ``LR-flipping'' have no effect, while for $G=(10)[1]$ ``LR-flipping'' has no effect. 
    For $G=[6][2]$, we obtain a new embedding with either type, but applying both ``flipping'' and ``LR-flipping''  
    removes the $2_L$ subgroup from $G$, so this embedding no longer contains the SM group and is discarded. We see from this discussion that the two types of flipping operations enabled us to derive exactly the embedding possibilities from Table~\ref{tab:embeddings}.
    \item Counting the generators of $G$ from its \text{subalgebra diagram}: each colored square in the L and R panels counts as $1$ generator, each colored square in the last panel counts as $6$ generators ($t^{\alpha}{}_{aa'}$ and $\bar{t}_{\alpha}{}^{aa'}$ for fixed $a$ and $a'$). Finally, one adds $8$ (for the color generators) to the total.
\end{itemize}

\subsection{Embeddings of intermediate symmetries through fermions\label{app:embeddings-fermions}}

The list of embeddings of intermediate symmetry $G$ from Table~\ref{tab:embeddings} can be alternatively understood in terms of how the fermions in each generation of the $\mathbf{27}$ from Eqs.~\eqref{eq:fermion-content-27}--\eqref{eq:exotic-fermions} are embedded into $G$-irreps. 

\begin{table}[htb]
\centering
\caption{The identification of fermions in $(15,1)^{ab}$ and $(\bar{6},2)_{a}{}^{i}$ under different embeddings of $\SU(6)\times\SU(2)$, where the upper index $i$ is associated to the fundamental irrep $\mathbf{2}$ of $\SU(2)$ and the upper indices $a,b$ refer to the fundamental $\mathbf{6}$ of $\SU(6)$. We make use of subgroup inclusions $3_C\,3_L\subset 6_{CL}$ and $3_C\,3_R\subset 6_{CR}$ 
and the respective irrep branching rules $\mathbf{6}_{CL}\to (\mathbf{3}_C,\mathbf{1}_L)\oplus  (\mathbf{1}_C,\mathbf{3}_L)$ and $\mathbf{6}_{CR}\to  (\mathbf{\bar{3}}_C,\mathbf{1}_R)\oplus  (\mathbf{1}_C,\mathbf{\bar{3}}_R)$ to form a basis for the irrep $\mathbf{6}$, as indicated 
for each embedding. \label{tab:fermion-embeddings-62}}
\begin{tabular}{llcc}
\toprule
$[6][2]$ embedding & Basis for $\mathbf{6}$&$(\mathbf{15},\mathbf{1})$ & $(\mathbf{\bar{6}},\mathbf{2})$\\
\midrule
$6_{CL}\,2_R$& $(\mathbf{3}_C,\mathbf{3}_L)$&
    $\frac{1}{\sqrt{2}}\;\begin{pmatrix}
        0 & d'^c_3 & -d'^c_2 & u_1 & d_1 & d'_1 \\
        - d'^c_3 & 0 & d'^c_1 & u_2 & d_2 & d'_2 \\
        d'^c_2 & -d'^c_1 & 0 & u_3 & d_3 & d'_3 \\
        -u_1 & -u_2 & -u_3 & 0 & n & \nu  \\
        -d_1 & -d_2 & -d_3 & -n & 0 & e \\
        - d'_1 & -d'_2 & -d'_3 & -\nu  & -e & 0 \\
    \end{pmatrix}$&
    $\begin{pmatrix}
        d^c_1& u^c_1\\
        d^c_2& u^c_2 \\
        d^c_3& u^c_3 \\
        \nu'^{c} & e' \\
        e'^c & -\nu'  \\
        e^c & \nu^c \\
    \end{pmatrix}$\\
    \addlinespace[8pt]
$6_{CL}\,2_{R'}$& $(\mathbf{3}_C,\mathbf{3}_L)$&
    $\frac{1}{\sqrt{2}}\;\begin{pmatrix}
        0 & u^c_3 & -u^c_2 & u_1 & d_1 & d'_1 \\
        -u^c_3 & 0 & u^c_1 & u_2 & d_2 & d'_2 \\
         u^c_2 & -u^c_1 & 0 & u_3 & d_3 & d'_3 \\
        -u_1 & -u_2 & -u_3 & 0 & e^c & -e'^c \\
        -d_1 & -d_2 & -d_3 & -e^c & 0 & \nu'^c \\
        -d'_1 & -d'_2 & -d'_3 & e'^c & -\nu'^c & 0 \\
    \end{pmatrix}$&
    $\begin{pmatrix}
        d'^c_1 & d^c_1 \\
        d'^c_2 & d^c_2 \\
        d'^c_3 & d^c_3 \\
        e' & e \\
        -\nu'  & -\nu  \\
        \nu^c & n \\
    \end{pmatrix}$\\
    \addlinespace[8pt]
$6_{CR}\,2_{L}$& $(\mathbf{\bar{3}}_C,\mathbf{\bar{3}}_R)$&
    $\frac{1}{\sqrt{2}}\;\begin{pmatrix}
        0 & d'_{3} & -d'_{2} & u^{c}_{1} & -d^c_{1} & d'^c_{1} \\
        -d'_{3} & 0 & d'_{1} & u^{c}_{2} & -d^c_{2} & d'^c_{2} \\
       d'_{2} & -d'_{1} & 0 & u^{c}_{3} & -d^c_{3} & d'^c_{3} \\
        -u^{c}_{1} & -u^{c}_{2} & -u^{c}_{3} & 0 & n & -\nu ^c \\
        d^c_{1} & d^c_{2} & d^c_{3} & -n & 0 & e^c \\
        -d'^c_{1} & -d'^c_{2} & -d'^c_{3} & \nu ^c & -e^c & 0 \\
        \end{pmatrix}$&
        $\begin{pmatrix}
            u_1 & d_1 \\
            u_2 & d_2 \\
            u_3 & d_3 \\
            -e'^c & \nu'^c \\
            \nu' & e'  \\
            \nu & e  \\
        \end{pmatrix}$
        \\
    \bottomrule
\end{tabular}
\end{table}

The fermions are embedded in the following way, as cross-checked by explicit action of generators present in the subalgebra diagarams of Table~\ref{tab:embeddings}:
\begin{itemize}
\item
    In trinification $3_C\,3_L\,3_R$, the fermions decompose according to Eq.~\eqref{eq:decomposition-to333-27}, and their embedding into irreps is already given by Eq.~\eqref{eq:particles-in-27}. 
\item
    Under $\SU(6)\times\SU(2)$, the fermions of one generation are according to Eq.~\eqref{eq:decomposition-to62-27} located in $(\mathbf{15},\mathbf{1})$ and $(\mathbf{\bar{6}},\mathbf{2})$. How they are distributed among the two irreps depends on the embedding (standard, flipped or LR-flipped) and is specified in Table~\ref{tab:fermion-embeddings-62}. The minus signs for fermions are consistent with the convention of Eq.~\eqref{eq:particles-in-27} and can arise when  the fundamental irrep $\mathbf{2}$ of some $\SU(2)$ subgroup is present in its equivalent form of $\mathbf{\bar{2}}$.
    
\item 
    Under $\SO(10)\times\UU$, the fermions decompose according to Eq.~\eqref{eq:decomposition-to101-27}.
    The standard embedding $(10)\,1_{\psi}$ gives the decomposition
    \begin{align}
        (\mathbf{1},+4)&=\{n \},\label{eq:fermion-embedding-101-1}\\
        (\mathbf{10},-2)&= \{ d'_i, d'^c_{i}, \nu',e', \nu'^c, e'^c\}, \label{eq:fermion-embedding-101-10}\\
        (\mathbf{16},+1)&=\{u_i, d_i, d^c_i, u^c_i, \nu, e, \nu^c, e^c\}, \label{eq:fermion-embedding-101-16}
    \end{align}
    while the flipped embedding $(10)'\,1_{\psi'}$ gives
    \begin{align}
        (\mathbf{1},+4)&=\{e^c\},\label{eq:fermion-embedding-101p-1}\\
        (\mathbf{10},-2)&= \{d'_i,  u^c_i, \nu',e', \nu,e \}, \label{eq:fermion-embedding-101p-10}\\
        (\mathbf{16},+1)&=\{u_i, d_i, d^c_i,d'^{c}_i, \nu'^c,e'^c, \nu ^c, n\}, \label{eq:fermion-embedding-101p-16}
    \end{align}
    where the index $i=1,2,3$ goes over colors.
\end{itemize}

\subsection{Projection matrices for embeddings\label{app:embeddings-projection-matrices}}

For computing how the irreps of the various $G$-embeddings in Table~\ref{tab:embeddings} decompose into irreps of the SM, projection matrices prove a very useful tool, see e.g.~\cite{Slansky:1981yr}. The projection matrix $P^{G}_{H}$ essentially projects the weight of $G$ to the weight of the subgroup $H\subset G$ in their respective Dynkin bases (also known as $\omega$ basis); the Dynkin basis is dual to the $\alpha$ basis, and the convention for ordering the $\alpha$ basis is based on the standard numbering of nodes in the Dynkin diagram. 

Although not unique, one consistent possibility for $P^{G}_{\textrm{SM}}$ is as follows:
\begin{align}
    P^{3_{C}\,3_{L}\,3_{R}}_{3_C\,2_L\,1_{Y}} &=
        \begin{pmatrix}
            1 & 0 & 0 & 0 & 0 & 0 \\
            0 & 1 & 0 & 0 & 0 & 0 \\
            0 & 0 & 1 & 1 & 0 & 0 \\
            0 & 0 & \tfrac{1}{6} & -\tfrac{1}{6} & \tfrac{2}{3} & \tfrac{1}{3} \\
        \end{pmatrix}, &
    P^{6_{CL}\,2_{R}}_{3_C\,2_L\,1_{Y}} &=
        \begin{pmatrix}
            1 & 0 & 0 & 0 & 0 & 0 \\
            0 & 1 & 0 & 0 & 0 & 0 \\
            0 & 0 & 0 & 0 & 1 & 0 \\
            \tfrac{1}{6} & \tfrac{1}{3} & \tfrac{1}{2} & 0 & 0 & \tfrac{1}{2} \\
        \end{pmatrix}, \label{eq:projection-matrices-begin} \\[6pt]
    P^{6_{CL}\,2_{R'}}_{3_C\,2_L\,1_{Y}} &=
        \begin{pmatrix}
            1 & 0 & 0 & 0 & 0 & 0 \\
            0 & 1 & 0 & 0 & 0 & 0 \\
            0 & 0 & 0 & 1 & 0 & 0 \\
            -\tfrac{1}{3} & -\tfrac{2}{3} & -1 & -\tfrac{1}{2} & 0 & 0 \\
        \end{pmatrix}, &
    P^{6_{CR}\,2_{L}}_{3_C\,2_L\,1_{Y}} &=
        \begin{pmatrix}
            -1 & 0 & 0 & 0 & 0 & 0 \\
            0 & -1 & 0 & 0 & 0 & 0 \\
            0 & 0 & 0 & 0 & 0 & 1 \\
            -\tfrac{1}{6} & -\tfrac{1}{3} & -\tfrac{1}{2} & 0 & \tfrac{1}{2} & 0 \\
        \end{pmatrix},\\[6pt]
    P^{(10)\,1_{\psi}}_{3_C\,2_L\,1_{Y}} &=
        \begin{pmatrix}
            0 & 1 & 1 & 0 & 1 & 0 \\
            1 & 1 & 1 & 1 & 0 & 0 \\
            0 & 0 & 1 & 1 & 1 & 0 \\
            \tfrac{1}{3} & 0 & \tfrac{1}{2} & -\tfrac{1}{6} & \tfrac{1}{6} & 0 \\
        \end{pmatrix}, &
    P^{(10)'\,1_{\psi'}}_{3_C\,2_L\,1_{Y}} &=   
        \begin{pmatrix}
            0 & 1 & 1 & 0 & 1 & 0 \\
            1 & 1 & 1 & 1 & 0 & 0 \\
            0 & 0 & 1 & 1 & 1 & 0 \\
            -\tfrac{1}{6} & 0 & 0 & \tfrac{1}{12} & -\tfrac{1}{12} & \tfrac{1}{4} \\
        \end{pmatrix}. \label{eq:projection-matrices-end}
\end{align}
The last row in the matrices represents the embedding of the hypercharge $Y$ in the usual SM normalization, i.e.~$Y(e^{c})=+1$. These projection matrices can be directly used in the \texttt{GroupMath} computational package~\cite{Fonseca:2020vke}. 

\section{Computing short-distance factors for proton decay \label{app:proton-decay-gamma-factors}}

We provide details on the computation of the short-distance factors $A_{SR}$ and $A_{SL}$ for proton decay in Eqs.~\eqref{eq:proton-decay-1}--\eqref{eq:proton-decay-2} of Section~\ref{sec:analysis-proton}. These factors come from RG running of proton decay operators from the GUT scale, where they are generated, down to the scale $\MZ$, from which point the long-distance factor $R_{L}$ takes into account the running from the EW scale down to the proton scale.

The running of the proton decay couplings in the SM is well known in the literature~\cite{Abbott:1980zj,Babu:2015bna}, but our situation is slightly more complicated due to the two-stage breaking. 

In the SM, the two $B$-violating operators generated by mediation of a heavy gauge boson are the well known~\cite{Weinberg:1979sa,Wilczek:1979hc}
\begin{align}
	O^{(1)}_{ABCD}&:= \Big( d_{R\alpha A} u_{R \beta B}\Big)\Big(Q_{Li\gamma C}L_{LjD}\Big)\;\epsilon^{\alpha\beta\gamma}\epsilon^{ij}, \label{eq:proton-decay-operator-1}\\
	O^{(2)}_{ABCD}&:= \Big(Q_{Li\alpha A}Q_{Lj\beta B}\Big)\Big(u_{R \gamma C}e_{RD}\Big)\;\epsilon^{\alpha\beta\gamma}\epsilon^{ij}, \label{eq:proton-decay-operator-2}
\end{align}
where $\{\alpha,\beta,\gamma\}$ are $\SU(3)_{C}$ indices going from $1$ to $3$, $\{i,j\}$ are $\SU(2)_L$ indices going from $1$ to $2$, $\{A,B,C,D\}$ are family indices going from $1$ to $3$, and L/R indicate whether the Weyl fermion is left- or right-handed. As per convention in the literature, the $A_{SR}$ and $A_{SL}$ short-distance factors correspond to the running of operators $O^{(1)}$ and $O^{(2)}$, respectively.

Suppose we write the coefficient in front of the operator $O^{(m)}$ as $C^{m}$, where family indices are suppressed. 
The RG running of the coefficients can be approximated as~\cite{Abbott:1980zj}
    \begin{align}
		\tfrac{d}{dt} C^{m}&=\tfrac{1}{16\pi^2}\sum_{i} (-2)\gamma^{m}{}_{i}\, g_{i}^2 \, C^{m} = -\tfrac{1}{2\pi}\sum_{i} \gamma^{m}{}_{i} \,\alpha_{i}\, C^{m}, \label{eq:proton-C-RGE-low}
	\end{align}
where $i$ goes over all gauge factors in the SM group $3_C\,2_L\,1_Y$ (in the order of the factors) and the SM $\gamma$-coefficients take the values
\begin{align}
    \gamma^{m}{}_{i}&= 
        \begin{pmatrix}
            2&\frac{9}{4}&\frac{11}{20} \\
	        2&\frac{9}{4}&\frac{23}{20} \\
        \end{pmatrix}. 
    \label{eq:gamma-SM}
\end{align}
In the last column, the coefficients refer to $\UU_{Y}$ and $\alpha_{1}$ in the GUT normalization.

In Eq.~\eqref{eq:proton-C-RGE-low}, only $1$-loop diagrams involving gauge bosons are considered. Unlike scalars, gauge bosons will not cause operator mixing or change family index, so the RGE is valid independently for each entry of the rank $4$ tensor in family space $C^{m}$.

We now consider the intermediate theory, whose quantities we shall label by a tilde. The relevant proton decay operators are $\tilde{O}^{(n)}_{ABCD}$, which contain the SM-operators $O^{(m)}_{ABCD}$. The coefficients of the intermediate operators are denoted by $\tilde{C}^{n}$, where we again suppress family indices, and the gauge contribution to their RG running can be written as 
    \begin{align}
		\tfrac{d}{dt} \tilde{C}^{n}&= -\tfrac{1}{2\pi}\sum_{j} \tilde{\gamma}^{n}{}_{j} \,\tilde{\alpha}_{j}\, \tilde{C}^{n} \label{eq:proton-C-RGE-high}
	\end{align}
in analogy with Eq.~\eqref{eq:proton-C-RGE-low}, where the index $j$ now runs over the gauge factors of the intermediate group $G$, and the coupling constants $\tilde{\alpha}_{j}$ are taken accordingly. 

The procedure for RG-running the operators from the GUT scale is to first run $\tilde{C}^{n}$ from the scale $\MGUT$ to $\MI$, match the operators of the intermediate theory to the SM ones at scale $\MI$ via
\begin{align}
    C^{m}(t_{I})&= P^{m}{}_{n}\,\tilde{C}^{n}(t_{I}),\label{eq:proton-C-matching}
\end{align}
and then run $C^{m}$ in the SM from the scale $\MI$ to $\MZ$. Only simple tree-level matching of operators in Eq.~\eqref{eq:proton-C-matching} was considered. The solutions of RG Eqs.~\eqref{eq:proton-C-RGE-high} and \eqref{eq:proton-C-RGE-low} between relevant scales can be written as
\begin{align}
	\tilde{C}^{n}(t_{I})& = \tilde{A}^{n}\;\tilde{C}^{n}(t_{U}), \\
	C^{m}(t_{Z})& = A^{m}\;C^{m}(t_{I}), 
\end{align}
where the running factors are
\begin{align}
    \tilde{A}^{n}&=
        \exp\left[\frac{1}{2\pi}\sum_{j}\tilde{\gamma}^{n}{}_{j}\int_{t_I}^{t_U}
		\tilde{\alpha}_{j}(t)\,dt\right] 
        \quad \approx 
        \prod_{j} \left[\frac{\tilde{\alpha}_{j}^{-1}(t_{I})}{\tilde{\alpha}_{j}^{-1}(t_U)}\right]^{\tilde{\gamma}^{n}{}_{j}/\tilde{a}_{j}}, \label{eq:proton-C-factor-high}\\
    A^{m} &=
        \exp\left[\frac{1}{2\pi}\sum_{i}\gamma^{m}{}_{i}\int_{t_Z}^{t_I}
		\alpha_{i}(t)\,dt\right] 
		\quad\approx 
        \prod_{i} \left[\frac{\alpha_{i}^{-1}(t_{I})}{\alpha_{i}^{-1}(t_2)}\right]^{\gamma^{m}{}_{i}/a_{i}}.
        \label{eq:proton-C-factor-low}
\end{align}
We have access to two-loop running of $\alpha$ from Section~\ref{sec:analysis-unification-procedure}, so we shall numerically integrate to obtain the factors $\tilde{A}^{n}$ and $A^{m}$; the literature usually cites the second expressions where the one-loop solution to the running was inserted, with $\tilde{a}_{j}$ and $a_i$ the intermediate-theory and SM one-loop beta coefficients found in Table~\ref{tab:minimal-models} and Eq.\eqref{eq:coefficients-ab-SM}, respectively. 

Collecting all the running together, including the matching at $\MI$, the complete short-distance factors between the GUT and $Z$ scales become
\begin{align}
    A_S^{(m)}&=  A^{m}\,\left(\sum_{n}P^{m}{}_{n}\;\tilde{A}^{n}\right),
\end{align}
where $A^{m}$ and $\tilde{A}^{n}$ are obtained from Eqs.~\eqref{eq:proton-C-factor-low} and \eqref{eq:proton-C-factor-high},
the SM $\gamma$-factors are in Eq.~\eqref{eq:gamma-SM}, and the $G$-specific quantities $\tilde{\gamma}^{n}{}_{i}$ and $P^{m}{}_{n}$ were computed for our intermediate theories and are given in Table~\ref{tab:proton-gamma-factors}. The translation to the conventional notation used in Eqs.~\eqref{eq:proton-decay-1} and \eqref{eq:proton-decay-2} in Section~\ref{sec:analysis-proton-general} is
\begin{align}
    A_{SR}&\equiv A_{S}^{(1)}, &
    A_{SL}&\equiv A_{S}^{(2)}.
\end{align}

\begin{table}[htb]
    \centering
        \caption{The operators $\tilde{O}^{(n)}$ relevant for proton decay in the intermediate theories with symmetry $G$ (with all index contractions suppressed), along with the $\gamma$-coefficients $\tilde{\gamma}^{n}{}_{j}$ appearing in Eq.~\eqref{eq:proton-C-RGE-high} and \eqref{eq:proton-C-factor-high}. $P^{m}{}_{n}$ is the matching matrix to SM operators.
        \label{tab:proton-gamma-factors}}
    \begin{tabular}{lp{4cm}p{2cm}p{2cm}}
         \toprule
         $G$-embedding& operators $\tilde{O}^{(n)}$ & $\tilde{\gamma}^{n}{}_{j}$ & $P^{m}{}_{n}$ \\ 
         \midrule
         $3_C\,3_L\,3_R$&
            $\tilde{O}^{(1)}=(\mathbf{N}^\ast \mathbf{N}^\ast)(\mathbf{L}\mathbf{M})$\par 
            $\tilde{O}^{(2)}=(\mathbf{L}\mathbf{L})(\mathbf{N}^\ast \mathbf{M}^\ast)$&
            $\begin{pmatrix} 2&\tfrac{4}{3}&2\\ 2&2&\tfrac{4}{3}\\\end{pmatrix}$&
            $\begin{pmatrix} 1&0\\ 0&1\\\end{pmatrix}$ \\[12pt]
         $6_{CL}\,2_{R}$&
            $\tilde{O}^{(1)}=\tfrac{1}{2}(\mathbf{D}^\ast \mathbf{D}^\ast)(\mathbf{S}\mathbf{S})$&
            $\begin{pmatrix} \frac{35}{4}& \frac{9}{4}\\\end{pmatrix}$&
            $\begin{pmatrix} 1\\1\\\end{pmatrix}$ \\[12pt]
         $6_{CR}\,2_{L}$&
            $\tilde{O}^{(1)}=\tfrac{1}{2}(\mathbf{D}\mathbf{D})(\mathbf{S}^\ast \mathbf{S}^\ast)$&
            $\begin{pmatrix} \frac{35}{4}& \frac{9}{4}\\\end{pmatrix}$&
            $\begin{pmatrix} 1\\1\\\end{pmatrix}$ \\
         \bottomrule
    \end{tabular}
\end{table}

\noindent
We conclude with some remarks regarding Table~\ref{tab:proton-gamma-factors}:
\begin{itemize}
    \item Only relevant $G$-embeddings are considered. Since only diagrams with gauge bosons are considered in the running of $\tilde{O}^{(n)}$, parities play no role and all trinification cases give expressions with the same coefficients $\tilde{\gamma}^{n}{}_{j}$ and $P^{m}{}_{n}$, the difference manifesting only in the different running of gauge couplings. The case $(10)'\,1_{\psi'}$ is not listed, because the proton-decay mediators appear already at the intermediate scale and thus only the factors $A^{m}$ from RG running in SM need to be considered, i.e.~$A^{(m)}_S=A^{m}$ for $G=(10)'\,1_{\psi'}$.  Table~\ref{tab:proton-gamma-factors} together with the stated considerations account for all cases of $G$-theories in Table~\ref{tab:minimal-models}.
    \item In the trinification case, there are two relevant operators for gauge-mediated proton decay. Using $\{\mathbf{L},\mathbf{M},\mathbf{N}\}$ as the notation for fermion irreps from Eq.~\eqref{eq:particles-in-27}, the operators can be written explicitly as 
        \begin{align}
            \tilde{O}^{(1)}_{ABCD}&= 
                (\mathbf{N}^{\ast}{}^{a'\alpha}{}_{A}\, \mathbf{N}^{\ast}{}^{b'\beta}{}_{B})
                (\mathbf{L}^{\gamma a}{}_{C}\, \mathbf{M}_{a}{}^{c'}{}_{D})
                \,\epsilon_{\alpha\beta\gamma}\,\epsilon_{a'b'c'},\\
            \tilde{O}^{(2)}_{ABCD}&=
                (\mathbf{L}^{\alpha a}{}_{A}\, \mathbf{L}^{\beta b}{}_{B})
                (\mathbf{N}^{\ast}{}^{a'\gamma}{}_{C} \, \mathbf{M}^{\ast}{}^{c}{}_{a'D})
                \,\epsilon_{\alpha\beta\gamma}\,\epsilon_{abc},
        \end{align}
    where Weyl indices are suppressed and contraction is indicated by parentheses; complex conjugation implies right-handed fermions. The used indices are $\{\alpha,\beta,\gamma\}$ for $\SU(3)_C$, $\{a,b,c\}$ for $\SU(3)_L$, and $\{a',b',c'\}$ for $\SU(3)_R$, with all indices taking values between $1$ and $3$, while $\{A,B,C,D\}$ are family indices. The Levi-Civita anti-symmetric tensor is denoted by $\epsilon$. The matching matrix $P^{m}{}_{n}$ can be derived by noting from Eq.~\eqref{eq:particles-in-27} that from left to right each field in $O^{(1,2)}$ is located in $\tilde{O}^{1,2}$ factor-by-factor, with overall coefficients matching up to an irrelevant sign. 
    \item For the cases $6_{CL}\,2_{R}$ and $6_{CR}\,2_{L}$, a single effective operator of the intermediate theory generates both SM operators $O^{(m)}$ in Eqs.~\eqref{eq:proton-decay-operator-1} and \eqref{eq:proton-decay-operator-2}. We denote $\mathbf{S}^{ab}\sim (\mathbf{15},\mathbf{1})$ and $\mathbf{D}_{a}{}^{i}\sim (\mathbf{\bar{6}},\mathbf{2})$ for the singlet and doublet in $\SU(6)\times\SU(2)$, so the operator $\mathcal{O}^{(1)}$ in $6_{CL}\,2_{R}$ is written explicitly as 
        \begin{align}
            \mathcal{O}^{(1)}_{ABCD} &= \tfrac{1}{2}
                (\mathbf{D}^{\ast}{}_{abA} \,\mathbf{D}^{\ast}{}_{cdB})
                (\mathbf{S}_{e}{}^{i}{}_{C} \,\mathbf{S}_{f}{}^{j}{}_{D})\,
                \epsilon^{abcdef}\,\epsilon_{ij},
        \end{align}
    where $\{a,b,c,d,e,f\}$ are $\SU(6)$ indices running from $1$ to $6$, $\{i,j\}$ are $\SU(2)$ fundamental indices, and $\{A,B,C,D\}$ are family indices. An analogous contraction happens in the operator for $6_{CR}\,2_{L}$. Table~\ref{tab:fermion-embeddings-62} confirms factor-by-factor that $\tilde{O}^{(1)}$ generates $O^{(1),(2)}$ for both $6_{CL}\,2_{R}$ and $6_{CR}\,2_{L}$, and the prefactor $1/2$ in the definition of $\tilde{O}^{(1)}$ ensures the SM operators are generated from it with coefficient $\pm 1$. 
    \item The $\gamma$-factors in Table~\ref{tab:proton-gamma-factors} for intermediate theories were computed along the lines of the SM calculation~\cite{Abbott:1980zj}. The generalized procedure ultimately boils down to a simple algorithm, which we now describe for the sake of concreteness and its potential usefulness to the reader.
    \par
    Suppose we have a semi-simple gauge group $G=\prod_{j} \SU(N_j)$ and are interested in proton decay mediated by gauge bosons of $G$. The four-fermion operators, constructed after the heavy gauge bosons are integrated out, will necessarily have two fermions transforming as left-handed Weyl spinors, i.e.~$(\tfrac{1}{2},0)$ under Lorentz, and two fields transforming as right-handed Weyl spinors. After the gamma matrices from the gauge currents are Fierzed away, the operator will be written as $(\mathbf{R}_{1}\mathbf{R}_{2})(\mathbf{R}_{3}^\ast \mathbf{R}_{4}^\ast)$, where each $\mathbf{R}_{k}$ is an irreducible irrep of $G$, i.e.~is a product of irreps for each $\SU(N)$ factor in the form $\mathbf{R}_{k}=\otimes_j \mathbf{R}_{kj}$. Assume also that there is a single gauge-invariant operator one can write down from these representations, so that no operator mixing is induced by one-loop diagrams involving gauge bosons. 
    The $\gamma$ factors for the operator are then computed via the expression
    \begin{align}
        \gamma_{j}&= 
            -\tfrac{1}{2} \sum_{k=1}^{4} C_{j}(\mathbf{R}_{k})
            + 2 \sum_{(k,l)\in \mathcal{S}_{\rightrightarrows}} \!\! n_j(\mathbf{R}_{k})\,n_j(\mathbf{R}_{l})\,(1+1/N_{j})
            + \tfrac{1}{2} \sum_{(k,l)\in \mathcal{S}_{\rightleftarrows}} \!\! n_j(\mathbf{R}_{k})\,n_j(\mathbf{R}_{l})\,(1+1/N_{j}), \label{eq:gamma-factor-recipe}
    \end{align}
    where $C_{j}$ is the Casimir factor of the $G$-irrep with respect to the factor $\SU(N_{j})$, and the factor $n_j$ of an irrep is equal to the number of (anti)fundamental $\SU(N_{j})$ indices it carries, i.e.~$n_j$ takes values $0$, $1$, or $2$ for an irrep which is a singlet, fundamental, or two-index antisymmetric under $\SU(N_j)$.     
    The sets $\mathcal{S}_{\rightrightarrows}$ and $\mathcal{S}_{\rightleftarrows}$ refer to those index-pairs connecting 
    the same or opposite type of fermion arrows; explicitly
    \begin{align}
        \mathcal{S}_{\rightrightarrows}&:=\{(1,2),(3,4)\},\\
        \mathcal{S}_{\rightleftarrows}&:=\{(1,3),(1,4),(2,3),(2,4)\}.
    \end{align}
    The three types of contributions in Eq.~\eqref{eq:gamma-factor-recipe} are coming in order from the three types of diagrams schematically shown in Figure~\ref{fig:operator-diagrams}. The gauge boson connects either to the same fermion leg (left diagram), to different fermion legs of same-arrow type (middle diagram), or to different legs of opposite-arrow type (right diagram).
    \par
    For all cases from Table~\ref{tab:proton-gamma-factors}, the assumptions of Eq.~\eqref{eq:gamma-factor-recipe} apply; these include having $\SU(N)$ factors with use of particular irreps, and that a single gauge-invariant operator can be constructed out of the four fermion irreps. The same expression can also be cross-checked on the SM result in Eq.~\eqref{eq:gamma-SM} for the operators in Eqs.~\eqref{eq:proton-decay-operator-1} and \eqref{eq:proton-decay-operator-2}.  
    For the $\UU_Y$ factor in the SM group, Eq.~\eqref{eq:gamma-factor-recipe} is still valid when taking the replacements $C_{j}(\mathbf{R}_k)\mapsto Y_{k}^{2}$, $n_j(\mathbf{R}_k)=Y_k$ and $N_{j}\mapsto 1$, where $Y_{k}$ denotes the hypercharge (in the GUT normalization) of the irrep $\mathbf{R}_{k}$.
\end{itemize}

\begin{figure}[htb]
    \centering
    \includegraphics[width=0.9\linewidth]{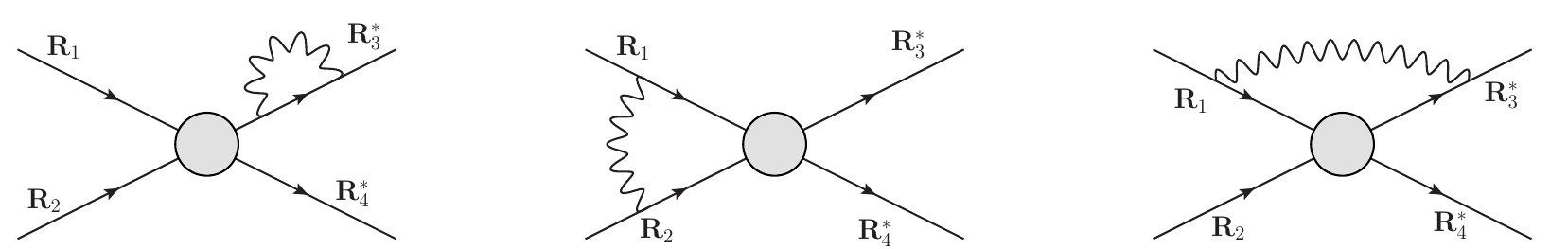}
    \caption{The three types of diagrams contributing to the RG running of a proton decay operator in Eq.~\eqref{eq:gamma-factor-recipe}. \label{fig:operator-diagrams}}
\end{figure}

\bibliographystyle{JHEP}
\providecommand{\href}[2]{#2}\begingroup\raggedright\endgroup


\end{document}